\documentclass[aps,prb,reprint,10pt,superscriptaddress,amssymb,amsfonts]{revtex4-1}
\usepackage{amsmath,amssymb,amsthm,amscd,latexsym,epsfig,bm}

\usepackage[colorlinks,bookmarks=false,citecolor=blue,linkcolor=red,urlcolor=blue]{hyperref}
\usepackage{verbatim}

\newcommand{\zz}{\mathbb{Z}_2}
\newcommand{\z}{\mathbb{Z}}

\def\cC{\mathcal{C}}

\def\cQ{\mathcal{Q}}
\def\tcQ{\tilde{\cQ}}

\theoremstyle{remark}
\newtheorem*{rem*}{\protect\remarkname}
\newtheorem{claim}{\protect\claimname}
\theoremstyle{plain}

\theoremstyle{plain}

\theoremstyle{plain}

\theoremstyle{plain}

  \providecommand{\claimname}{Claim}
  \providecommand{\lemmaname}{Lemma}
  \providecommand{\propositionname}{Proposition}
\providecommand{\theoremname}{Theorem}
\providecommand{\theoremtextname}{Theorem}
  \providecommand{\remarkname}{Remark}

\begin{document}

\title{Building crystalline topological phases from lower-dimensional states}

\author{Sheng-Jie Huang} \thanks{These authors contributed equally to this work.}
\affiliation{Department of Physics, University of Colorado, Boulder, Colorado 80309, USA}
\affiliation{Center for Theory of Quantum Matter, University of Colorado, Boulder, Colorado 80309, USA}
\author{Hao Song} \thanks{These authors contributed equally to this work.}
\affiliation{Departamento de F\'{\i}sica Te\'{o}rica I, Universidad Complutense, 28040 Madrid, Spain}
\author{Yi-Ping Huang}
\affiliation{Department of Physics, University of Colorado, Boulder, Colorado 80309, USA}
\affiliation{Center for Theory of Quantum Matter, University of Colorado, Boulder, Colorado 80309, USA}
\author{Michael Hermele}
\affiliation{Department of Physics, University of Colorado, Boulder, Colorado 80309, USA}
\affiliation{Center for Theory of Quantum Matter, University of Colorado, Boulder, Colorado 80309, USA}
\date{\today}

\begin{abstract}
We study the classification of symmetry protected topological (SPT) phases with crystalline symmetry (cSPT phases).  Focusing on bosonic cSPT phases in two and three dimensions, we introduce a simple family of cSPT states, where the system is comprised of decoupled lower-dimensional building blocks that are themselves SPT states.  We introduce a procedure to classify these block states, which surprisingly reproduces a classification of cSPT phases recently obtained by Thorngren and Else using very different methods, for all wallpaper and space groups.  The explicit constructions underlying our results clarify the physical properties of the phases classified by Thorngren and Else, and expose additional structure in the classification.  Moreover, the states we classify can be completely characterized by point group SPT (pgSPT) invariants and related weak pgSPT invariants that we introduce. In many cases, the weak invariants can be visualized in terms of translation-symmetric stacking of lower-dimensional pgSPT states.  We apply our classification to propose a Lieb-Shultz-Mattis type constraint for two-dimensional spin systems with only crystalline symmetry, and establish this constraint by a dimensional reduction argument.  Finally, the surprising matching with the Thorngren-Else classification leads us to conjecture that all SPT phases protected only by crystalline symmetry can be built from lower-dimensional blocks of invertible topological states.  We argue that this conjecture holds if we make a certain physically reasonable but unproven assumption.
\end{abstract}

\maketitle

\tableofcontents

\section{Introduction}
\label{sec:intro}

\subsection{Background and overview}
\label{subsec:mainintro}

Symmetry protected topological (SPT) phases \cite{gu09tensor,pollmann10,fidkowski11,turner11,chen11a,schuch11,chen13symmetry,levin12braiding}  are a generalization of topological band insulators \cite{hasan10,hasan11,qi11} and other free-fermion topological phases\cite{kitaev09,ryu10} to  interacting systems.  SPT phases have an energy gap and a unique ground state with periodic boundary conditions, lack spontaneous symmetry breaking, and are adiabatically connected to a trivial product wave function if the symmetries of the system are broken explicitly.  There are many SPT phases requiring strong interactions to exist.\cite{chen13symmetry} Following rapid progress over the past few years, much is now understood about the classification and characterization of SPT phases protected by internal symmetry, such as charge conservation, ${\rm SU}(2)$ spin rotation, or time reversal.\cite{turner13review,senthil15review}

Another important class of symmetries are those of crystalline lattices, which play a crucial role in many phenomena in solids.  However, compared to their internal symmetry cousins, SPT phases protected by crystalline symmetry, which we dub crystalline SPT (cSPT) phases, are much less understood.  An important exception are cSPT phases in non-interacting fermion systems, including topological crystalline insulators (see~[\onlinecite{ando15topological}] and references therein).  A number of works have studied  examples or families of interacting cSPT phases,\cite{chen13symmetry,chsieh14symmetry,you14symmetry,chsieh14CPT,isobe15,cho15,yoshida15bosonic,kapustin15fermionic,ware15topological,yqi15anomalous,yoshida15correlation,morimoto15breakdown,fuji15distinct,cheng16translational,lapa16interaction,chsieh16global,lake16anomalies}
 and there is a general theory of cSPT phases in one spatial dimension ($d=1$).\cite{gu09tensor,pollmann10,chen11a,schuch11}  However, until very recently, general approaches to interacting cSPT phases have been lacking.

This situation is now changing, and recent works have made progress in  classifying and characterizing general cSPT phases.  Enabled by ideas introduced by Isobe and Fu to study surfaces of interacting topological crystalline insulators,\cite{isobe15} in Ref.~\onlinecite{song17topological}, some of the authors of this paper (H. S., S.-J. H. and M. H.) and Fu devised an approach to classify SPT phases protected by crystalline point group symmetry, or point group SPT (pgSPT) phases.\footnote{By point group symmetry, we more precisely mean a group of symmetries leaving a single point in space fixed.  Such symmetries are more properly called site symmetries, but we abuse terminology slightly to employ the more evocative term ``point group.''}  It was shown that a pgSPT ground state can be adiabatically continued to a state defined on a lower-dimensional space, where point group operations act as internal symmetries.  This observation was used to classify pgSPT phases for a few examples of point group symmetry, and  implies that any pgSPT phase can be built out of lower-dimensional topological states, on which certain point group operations act as internal symmetries.   These ideas were extended to treat glide reflection symmetry by Lu, Shi and Lu.\cite{lu17classification}  A discussion of non-interacting topological crystalline insulators with some connections to the above works appeared in Ref.~\onlinecite{fulga16coupled}.

The approach of Ref.~\onlinecite{song17topological} cannot be directly applied for space group\footnote{In this paper, we use the term space group to refer to the symmetry group of a crystalline lattice in an arbitrary number of spatial dimensions, as well as more specifically for three-dimensional crystals.  When we specifically discuss two-dimensional crystals, the symmetry groups are referred to as wallpaper groups.} symmetry, for reasons discussed in Sec.~\ref{sec:general}.
  However, Ref.~\onlinecite{song17topological} did discuss how to use pgSPT classification to give constructions and partial classifications of non-trivial space group cSPT phases.

Even more recently, in a remarkable development, Thorngren and Else extended the idea of gauging symmetry to crystalline symmetries.\cite{thorngren16gauging}  This idea has been very powerful in the study of internal-symmetry topological phases, but it had not been clear if it could be generalized to spatial symmetry.  Ref.~\onlinecite{thorngren16gauging} argued that many bosonic cSPT phases in $d$ dimensions are classified by the group cohomology $H^{d+1}(G, {\rm U}(1))$, where orientation-reversing operations in $G$ act non-trivially on the ${\rm U}(1)$ coefficients.  This agrees with results from a tensor-network approach to construction of SPT states in Ref.~\onlinecite{jiang17anyon}.  Thorngren and Else gave classifications of bosonic cSPT phases for all 17 wallpaper groups in two dimensions ($d=2$), and almost all 230 space groups in three dimensions ($d=3$).  They also discussed some examples of fermionic cSPT phases.  While the ideas underlying the Thorngren-Else classification are quite physical, the classification procedure itself is rather formal, and the physical properties of the states classified are not yet clear.

In this paper, we tie these developments together, focusing on bosonic cSPT phases.  For simplicity, we focus on ``integer spin'' bosonic systems, meaning more precisely that we take the microscopic degrees of freedom to transform linearly (\emph{i.e.}, not projectively) under the crystalline symmetry.  We consider a particularly simple family of cSPT states, where the system is comprised of decoupled lower-dimensional ``building blocks,'' which are themselves lower-dimensional invertible topological phases.  The cSPT ground state is obtained by taking the product of ground states for the individual blocks.

Focusing on the case where the building blocks are lower-dimensional SPT states, we introduce a procedure to classify cSPT block states for all wallpaper and space groups, and reproduce the Thorngren-Else classification.\footnote{We reproduce the Thorngren-Else classification in the sense that we obtain the same result for all 17 wallpaper groups and all 230 space groups.  We have not established a direct link between our approach and that of Ref.~\onlinecite{thorngren16gauging}, beyond checking case-by-case; for instance, we have not shown directly that our approach gives a $H^{d+1}(G, {\rm U}(1))$ classification.  It would be interesting to find such a link, a problem we leave for future work.}  This leads us to conjecture that all cSPT phases protected only by crystalline symmetry can be obtained from lower-dimensional building blocks.  This conjecture is further supported by a general argument that rests on a physically reasonable but unproven hypothesis.

More generally, the building blocks can be ground states of an invertible topological phase, which need not be a SPT phase.  In particular, two-dimensional building blocks of three-dimensional cSPT phases can be $E_8$ states.\cite{song17topological}  The $E_8$ state is an analog of an integer quantum Hall state for bosonic systems, and is characterized by a unique ground state on the torus, the absence of bulk anyon excitations, and edge modes with chiral central charge $c = 8$.\cite{kitaev11KITP}  Non-trivial cSPT phases can be obtained for instance by placing $E_8$ states on mirror\cite{song17topological} or glide\cite{lu17classification} planes, and these phases are beyond the Thorngren-Else classification.  We leave discussion of these cSPT phases for future work.

Our results clarify the physical nature of the Thorngren-Else states, all of which are adiabatically connected to cSPT block states. This provides a starting point for future analysis of physical properties.  One application discussed here is a Lieb-Schultz-Mattis (LSM) type constraint \cite{lieb61} applicable to two-dimensional spin systems.  Our LSM constraint goes beyond other related results \cite{oshikawa00commensurability, misguich02degeneracy, hastings04Lieb, chen11a, parameswaran13topological,roy12space,watanabe15filling,cheng16translational, hsieh16majorana,po17lattice, lu17Lieb, yang17dyonic} in that it \emph{only} involves crystalline symmetry, as opposed to an interplay between internal and crystal symmetry.\footnote{LSM constraints discussed in Refs.~\onlinecite{hsieh16majorana} and~\onlinecite{lu17Lieb} involve the interplay between fermion parity and crystal symmetry.  Strictly speaking, fermion parity should not be viewed as a symmetry; it is a property of any fermion system and cannot be broken.  However, from a formal point of view it acts like an internal symmetry.}   Following ideas of Ref.~\onlinecite{cheng16translational}, systems where our LSM constraint holds can be viewed as two-dimensional symmetry-preserving surfaces of $d=3$ cSPT states built from one-dimensional blocks. We note that Qi, Fang and Fu have independently obtained the same LSM constraint.\cite{yangqiLSM}

The classifications we obtain for bosonic cSPT phases in $d=2,3$ can be fully understood in terms of point group SPT phases.  For each wallpaper or space group, the classification can be decomposed into pgSPT invariants and other invariants we dub \emph{weak pgSPT invariants}.   Each pgSPT invariant is simply the SPT invariant associated with a given site symmetry subgroup of the full space group.  The weak pgSPT invariants can be understood by making one or more directions in space finite, viewing the system as a lower-dimensional pgSPT phase, and computing the resulting pgSPT invariant as a function of system size in the finite directions.  In many cases, this can be visualized as a stacking of lower-dimensional pgSPT states, with translation symmetry in the stacking direction.  In most cases, the cSPT classification can be factored into pgSPT and weak pgSPT invariants, but the general structure of the decomposition is more subtle than a simple factorization.

\subsection{Block states for crystalline SPT phases}
\label{subsec:blockintro}

Block states play a central role in this paper, so we now describe them in more detail. A block state $| \Psi \rangle$ is a  state of the form
\begin{equation}
| \Psi \rangle = \bigotimes_{b \in B} | \psi_b \rangle \text{,}
\end{equation}
where $B$ is a set of blocks.  Each block $b$ is a $d_b$-dimensional quantum system embedded in $d$-dimensional space, with $d_b < d$.  Zero-dimensional blocks are allowed and play an important role.  Blocks with $d_b \geq 1$ can be of finite extent, semi-infinite, or infinite.  For the purposes of this paper, we will see that it is sufficient to consider infinite blocks when $d_b \geq 1$.

The blocks form a spatial pattern invariant under the crystalline symmetry group $G$, which is a point group or space group.  The action of $g \in G$ on a block $b$ is denoted by $g b$.  Each block is associated with a subgroup $G_b \subset G$, that we call the \emph{effective internal symmetry} of $b$.  When $b$ is a point, $G_b$ is the same as the site symmetry of $b$.  In general, $G_b$ is defined to consist of all elements $g \in G$ that, when restricted to $b$, act as the identity rigid motion.  That is, if $g \in G_b$, then $g$ takes any point lying in $b$ to itself.\footnote{Note that while $g \in G_b$ implies $g b = b$, the converse is not true.}  For example, if $b$ is a two-dimensional block lying on a mirror plane, then $G_b$ is generated by the mirror reflection and is isomorphic to $\zz$.

We assume $|\psi_b \rangle$ is in a $d_b$-dimensional SPT phase protected by the effective internal symmetry $G_b$.  For zero-dimensional blocks, this means that $| \psi_b \rangle$ can carry $G_b$ charge; that is, it transforms in some one-dimensional representation of $G_b$.  Different $G_b$ charges can be viewed as different ``zero-dimensional SPT phases.''

The general structure of our results on cSPT classification can be summarized as follows.  We let $G$ be some crystalline symmetry group, and $\cC(G)$ the corresponding classification of those bosonic cSPT phases that are adiabatically connected to a block state built from lower-dimensional SPT states.  We obtain $\cC(G)$ by classifying cSPT block states using \emph{block-equivalence} operations that we introduce.  Block-equivalence operations are closely related to the lattice homotopy operations introduced in Ref.~\onlinecite{po17lattice} to study LSM constraints (see Sec.~\ref{sec:discussion} for a discussion of the relationship).    We find that $\cC(G)$ agrees with the Thorngren-Else classification.  In three dimensions, $\cC(G)$ is not a complete classification, at least for some symmetries, because it excludes cSPT phases built from $E_8$ states.

We define $\cC_{d_b}(G)$ to be the classification of $G$-symmetric cSPT phases built only from $d_b$-dimensional SPT blocks, and we say such cSPT phases have block-dimension $d_b$.  For $d \leq 3$, we find
\begin{equation}
\cC(G) = \cC_0(G) \times \cdots \times \cC_{d-1}(G) \text{.} \label{eqn:block-factorization}
\end{equation}
In $d=2$, $\cC_1(G)$ is always trivial, and we find one- and two-dimensional bosonic cSPT phases all have block-dimension zero.  More generally, in settings beyond bosonic cSPT phases with only crystalline symmetry and $d \leq 3$, $\cC(G)$ need not factorize by block dimension as in Eq.~(\ref{eqn:block-factorization}); the general structure is not a product but a sequence of subgroups, as explained in Appendix~\ref{app:block-dim-factor}.

The physical properties of cSPT phases at symmetry-preserving surfaces depend on the block dimension.  States built from zero-dimensional blocks can be viewed as product states, and it follows that none of these states have any anomalous boundary properties.  However, these states can have non-trivial entanglement protected by site symmetry, but only when no degrees of freedom lie precisely at the relevant site.\cite{song17topological} In that situation, a non-trivial block-dimension zero state is not a product state of the microscopic degrees of freedom, even though it can be viewed as a product state of larger effective degrees of freedom.   We note that all bosonic cSPT phases in one and two dimensions are of block-dimension zero.  Moreover, in three dimensions, for space groups with only orientation-preserving operations, we also find only block-dimension zero cSPT phases.

In contrast to the block-dimension zero phases, cSPT phases of higher block dimension have anomalous surface properties, which is a sign of non-trivial symmetry-protected entanglement. Surfaces of $d=3$ cSPT states built from one-dimensional blocks are equivalent to ``half-integer spin'' bosonic systems in two dimensions, with microscopic degrees of freedom that transform projectively. Seeing properties characteristic of such a system at the surface of an ``integer spin'' system is a sign of a non-trivial bulk SPT phase.  Moreover, these ``half-integer spin'' surfaces lead us to obtain LSM type constraints for two-dimensional systems, as mentioned in Sec.~\ref{subsec:mainintro} and discussed in Sec.~\ref{sec:LSM}.

Finally, symmetry-preserving surfaces of $d=3$ cSPT states built from two-dimensional blocks are truly anomalous, in that the surface physics cannot occur in an isolated two-dimensional system.\cite{isobe15,yqi15anomalous,chsieh16global,song17topological,lake16anomalies} 
  Therefore, we believe that block-dimension two cSPT phases have the greatest potential for interesting experimentally observable surface phenomena.
  
In light of the product-state nature of block-dimension zero cSPT states, it is important to discuss what we mean by a trivial SPT phase.  In many discussions of SPT phases, product states and trivial states are synonymous.  Here, distinct block-dimension zero cSPT phases can be viewed as product states that are not adiabatically connected if symmetry is preserved.  It has been observed before that such SPT states occur for crystalline symmetry.\cite{chen11a,fuji15distinct,song17topological}

Among the block-dimension zero cSPT phases, we define the trivial phase to be the unique block-equivalence class containing states where all blocks carry trivial charge, \emph{i.e.} they transform as the trivial representation under site symmetry.  There are some subtleties associated with this definition; readers not interested in them can skip this paragraph, as nothing else in the paper depends on it.  The key issues were discussed by some of us and Fu in Appendix~A of Ref.~\onlinecite{song17topological}, which should be consulted for more details.  That discussion was for reflection pgSPT phases in $d=1$, but much of it is expected to apply to general block-dimension zero cSPT phases, as we now describe.  When microscopic degrees of freedom lie precisely at symmetry centers, the symmetry operations can be redefined to arbitrarily change the charge at the blocks.  In this case, it becomes meaningless to ask whether any particular block-dimension zero phase is trivial, but differences between these phases remain well defined.  The $\cC_0(G)$ classification thus still applies, but it should be interpreted as a torsor rather than as a group.  However, such redefinitions of the symmetry operations are not necessarily legitimate, \emph{e.g.} if one views the lattice model as an approximate description of a continuum system, and the site symmetry charges originate from transformation properties of Wannier orbitals.\cite{ElsePrivateComm}  Finally, in a lattice model where the degrees of freedom lie away from symmetry centers, such redefinitions are not possible.  In that case, block-dimension zero cSPT phases are expected to be distinguished by entanglement spectrum signatures, and there is no arbitrary choice involved in the definition of trivial phase.

\subsection{Outline}
\label{subsec:outline}

As an intermediate step toward classifying general cSPT phases, we first review the classification of pgSPT phases in Sec.~\ref{subsec:pgblockstates}.  Further review of the dimensional reduction approach underlying pgSPT classification is given in Appendix~\ref{app:pgreview}.  Section~\ref{subsec:pgspt2d} classifies $d=2$ pgSPT phases for all crystallographic point groups, and Sec.~\ref{subsec:pgspt3d} does the same in $d=3$, excluding pgSPT phases built from $E_8$ states, which lie outside the focus of this paper.  In Appendix~\ref{app:block-dim-factor}, we explain that the classification of pgSPT and cSPT phases does not, in general, factor over block dimension.  However, for the bosonic pgSPT and cSPT phases we consider in $d=3$, such a factorization does hold, \emph{i.e.} $\cC(G) = \cC_0(G) \times \cC_1(G) \times \cC_2(G)$.

Section~\ref{sec:2dcSPT} describes the classification of $d=2$ cSPT phases protected by wallpaper group symmetry.  The block-equivalence operations used to classify these phases are introduced, several example wallpaper groups are discussed, and the cSPT classification is given for all 17 wallpaper groups.  In addition, weak pgSPT invariants are introduced via examples.  The classification $\cC(G) = \cC_0(G)$ for each wallpaper group factors into a subgroup of pgSPT invariants and another of weak pgSPT invariants, and this factorization is given.  This factorization shows that the block equivalence operations give a classification of distinct cSPT phases in two dimensions.

The classification of $d=3$ cSPT phases is discussed in Sec.~\ref{sec:3dcSPT}.  It is simple to obtain $\cC_1(G)$ and $\cC_2(G)$.  A more involved computational procedure based on the block equivalence operations is introduced to obtain $\cC_0(G)$, and applied to two illustrative examples.  More details of this procedure are given in Appendices~\ref{app:splitting} and~\ref{app:twisting}.  The classification $\cC(G) = \cC_0(G) \times \cC_1(G) \times \cC_2(G)$ is given for all 230 space groups in Appendix~\ref{app:3dclassification}.

Section~\ref{sec:3dcSPT} also explains that both $\cC_1(G)$ and $\cC_2(G)$ factor into pgSPT invariants.  Appendix~\ref{app:decomposition} shows that  states corresponding to different elements of $\cC_0(G)$ are completely characterized by pgSPT and weak pgSPT invariants, so $\cC_0(G)$ can be decomposed into pgSPT and weak pgSPT invariants.  It follows that $\cC(G) = \cC_0(G) \times \cC_1(G) \times \cC_2(G)$ is a classification of distinct cSPT phases.

In Sec.~\ref{sec:LSM}, we use our classification of block-dimension one cSPT phases in $d=3$ to obtain a Lieb-Schultz-Mattis (LSM) type constraint for $d=2$ spin systems with wallpaper group symmetry, via a type of bulk-boundary correspondence.  We then present an independent argument for this constraint based on dimensional reduction.  Our LSM constraint is simple to state:  if a $d=2$ spin system contains a spin transforming projectively under its crystalline site symmetry, then a symmetry-preserving, gapped, short-range entangled ground state is impossible.  It is interesting to note that, although the considerations leading to this constraint take full wallpaper group symmetry into account, only site symmetry plays an important role.

Section~\ref{sec:general} addresses the conjecture that all SPT phases protected only by crystalline symmetry can be built from lower-dimensional invertible topological states.  We argue that this conjecture holds provided we make a certain physically reasonable but unproven assumption.  We close the paper in Sec.~\ref{sec:discussion} with a discussion of possible extensions of our work, and some remarks on the relationship between block equivalence operations and the lattice homotopy operations introduced in Ref.~\onlinecite{po17lattice} to obtain LSM constraints.

Two appendices beyond those mentioned above contain additional technical details.  Appendix~\ref{app:coho} defines the first cohomology group $H^1(G, {\rm U}(1))$, and Appendix~\ref{app:zero} treats some details of states built from zero-dimensional blocks that are used throughout the paper.

\section{Point group SPT phases}
\label{sec:pgspt}

In this section, we consider point group SPT (pgSPT) phases, protected by a crystalline point group symmetry $G$.  We follow the approach of Ref.~\onlinecite{song17topological}, framing our discussion in terms of block states, to make contact with our results on more general crystalline SPT phases.  We classify pgSPT phases in $d=2,3$ for all crystallographic point groups.  In $d=3$, we classify only those pgSPT phases built from lower-dimensional SPT blocks; this is not complete for some point groups, because it misses pgSPT phases built from $E_8$ states.

In Sec.~\ref{subsec:pgblockstates}, we review the approach of Ref.~\onlinecite{song17topological}, and the classification of pgSPT phases with mirror reflection symmetry in $d=1,2,3$.   Further details of Ref.~\onlinecite{song17topological} are reviewed in Appendix~\ref{app:pgreview}.  Section~\ref{subsec:pgspt2d} develops the classification of two-dimensional pgSPT phases, first illustrating some key ideas via examples, then presenting the classification for arbitrary point groups.  Similarly, in Sec.~\ref{subsec:pgspt3d} we first discuss the illustrative example of $D_{2h}$ symmetry, then proceed to describe the general classification procedure, and present the classification for all the crystallographic point groups.

\subsection{Block states for point group SPT phases}
\label{subsec:pgblockstates}

Ref.~\onlinecite{song17topological} showed that all pgSPT phases can be built from lower-dimensional topological states, and obtained the classification of such phases for a few simple point groups.  For general point groups, the approach of Ref.~\onlinecite{song17topological} can be cast as a step-wise dimensional reduction procedure, which we review in Appendix~\ref{app:pgreview}.  Here, we focus on bosonic pgSPT phases with crystallographic point group symmetry $G$.  Moreover, we are not interested in completely general bosonic pgSPT phases, but instead we make the assumption that $E_8$ states do not appear at any step of the dimensional reduction process (see Appendix~\ref{app:pgreview} for a further explanation of this statement).

The pgSPT phases of interest can be represented as block states, and our assumption that $E_8$ states do not appear in the dimensional reduction process implies that the blocks are lower-dimensional SPT phases protected by $G_b$ effective internal symmetry.  Working in infinite $d$-dimensional space ${\mathbb R}^d$, all the blocks can be taken to lie the subset $S \subset {\mathbb R}^d$ defined as the union of all points in space fixed by at least one non-trivial point group operation $g \in G$.  (Points lying outside $S$ have no effective internal symmetry.)

Using block states, we review the classification of pgSPT phases protected by a mirror reflection $\sigma$ in $d=1,2,3$, which was discussed in Ref.~\onlinecite{song17topological} for $d = 1,3$.   In $d=3$, this is the point group $C_s$.  In all dimensions, this is a symmetry where one spatial coordinate is reversed, \emph{e.g.} $(x,y) \to (-x,y)$ in $d=2$.

We start with $d=1$, where mirror symmetry is the same as inversion symmetry.  There, $S$ is just a single point at the origin.  We place a single zero-dimensional block $b_0$ at the origin, and consider the state
\begin{equation}
| \Psi \rangle = | \psi_{b_0} \rangle \text{.}
\end{equation}
The effective internal symmetry of $b_0$ is $G_{b_0} \simeq \zz$, and there are two possible states depending on the $\zz$ charge,
\begin{equation}
U_{\sigma} | \psi_{b_0} \rangle = \pm | \psi_{b_0} \rangle \text{,}
\end{equation}
where $U_{\sigma}$ is the unitary operator implementing mirror reflection.

Ref.~\onlinecite{song17topological} introduced an equivalence operation referred to as adjoining.  Dimensional reduction adiabatically connects a general pgSPT state to a state on a thickened version of $S$, but the thickness is arbitrary.  Adjoining corresponds to increasing the thickness of this region, which has the effect of adding extra degrees of freedom to a state defined on $S$.   In the present case, the adjoining operation is realized by sending
\begin{equation}
| \Psi \rangle \to | l \rangle \otimes | \Psi \rangle \otimes | r \rangle \text{,}
\end{equation}
where $| l \rangle$ and $| r \rangle$ zero-dimensional blocks to the left and right of $b_0$, respectively.  We can choose reflection to act by $U_{\sigma} | l \rangle = | r \rangle$ and $U_{\sigma} | r \rangle = | l \rangle$.
This operation has no effect on the $U_{\sigma}$ charge, and this can be anticipated, because the adjoined blocks themselves have no effective internal symmetry, and must therefore be trivial.  We thus obtain a $\zz$ classification, where the two states are labeled by different $U_{\sigma}$ charges.  As discussed further in Ref.~\onlinecite{song17topological} the $\zz$ classification agrees with earlier works that employed different approaches.\cite{gu09tensor,pollmann10,chen11a,schuch11}

The discussion of adjoining above illustrates a general principle:
\begin{itemize}
\item[] \emph{The adjoining operation can only have an effect on the classification when the adjoined blocks are themselves non-trivial.}
\end{itemize}

Moving on to mirror symmetry in $d=2$, $S$ is the one-dimensional reflection axis. We can place a single $d_b = 1$ block on the axis, which has an effective $\zz$ internal symmetry.  One-dimensional bosonic systems with $\zz$ symmetry have only a trivial topological phase,\cite{chen11a} so the resulting state is trivial.  We can also place zero-dimensional blocks along the axis carrying reflection charge, but since the axis is infinite and has no symmetries such as translation, these blocks can always be grouped together to carry trivial reflection charge.  Therefore we conclude that there is only a trivial pgSPT phase for mirror reflection in two dimensions.

In three dimensions, $S$ is the two-dimensional mirror plane, and we consider blocks lying in this plane with effective $\zz$ internal symmetry.  Zero- or one-dimensional blocks can always be grouped together (and the one-dimensional blocks are themselves trivial).  However, covering the mirror plane with a single two-dimensional block $b_2$ and considering states $| \Psi \rangle  = | \psi_{b_2} \rangle$  leads to a non-trivial pgSPT phase when
$| \psi_{b_2} \rangle$ is in the non-trivial $d=2$ SPT phase with $\zz$ symmetry, which we refer to as the Ising SPT phase.\cite{chen13symmetry,levin12braiding}  This leads to the classification
\begin{equation}
\cC(C_s) = \zz \text{,}
\end{equation}
which was obtained in Ref.~\onlinecite{song17topological}.  In fact, the block $b_2$ can also be an $E_8$ state, which leads to a $\zz \times \zz$ classification.\cite{song17topological}  However, we are not considering pgSPT phases built from $E_8$ states here.

\subsection{Point group SPT phases in two dimensions}
\label{subsec:pgspt2d}

\begin{figure}
\includegraphics[width=0.5\columnwidth]{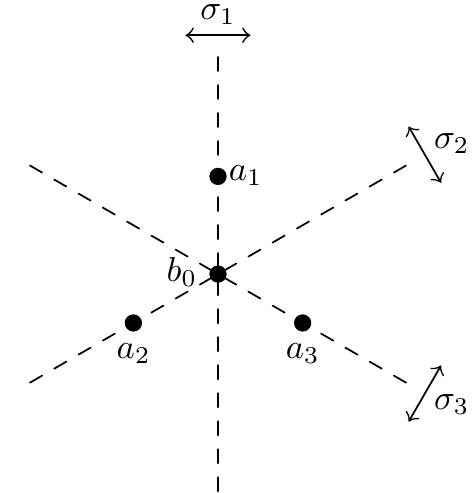}
\caption{The two-dimensional point group $D_3$ is generated by three mirror reflections (dashed lines). To classify states of block dimension zero, we place a single block $b_0$ at the origin, and consider the effect of adjoining three symmetry-related blocks lying on the reflection axes ($a_1,a_2,a_3$).}
\label{fig:D3}
\end{figure}

We now consider general crystallographic point groups in two dimensions.  We begin with the illustrative example $G = D_3$, which is the symmetry group of a regular triangle.  This group is generated by mirror reflections about three axes as shown in Fig.~\ref{fig:D3}, which together comprise the space $S$.

To obtain a classification of pgSPT phases, we consider different possible block states.  While we can place one-dimensional blocks on the reflection axes, these are one-dimensional systems with $\zz$ effective internal symmetry, and are thus trivial.  We can place a zero-dimensional block $b_0$ at the origin, which has effective $D_3 \simeq \z_3 \rtimes \zz$ internal symmetry.  Possible $D_3$ charges of $| \psi_{b_0} \rangle$ are one-dimensional representations of $D_3$, and there are two such representations labeled by the elements of the first cohomology group $H^1(D_3, {\rm U}(1)) =  \zz$.  The first cohomology group $H^1(G, {\rm U}(1))$ is the group formed by the one-dimensional representations of $G$ under the tensor product operation, and is defined in Appendix~\ref{app:coho}.  The non-trivial representation is characterized by $U_{\sigma} | \psi_{b_0} \rangle = - | \psi_{b_0} \rangle$, for $\sigma$ any of the three reflections.

Na\"{\i}vely this would seem to imply a $\zz$ classification, but this is not the end of the story.  This is because we can adjoin zero-dimensional blocks lying on the reflection axes, as shown in Fig.~\ref{fig:D3}.  Labeling these blocks by $a_{i}$, where $i = 1,2,3$ labels the three axes, this modifies the state by
\begin{equation}
| \psi_{b_0} \rangle \to  | \psi_{b_0} \rangle \otimes \Big[ \bigotimes_{i=1}^3  | \psi_{a_i} \rangle \Big] \text{.} \label{eqn:d3-adjoining}
\end{equation}
To be consistent with $D_3$ symmetry, the three blocks $a_i$ must all carry the same reflection charge, and if this charge is non-trivial, the adjoining operation of Eq.~(\ref{eqn:d3-adjoining}) changes the overall $D_3$ charge of the state, which can be shown following the discussion of Appendix~\ref{app:zero}.
 Therefore, the $\zz$ classification is not stable under the adjoining operation, and it collapses to trivial classification, \emph{i.e.} $\cC(D_3)$ is trivial.

\begin{figure}
\includegraphics[width=0.5\columnwidth]{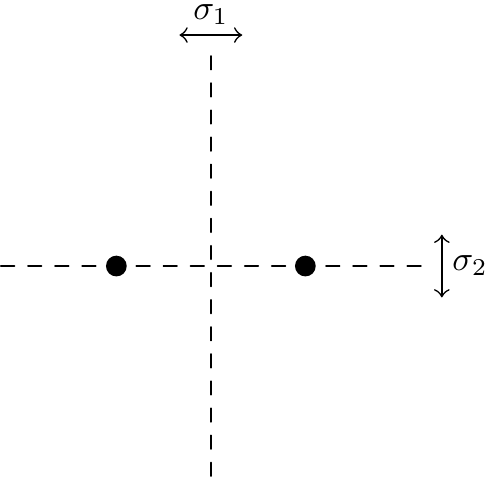}
\caption{The two-dimensional point group $D_2$ is generated by two perpendicular mirror reflections (dashed lines). For states of block dimension zero, adjoining operations involve pairs of symmetry-related blocks (filled circles) on one of the reflection axes. The reflection charges of these blocks cancel out, so the adjoining operation is trivial for the classification of $D_2$-symmetric pgSPT phases.}
\label{fig:D2}
\end{figure}

The adjoining operation turns out to have a trivial effect for all $d=2$ point groups except $D_3$.  For example, for $D_2$ symmetry, we can adjoin zero-dimensional blocks carrying reflection charge as shown in Fig.~\ref{fig:D2}.  Because these blocks must always be added in pairs to preserve the $D_2$ symmetry, adjoining them does not change the one-dimensional representation of the block at the origin, and the $D_2$ pgSPT classification is given by $H^1(D_2, {\rm U}(1)) = \zz \times \zz$.  These statements can be verified following the more general and systematic discussion of the adjoining operation for states of block dimension zero, given  in Appendix~\ref{app:splitting}.

In general, except for $G = D_3$ and for the case of mirror reflection ($G = D_1$), the two-dimensional pgSPT classification is given by 
\begin{equation}
\cC(G) = \cC_0(G) = H^1(G, {\rm U}(1) ) \text{.}
\end{equation}
Table~\ref{tab:2dpgSPT} gives the classification of pgSPT phases for the nine non-trivial crystallographic point groups in two dimensions.  These groups are $n$-fold rotation ($C_n$) for $n=2,3,4,6$, and the dihedral group $D_n$ for $n=1,2,3,4,6$.  $D_1$ is generated by a single mirror reflection, $D_2$ is the symmetry group of a rectangle, and, for $n \geq 3$, $D_n$ is the symmetry of a regular $n$-sided polygon.

\begin{table}
\begin{tabular}{c|c}
\parbox[t]{2.5cm}{Two-dimensional \\ point group $G$} & Classification $\cC(G)$ \\
\hline
$C_n$ ($n=2,3,4,6$) & $\z_n$   \\
\hline
$D_1$, $D_3$ & Trivial  \\
\hline
$D_n$ ($n=2,4,6$) & $\zz \times \zz$ \\
\hline
\end{tabular}
\caption{Classifications of two-dimensional pgSPT phases for all nine non-trivial crystallographic point groups.
 \label{tab:2dpgSPT}}
\end{table}

\subsection{Point group SPT phases in three dimensions}
\label{subsec:pgspt3d}

Here, we discuss the classification of pgSPT phases in three-dimensions.  We emphasize that we consider only those pgSPT phases built from lower-dimensional SPT states; because there can also be $d=3$ pgSPT phases built from $E_8$ states, the resulting classifications are not complete for all point groups.  We start by considering the illustrative example of $D_{2h}$ symmetry, then describe a general procedure to classify $d=3$ pgSPT phases, relegating some of the more formal aspects to Sec.~\ref{sec:2dcSPT} and Appendices~\ref{app:zero} and \ref{app:splitting}.  The classification for all crystallographic point groups is presented in Table~\ref{tab:3dpgSPT}.

For the point groups considered thus far, all the non-trivial SPT phases can be represented with blocks of a fixed dimension.  That is, $\cC(G) = \cC_{d_b}(G)$ for some fixed $d_b$.  Indeed, this holds for all $d=1,2$ point groups, with $d_b = 0$.  The situation changes in $d=3$, where
\begin{equation}
\cC(G) = \cC_0(G) \times \cC_1(G) \times \cC_2(G) \text{.}
\end{equation}
The general structure is not a product over block dimensions but a sequence of subgroups.  This is explained in Appendix~\ref{app:block-dim-factor}, where the factorization is argued to hold for $d=3$ pgSPT and space group SPT phases built from lower-dimensional SPT blocks.

\begin{figure}
\begin{minipage}[t]{0.4\columnwidth}%
\includegraphics[width=1\columnwidth]{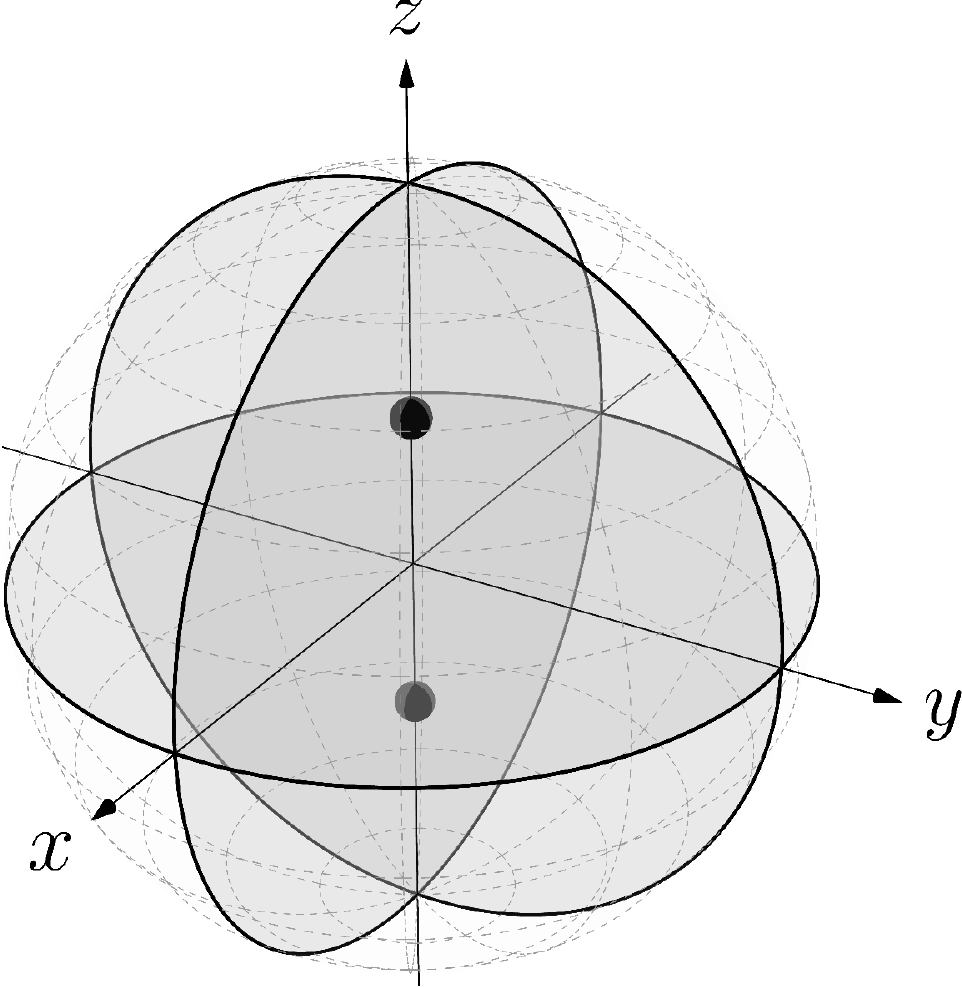}

(a)%
\end{minipage}%
\begin{minipage}[t]{0.4\columnwidth}%
\includegraphics[width=1\columnwidth]{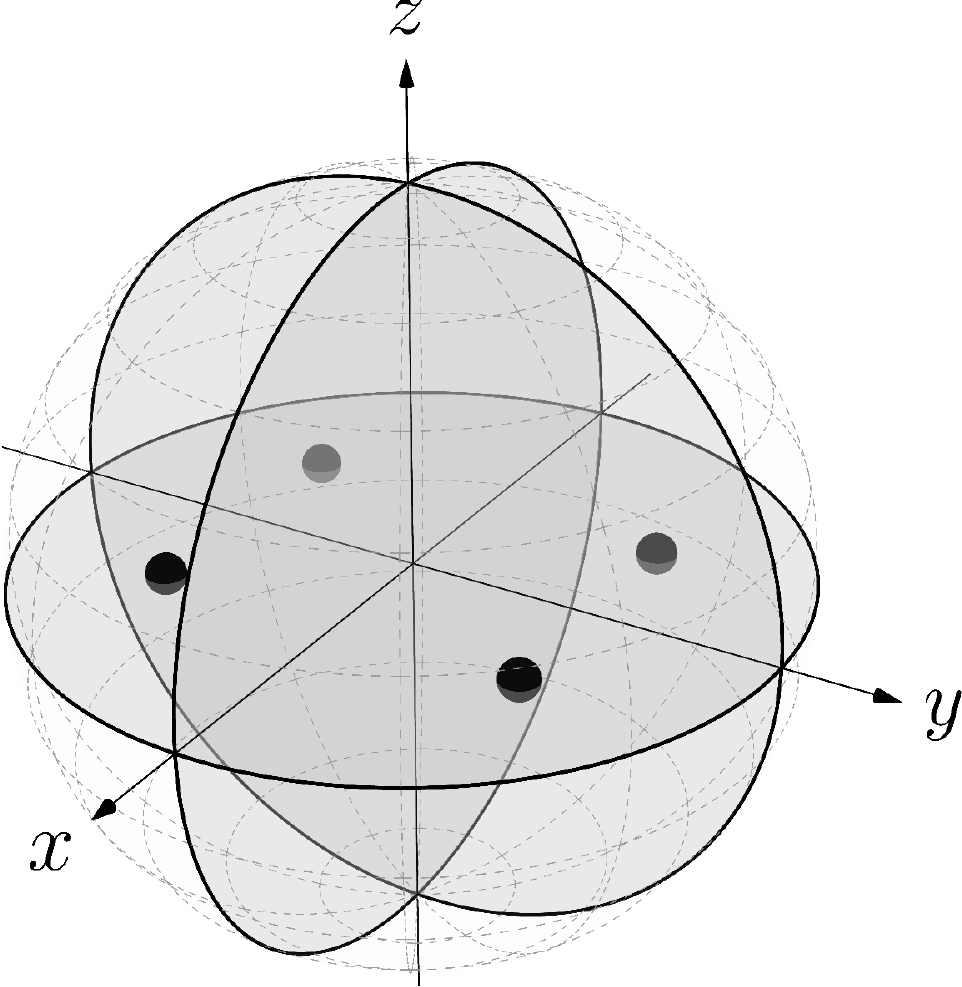}

(b)%
\end{minipage}
\caption{The three-dimensional point group $D_{2h}$ is generated by three perpendicular mirror reflections (gray shaded circles).  Two mirror planes intersect on the $x$, $y$ and $z$ axes, so that points along each are fixed by a $C_{2v}$ subgroup of $D_{2h}$.  For states of block dimension zero, two kinds of adjoining operations are possible, with the adjoined blocks shown as filled circles.  In (a), two $C_{2v}$ charges are adjoined on one of the $C_{2v}$ axes.  In (b), four mirror reflection charges are adjoined, lying at symmetry related positions in a single mirror plane.  In both cases, the adjoining operation does not alter the total $D_{2h}$ charge, and thus has no effect on the classification of $D_{2h}$ pgSPT phases.}
\label{fig:D2h}
\end{figure}

We illustrate this by considering the point group $D_{2h}$, which is generated by three perpendicular mirror reflections (Fig.~\ref{fig:D2h}), and where we find $\cC_{d_b}(G)$ to be non-trivial for each of $d_b = 0,1,2$.   The space $S$ is given by the union of the three mirror planes.

We start with the block dimension zero states, \emph{i.e.} those in $\cC_{0}(G)$.  It is enough to put a single zero-dimensional block at the origin, on which $D_{2h}$ acts as an effective $\zz^3$ internal symmetry, and possible $D_{2h}$ charges are labeled by elements of $H^1(D_{2h}, {\rm U}(1)) = \zz^3$.  There are two kinds of adjoining operations to consider, both of which are illustrated in Fig.~\ref{fig:D2h} and have no effect on the classification.  Therefore, we find $\cC_0(D_{2h}) = \zz^3$.

Next, we consider states of block dimension $d_b = 1$, which can be constructed by placing one-dimensional blocks on the $C_{2v}$ axes ($x$, $y$ or $z$ axis).  Each axis has an effective $\zz^2$ internal symmetry, and there is a single non-trivial one-dimensional SPT phase with this symmetry, the Haldane phase.\cite{haldane83a,haldane83b,affleck87rigorous,gu09tensor,pollmann10,chen11a,schuch11}

We first consider the $x$ axis.  In principle, we should divide the axis into two semi-infinite one-dimensional blocks, one for $x > 0$ and one for $x < 0$, since then each block has \emph{only} the effective internal $\zz^2$ symmetry.  These two blocks are related to one another by $x \to -x$ mirror reflection, so they must be in the same $d=1$ SPT phase, and they can be sewn together at the origin to form a single block.  This works precisely because reflection acts trivially on the $\zz$ SPT index characterizing the Haldane phase.  For each $C_{2v}$ axis, we get a $\zz$ classification of pgSPT phases, so considering all three axes we have found $\cC_1(D_{2h}) = \zz^3$.

This discussion illustrates a general property of all the cSPT phases arising in this paper:
\begin{itemize}
\item[] \emph{For a general $d=3$ point group or space group $G$, cSPT phases of block dimension one are built from  $d=1$ SPT phases classified by a $\zz$ invariant, so we can always represent phases in $\cC_1(G)$ using states with  infinite one-dimensional blocks.}
\end{itemize}

Finally we consider states of block dimension $d_b = 2$, constructed by placing two-dimensional blocks on the mirror planes.  As discussed in Sec.~\ref{subsec:pgblockstates} for a single mirror reflection in $d=3$, there is a single non-trivial $d_b = 2$ SPT state, the Ising SPT phase.  (Again, we do not consider two-dimensional blocks that are $E_8$ states.) Reflection and other spatial symmetries act trivially on the $\zz$ SPT invariant characterizing the Ising SPT phase.  Therefore, just as for states of block dimension one, we can represent these states using infinite two-dimensional blocks.  We thus get a $\zz$ SPT index for each mirror plane, and all together we find $\cC_2 (D_{2h}) = \zz^3$. 

Again, this illustrates a general statement:
\begin{itemize}
\item[]  \emph{For a general $d=3$ point group or space group $G$, cSPT phases in $\cC_2(G)$ and built from SPT blocks can be represented using infinite two-dimensional blocks. }
\end{itemize}

\begin{table}
\begin{tabular}{c|c|c|c|c}
Point group $G$ & $\cC(G)$ & $\cC_0(G)$ & $\cC_1(G)$ & $\cC_2(G)$ \\
\hline
$C_i$ & $\zz$ &  $\zz$ & -- & --   \\
\hline \hline
$C_2$  & -- & -- & -- & --  \\
\hline 
 $C_s$ & $\zz$ & -- & -- & $\zz$  \\
\hline
$C_{2h}$ & $\zz^3$ & $\zz^2$ & -- & $\zz$ \\
\hline \hline
 $D_2$  & $\zz^2$ &   $\zz^2$  & -- & -- \\
\hline
 $C_{2v}$ & $\zz^3$ & -- & $\zz$ & $\zz^2$ \\
\hline
 $D_{2h}$  & $\zz^9$ & $\zz^3$ & $\zz^3$ & $\zz^3$ \\
\hline \hline
$C_4$ & -- & -- & -- & --  \\
\hline
$S_4$  & $\zz$  & $\zz$ & -- & --  \\
\hline
$C_{4h}$  & $\zz^3$ & $\zz^2$ & -- & $\zz$ \\
\hline
$D_4$ & $\zz^2$ & $\zz^2$ & -- & -- \\
\hline
 $C_{4v}$ & $\zz^3$ & -- & $\zz$ & $\zz^2$ \\
\hline
 $D_{2d}$ & $\zz^3$ & $\zz$ & $\zz$ & $\zz$  \\
\hline
$D_{4h}$ & $\zz^9$ & $\zz^3$  & $\zz^3$ & $\zz^3$  \\
\hline \hline
 $C_3$ & -- & -- & -- & --  \\
\hline
 $C_{3i}$ & $\zz$ & $\zz$ & -- & -- \\
\hline
 $D_3$ & -- & -- & -- & --   \\
\hline
 $C_{3v}$ & $\zz$  & --  & -- & $\zz$  \\
\hline
 $D_{3d}$ & $\zz^3$ &  $\zz^2$ & -- & $\zz$   \\
\hline \hline
$C_6$ & --  & -- & -- & --   \\
\hline
 $C_{3h}$ & $\zz$  & --  & -- & $\zz$  \\
\hline
$C_{6h}$ & $\zz^3$ & $\zz^2$ & -- & $\zz$    \\
\hline
 $D_6$ & $\zz^2$ & $\zz^2$ & -- & --   \\
\hline
 $C_{6v}$ & $\zz^3$  & -- & $\zz$ & $\zz^2$   \\
\hline
 $D_{3h}$ & $\zz^3$ & -- & $\zz$ & $\zz^2$  \\
\hline
 $D_{6h}$ & $\zz^9$ & $\zz^3$ & $\zz^3$ & $\zz^3$  \\
\hline \hline
$T$  &-- & -- & -- & --  \\
\hline
$T_h$  & $\zz^3$ &  $\zz$ & $\zz$ & $\zz$  \\
\hline
$O$ & $\zz$  &  $\zz$ & -- & -- \\
\hline
$T_d$ & $\zz^2$ & -- & $\zz$ & $\zz$     \\
\hline
 $O_h$ & $\zz^6$ & $\zz^2$ & $\zz^2$ & $\zz^2$   
\end{tabular}
\caption{Classification of those $d=3$ pgSPT phases built from lower-dimensional SPT states protected by effective internal symmetry.  We give the classification $\cC(G)$, and its factorization $\cC(G) = \cC_0(G) \times \cC_1(G) \times \cC_2(G)$.  These classifications are incomplete for some point groups because we neglect pgSPT phases built from $E_8$ states.
 \label{tab:3dpgSPT}}
\end{table}

We now summarize and formalize the above discussion to describe more generally the classification of pgSPT phases in three dimensions.

We start with states of block dimension two, and work our way down in block dimension.
Phases in $\cC_2(G)$ are obtained by placing either Ising SPT states or trivial states on mirror planes.  For each set of symmetry-equivalent  mirror planes, we have a $\zz$ SPT index, and  $\cC_2(G)$ is just a product of these $\zz$ factors.  Each $\zz$ index can be interpreted as a pgSPT index for mirror reflection symmetry alone, by focusing on an appropriate mirror plane and ignoring the rest of the $G$ symmetry.

To classify states of block dimension one, we first need to identify one-dimensional axes with effective internal symmetry.  There are two types of such axes, those with $C_n$ symmetry, and those with $C_{nv}$ symmetry ($n = 2,3,4,6$ in both cases).  The first type of axis has a $\z_n \simeq C_n$ effective internal symmetry, and thus only hosts trivial one-dimensional SPT states, because $H^2(\z_n, {\rm U}(1))$ is trivial.\footnote{See, for instance, Ref.~\onlinecite{chen13symmetry} for a definition of the second cohomology group $H^2$ and a discussion of its role in the classification of $d=1$ SPT phases.}  In the second case, we note that $C_{nv} \simeq \z_n \rtimes \zz$.  The classification of one-dimensional SPT phases with this symmetry is
\begin{equation}
H^2(\z_n \rtimes \zz, {\rm U}(1)) = \left\{ \begin{array}{ll}
{\rm Trivial}, & n \text{ odd} \\
\zz , & n \text{ even}
\end{array}\right. \text{.}
\end{equation}
Therefore $C_{3v}$ axes are trivial, but each set of symmetry-equivalent $C_{nv}$ axes with $n=2,4,6$ carries a $\zz$ SPT invariant, and $\cC_1(G)$ is a product of these $\zz$ factors.  The adjoining operation behaves trivially, because lines nearby and parallel to a $C_{nv}$ axis have at most $\zz$ effective internal symmetry (if they lie in a mirror plane), which is not enough to protect non-trivial one-dimensional states.

 Finally, we consider block dimension zero states. Given a point group $G$, we define its fixed space to be the subset of ${\mathbb R}^3$ fixed by all group operations.   The fixed space can be a single point at the origin, a line, or a plane.  When the fixed space is a line or a plane, $\cC_0(G)$ is trivial.  This is because zero-dimensional blocks lying on the fixed space can always grouped together into composites with no $G$ charge.  We can therefore focus on point groups whose fixed space is a single point.

To proceed requires a more detailed description as compared to $d_b = 1,2$ states, because the adjoining operation is non-trivial for $d_b = 0$ states of some point groups.  This is addressed in a general treatment of block dimension zero states given in Sec.~\ref{sec:2dcSPT}, with further details in Appendices~\ref{app:zero} and \ref{app:splitting}.  In Appendix~\ref{app:splitting}, we obtain the result
\begin{equation}
\cC_0(G) = \frac{H^1(G, {\rm U}(1))}{{\rm Adj}(G)}\text{.}
\end{equation}
Here, $H^1(G, {\rm U}(1))$ labels one-dimensional representations of $G$ (see Appendix~\ref{app:coho}), and ${\rm Adj}(G)$ is a subgroup of $H^1(G, {\rm U}(1))$ containing all  one-dimensional representations that can be obtained by the adjoining operation, starting with a trivial block at the origin.  Taking the quotient precisely captures the information in $H^1(G, {\rm U}(1))$ that is stable under the adjoining operation.  The computation of ${\rm Adj}(G)$ is described in Appendix~\ref{app:splitting}.

Following the above discussion, the classification $\cC(G) = \cC_0(G) \times \cC_1(G) \times \cC_2(G)$ is given in Table~\ref{tab:3dpgSPT} for all $d=3$ crystallographic point groups.

\section{Crystalline SPT phases: two dimensions}
\label{sec:2dcSPT}

Here, we consider cSPT phases protected by the 17 wallpaper groups in two dimensions.  We introduce our procedure to classify cSPT block states, and give a classification for each wallpaper group.  In and of itself, our procedure not guaranteed to produce a classification of distinct SPT phases.  The classifications we obtain must therefore be further justified, which can be done in two ways: (1) Our results match the Thorngren-Else classification, which is obtained by very different methods.  (2) For each wallpaper group, our classification can be factored into $d=2$ pgSPT invariants and weak pgSPT invariants.  The weak invariants, which we introduce below via specific examples, are defined by compactifying one spatial dimension to obtain $d=1$ pgSPT states, and examining the dependence of the $d=1$ pgSPT invariant on the length in the finite dimension.  These invariants can be understood as the $d=1$ pgSPT index per layer of a stack of $d=1$ pgSPT states, with translation symmetry along the stacking direction.

In general, a cSPT block state in two dimensions can be built from zero- and one-dimensional blocks.  However, any one-dimensional blocks are always trivial.  The effective internal symmetry of a one-dimensional block is at most $\zz$ (for a reflection axis), and this is not enough to protect non-trivial invertible topological phases in one dimension.\cite{chen11a}  Therefore, it is enough to consider dimension zero block states, \emph{i.e.} those with only zero-dimensional blocks.  This discussion can be summarized by the statements that $\cC_1(G)$ is trivial for wallpaper groups in two dimensions, and $\cC(G) = \cC_0(G)$.

To describe and classify block dimension zero states for wallpaper groups, we first introduce some notation.  The same discussion applies in three dimensions, so for the moment we keep the spatial dimension $d$ arbitrary.  We let ${\cal B}_0$ be the set of block dimension zero states.  A state $\Psi \in {\cal B}_0$ is specified by the following data:
\begin{enumerate}
\item  A discrete set of points $\Lambda \subset {\mathbb R}^d$ which is invariant under the action of $G$.  A point $p \in \Lambda$ is fixed by its site-symmetry group $G_p \subset G$.

\item  We place a zero-dimensional block at each point $p$, and $G_p$ is the effective internal symmetry of this block. We denote the $G_p$ charge at $p$ by $q_p \in H^1(G_p, {\rm U}(1))$.  Knowing the charge at one point determines the charge at all symmetry-related points, as we discuss below.
\end{enumerate}

This data is manifest physically in the wave function
\begin{equation}
| \Psi \rangle = \bigotimes_{p \in \Lambda} | \psi_p \rangle \text{.}
\end{equation}
The action of $g \in G$ is given by
\begin{equation}
U_g | \psi_p \rangle = \lambda(g, p) | \psi_{gp} \rangle \text{,}
\end{equation}
where $\lambda(g,p)$ is a phase factor.
We assume that the degrees of freedom transform linearly (\emph{i.e.} not projectively) under the symmetry, which means that $U_{g_1} U_{g_2} | \psi_p \rangle = U_{g_1 g_2} | \psi_p \rangle$, implying the condition
\begin{equation}
\lambda(g_1 g_2, p) = \lambda(g_1, g_2 p) \lambda(g_2, p) \text{.} \label{eqn:lambda}
\end{equation}
For a fixed $p$ and restricting to $g \in G_p$, this equation just says that $\lambda(g,p)$ is a one-dimensional representation of $G_p$, and we choose it to be the representation given by $q_p$.  Formally, for $g \in G_p$, we write
\begin{equation}
\lambda(g,p) = D_{q_p} (g) \text{,}
\end{equation}
where $D_{q_p}$ is the one-dimensional representation of $G_p$ labeled by the charge $q_p$.

It may appear that there is physical information in $\lambda(g,p)$ beyond the charges $q_p$, but this is not the case:  knowing $q_p$ for all $p \in \Lambda$ completely determines $\lambda(g,p)$ up to some gauge-like freedom arising from the freedom to adjust the phase of $| \psi_p \rangle$.  This is shown in Appendix~\ref{app:zero}, and justifies specifying only $q_p$ in the data characterizing a state.  

Charges at symmetry-related points are related.  Consider a point $p \in \Lambda$, and some operation $g \in G$ so that $g p \neq p$.  Then let $h \in G_p$, so that $h p = p$.  Using Eq.~(\ref{eqn:lambda}), we find
\begin{equation}
D_{q_{gp}}( g h g^{-1} ) = D_{q_p} (h) \text{.} \label{eqn:related-charges}
\end{equation}
The charges $q_p$ and $q_{gp}$ can be identified if we identify $G_p$ and $G_{gp}$ using the isomorphism induced by conjugation by $g$.  However, this identification is not always natural, and in general we can only say that the charges are related according to Eq.~(\ref{eqn:related-charges}).

Our goal is to obtain the classification $\cC_0(G)$ by studying states in ${\cal B}_0$.  We do this by introducing equivalence operations, referred to as block equivalence operations, that group ${\cal B}_0$ into classes that will turn out to correspond to cSPT phases.  Two states are considered block-equivalent when they are related by some combination of the following operations:
\begin{enumerate}
\item Continuously slide points around so that $G$ symmetry is always preserved.  We require that, for each point $p$, the site symmetry $G_p$ is constant throughout the sliding process.

\item A collection of points ``near'' $p$, where the collection has symmetry $G_p$, can be grouped together to a single new point at $p$.  The whole collection transforms in a one-dimensional representation of $G_p$ with charge $q_p$, which is a function of site symmetry charges of the points in the collection (see Appendix~\ref{app:zero}).  There is also an inverse operation, where a point $p$ can be split to a collection of nearby points respecting $G_p$ symmetry, with the restriction that the collection transforms in the representation labeled by $q_p$.

\item Points with trivial charge can be added or removed as long as $G$ symmetry is respected.
\end{enumerate}
These operations are closely related to the lattice homotopy operations introduced in Ref.~\onlinecite{po17lattice} to obtain LSM constraints; the relationship is discussed in Sec.~\ref{sec:discussion}

Two block-equivalent states are certainly in the same phase.  However, these operations only correspond to a special family of adiabatic paths between states, and, in principle, two inequivalent states could be in the same phase.  That is, block equivalence classes could be finer than the actual classification of phases. It turns out this is not the case, and these operations do give a classification of distinct cSPT phases in two dimensions.  As stated above, this statement is based on the facts that the block-equivalence classification matches the Thorngren-Else classification obtained by very different means, and that it can be factored into pgSPT and weak pgSPT invariants.

We now illustrate this general discussion with some examples.  First, we consider $G = p1$ (wallpaper group \#1), which consists only of translations.  Here, all points have trivial site symmetry, so all block states are trivial, and we find a trivial classification.

A more interesting example is $G = p2$ (wallpaper group \#2), which is generated by two primitive translations and $C_2$ rotation.  Within each primitive cell, there are four inequivalent points with $C_2$ site symmetry.  This information is readily obtained from the International Tables for Crystallography.\cite{inttables}  There, for each wallpaper group, a Wyckoff letter $w = a,b,\dots$ is assigned to each family of symmetry-equivalent points.  For each Wyckoff class, the site symmetry, unit cell coordinates, and multiplicity within the unit cell are given.  One of the Wyckoff classes always consists of points with no site symmetry; this class plays no role in our analysis and we ignore it.  The coordinates for each point in a Wyckoff class sweep out a space of either zero or one dimension, according to the number of free parameters, and we refer to this as the dimension $d_w$ of the Wyckoff class.  Points in $d_w = 1$ Wyckoff classes can be slid continuously, while those in $d_w = 0$ classes are fixed.

Returning to the present case of $G = p2$, the four Wyckoff classes with $C_2$ site symmetry have $d_w = 0$.  We can attach zero-dimensional blocks carrying definite $C_2 \simeq \zz$ charge to the points in each Wyckoff class, so for each class we obtain a $\zz$ invariant.  We thus find the classification $\cC_0(p2) = \zz^4$.  Here, each $\zz$ factor in the classification is a pgSPT invariant for a different $C_2$ subgroup of $p2$.  We can thus factor $\cC_0(p2)$ into  pgSPT invariants, which shows that all 16 states labeled by $\cC_0(p2)$ are truly distinct SPT phases.

We now turn to an example where the cSPT classification cannot be factored into pgSPT invariants, and where we need to consider weak pgSPT invariants instead.  We consider $G = pm$  (wallpaper group \#3), which is generated by a single mirror reflection and two primitive translations, which can be taken parallel and perpendicular to the reflection axis.  There are two Wyckoff classes, both of which are one-dimensional and have site symmetry $D_1$. Each class corresponds to a symmetry-equivalent family of reflection axes.  We can place zero-dimensional blocks carrying $D_1 \simeq \zz$ charge on the points in each Wyckoff class, which gives a $\zz^2$ classification.  (The block equivalence operations play a trivial role here.)

Here, the $\zz$ factors in the classification are weak pgSPT invariants.  We single out a particular reflection axis, choosing coordinates so the axis runs along the $y$-direction and lies at $x=0$.  We focus on $x \to -x$ reflection, and translation in the $y$-direction, ignoring other symmetries.  We choose the system to have finite length $L$ in the $y$-direction, with periodic boundary conditions.  The resulting system can be viewed as a $d=1$ pgSPT state for the $x \to -x$ reflection, and is thus characterized by a $\zz$ SPT invariant. Due to translation symmetry along the $y$-direction, each time $L$ is increased by one lattice constant, the $d=1$ invariant either remains the same, or it flips.  The weak pgSPT invariant is defined to be the difference in $d=1$ invariant between systems with odd $L$ and even $L$.

The weak invariant can also be visualized as the $d=1$ pgSPT invariant per layer. The system can be viewed as a stack in the $y$-direction of $d=1$ pgSPT states with $x \to -x$ reflection symmetry.  Due to translation symmetry in the stacking direction, each stacked layer has the same $\zz$ $d=1$ pgSPT invariant, and this is the weak pgSPT invariant.  This simple visualization applies to block states, but the definition of the weak invariant above is more general.  We also note that in $d=3$ cSPT phases, there are cases where weak pgSPT invariants defined by compactifying one spatial dimension cannot be interpreted in terms of stacking of lower-dimensional SPT states (see Appendix~\ref{app:decomposition}).

\begin{table*}
\begin{tabular}{c|c|c|c}
 \parbox[t]{2cm}{Wallpaper \\ group \#} & $\cC(G) = \cC_0(G)$ & \parbox[t]{2cm}{pgSPT \\ invariants} & \parbox[t]{2cm}{Weak pgSPT \\ invariants}  \\
\hline
1 & -- & -- & --  \\
\hline
2 & $\zz^4$ & $\zz^4$ & --  \\
\hline
3 & $\zz^2$ & -- & $\zz^2$ \\
\hline
4 & -- & -- & --  \\
\hline
5 & $\zz$ & -- & $\zz$  \\
\hline
6 & $\zz^8$ & $\zz^8$ & --  \\
\hline
7 & $\zz^3$ & $\zz^2$ & $\zz$  \\
\hline
8 & $\zz^2$ & $\zz^2$ & --  \\
\hline
9 & $\zz^5$ & $\zz^5$ & -- \\
\hline
10 & $\z_4^2 \times \zz$ & $\z_4^2 \times \zz$ & --  \\
\hline
11 & $\zz^6$ & $\zz^6$ & --  \\
\hline
12 & $\z_4 \times \zz^2$ & $\z_4 \times \zz^2$ & --  \\
\hline
13 & $\z_3^3$ & $\z_3^3$ & --  \\
\hline
14 & $\zz$ & -- & $\zz$  \\
\hline
15 & $\z_3 \times \zz$ & $\z_3$ & $\zz$ \\
\hline
16 & $\z_6 \times \z_3 \times \zz$ & $\z_6 \times \z_3 \times \zz$ & --  \\
\hline
17 & $\zz^4$ & $\zz^4$ & --  \\
\hline
\end{tabular}
\caption{Classifications of two-dimensional cSPT phases.  The first column gives the number of each wallpaper group; the corresponding name can be found in the International Tables for Crystallography.\cite{inttables}   For each wallpaper group, we give the classification $\cC(G) = \cC_0(G)$, and the factorization of the classification into pgSPT and weak pgSPT invariants.
 \label{tab:2dcSPT}}
\end{table*}

\begin{figure}
\includegraphics[width=\columnwidth]{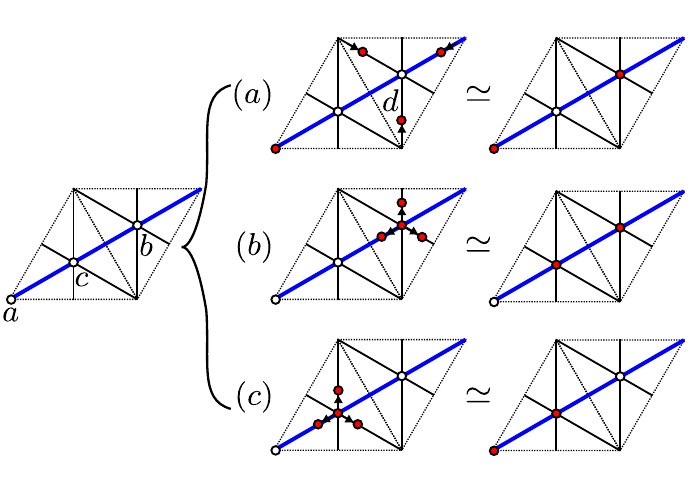}
\caption{The left panel shows a primitive cell of the wallpaper group $G = p3m1$ (wallpaper group \# 14), with the positions of the $a,b,c$ Wyckoff points shown.  The solid lines are reflection axes, on which $d$ Wyckoff points lie.  The right-hand side shows sequences of equivalence operations where the $(q_a,q_b,q_c) = (0,0,0)$ charge configuration is brought to $(1,1,0)$, $(0,1,1)$, and $(1,0,1)$, respectively in (a), (b) and (c).  One reflection axis, singled out using blue, is used to show that $\cC_0(p3m1) = \zz$ is a weak pgSPT invariant, as described in the text.}
\label{fig:p3m1}
\end{figure}

Another example involving a weak pgSPT invariant is $G = p3m1$ (wallpaper group \#14).  There are four non-trivial Wyckoff classes.  Three of these ($a,b,c$) are centers of $D_3$ symmetry, and are fixed, as shown in Fig.~\ref{fig:p3m1}.  The fourth Wyckoff class ($d$) is one-dimensional and has site symmetry $D_1$.  Points in this class lie on the reflection axes that join the $D_3$ centers.  The charge $q_p$ for a point in any of the Wyckoff classes is labeled by an element of $\zz$, since $H^1(D_1, {\rm U}(1)) = H^1(D_3, {\rm U}(1)) = \zz$.  Points in class $d$ can always be slid near one of the $D_3$ centers and joined to it, so we can focus on $D_3$ charge configurations, which are labeled by $(q_a,q_b,q_c) \in \zz^3$.

Given such a configuration, we can change $(q_a, q_b, q_c) \to (q_a +1, q_b+1, q_c)$ by the following sequence of equivalence operations illustrated in Fig.~\ref{fig:p3m1}a.  First, we can split each $a$ block into a new $a$ block and three $d$ blocks, each carrying non-trivial $D_1$ charge.  This changes $q_a \to q_a + 1$.  Then, we can slide the $d$ blocks near $b$, and group them together with $b$ blocks.  This eliminates all the $d$ blocks and changes $q_b \to q_b + 1$, as desired.  There is nothing special about the pair $a,b$, and this process can be done for any pair of $a,b,c$, as shown in Fig.~\ref{fig:p3m1}.  Under such equivalence operations, every configuration $(q_a, q_b, q_c)$ is equivalent either to $(0,0,0)$ or to $(1,1,1)$, so these operations collapse the $\zz^3$ down to a single $\zz$, and we find $\cC_0(p3m1) = \zz$.

This $\zz$ invariant cannot be a pgSPT invariant, because both $D_3$ and $D_1$ pgSPT phases have a trivial classification.  Instead, just like for $G = pm$ symmetry, it is a weak pgSPT invariant.  Focusing on a particular reflection, shown as the blue reflection axis in Fig.~\ref{fig:p3m1}, we see that the one-dimensional primitive cell along the axis contains a single point in each of the $a,b,c$ Wyckoff classes.  Therefore, the $(1,1,1)$ charge configuration has non-trivial reflection charge per axis primitive cell, and can be viewed as a stacking of non-trivial $d=1$ pgSPT states.  On the other hand, the $(0,0,0)$ charge configuration is a stacking of trivial states.

We used the formal procedure described in Sec.~\ref{sec:3dcSPT} to calculate $\cC_0(G)$ and thus classify cSPT phases for all wallpaper groups.   Additional technical details appear in Appendices~\ref{app:zero} and \ref{app:splitting}.  The results, which agree with those of Thorngren and Else,\cite{thorngren16gauging} are shown in Table~\ref{tab:2dcSPT}.  For each symmetry group, the classification factorizes into pgSPT and weak pgSPT invariants.

\section{Crystalline SPT phases: three dimensions}
\label{sec:3dcSPT}

Now we consider cSPT phases protected by space group symmetry in three dimensions.  We focus only on those states built from lower-dimensional SPT building blocks; that is, we do not consider two-dimensional $E_8$ state blocks. As argued in Appendix~\ref{app:block-dim-factor}, the classification has the structure $\cC(G) = \cC_0(G) \times \cC_1(G) \times \cC_2(G)$, where, in contrast to $d=2$, non-trivial contributions from all block dimensions ($d_b = 0,1,2$) can appear.  Following the discussion given here, and supplemented by technical details presented in Appendices~\ref{app:splitting} and \ref{app:twisting}, we obtained $\cC_{d_b}(G)$ ($d_b = 0,1,2$) for all 230 space groups.  The classifications are presented in Appendix~\ref{app:3dclassification}.  We find that $\cC(G)$ agrees with the Thorngren-Else classification, which was obtained in Ref.~\onlinecite{thorngren16gauging} for all space groups except numbers 227, 228 and 230.

The classifications obtained here are based on the block-equivalence operations described for $d=2$ cSPT phases in Sec.~\ref{sec:2dcSPT}.  Just like in two dimensions, block equivalence is not \emph{a priori} guaranteed to give a classification of distinct cSPT phases, and further justification is needed.  This is provided in part by the fact that our results match those of Thorngren and Else.  Moreover, we show that a state with non-trivial block-equivalence class (\emph{i.e.}, non-zero element of $\cC(G)$) has a non-trivial pgSPT or weak pgSPT invariant. This establishes that non-zero elements of $\cC(G)$ are non-trivial phases.  It also implies distinct elements are distinct phases with different sets of  pgSPT and weak pgSPT invariants; that is, that $d=3$ bosonic cSPT phases can be completely characterized in terms of pgSPT and weak pgSPT invariants.  In this section, we show that $\cC_1(G) \times \cC_2(G)$ factors into pgSPT invariants.  Establishing the above statements is more involved for the block-dimension zero states classified by $\cC_0(G)$, and this is done in Appendix~\ref{app:decomposition}.

We classify states of different block dimension separately.  For $d_b = 0$, we use the same block-equivalence operations described in Sec.~\ref{sec:2dcSPT}.  Below we describe a systematic computational procedure used to obtain $\cC_0(G)$ in both $d=2$ and $d=3$.

Obtaining $\cC_1(G)$ and $\cC_2(G)$ is much simpler; these factors in the classification can essentially be read off from the entry for $G$ in the International Tables for Crystallography,\cite{inttables} with no calculation required.  This occurs because the block-equivalence operations are trivial for $d_b = 1,2$. Sliding is trivial because $d_b =1,2$ blocks with enough effective internal symmetry to protect a non-trivial SPT state are always fixed in space; they cannot be slid without lowering their effective internal symmetry.  Splitting and grouping are also trivial. Whenever a $d_b = 1$ block has effective internal symmetry capable of supporting a non-trivial SPT phase ($C_{nv}$ with $n=2,4,6$), nearby parallel lines have at most $\zz$ effective internal symmetry, which is not enough to protect non-trivial one-dimensional states.  Similarly, for $d_b = 2$ blocks with mirror symmetry, nearby parallel planes have no symmetry and cannot host non-trivial two-dimensional SPT states.

We now describe how to obtain $\cC(G)$ from the information in the International Tables for Crystallography.\cite{inttables}  Just like for wallpaper groups, the entry for each $d=3$ space group includes information about crystal positions.  These are labeled by letters $w = a,b,c,\dots$ corresponding to Wyckoff classes, where the points in each Wyckoff class are related by symmetry, and points in different Wyckoff classes are not related by symmetry.  The site symmetry for each Wyckoff class is given, and we refer to these groups as $G_a, G_b, \dots$.  One of the Wyckoff classes always consists of general points with no site symmetry.  This class plays no role in our analysis, and we ignore it.

The points in each Wyckoff class have either zero, one or two free parameters.  (The latter two cases correspond to high symmetry axes and planes, respectively.)  We refer to this number as the dimension of the Wyckoff class and denote it by $d_w$, because as the free parameters are varied, each point sweeps out a space of the given dimension.

We first discuss the $d_b = 1,2$ factors in the classification, before proceeding to the more involved calculations for $d_b = 0$.  To obtain $\cC_2(G)$, we simply identify all the two-dimensional Wyckoff classes, which are always mirror planes.  Each such Wyckoff class gives a $\zz$ invariant associated with putting Ising SPT states on the symmetry-equivalent mirror planes.  $\cC_2(G)$ is simply a product of these $\zz$ invariants.

Similarly, $\cC_1(G)$ is obtained by identifying all one-dimensional Wyckoff classes with $C_{nv}$ site symmetry, for $n = 2,4,6$.  Each such class gives a $\zz$ invariant associated with non-trivial $d=1$ SPT states with $\z_n \rtimes \zz$ effective internal symmetry, and $\cC_1(G)$ is a product of these $\zz$ invariants.  We need not consider $C_{3v}$ axes, because the $\z_3 \rtimes \zz$ effective internal symmetry does not admit non-trivial $d=1$ SPT phases  (see Sec.~\ref{subsec:pgspt3d}).

It is easy to see that $\cC_1(G) \times \cC_2(G)$  factors into pgSPT invariants and thus gives a classification of distinct cSPT phases.  Each $\zz$ factor in $\cC_1(G)$ is a pgSPT invariant for a $C_{nv}$ subgroup of $G$.  Similarly, each $\zz$ factor in $\cC_2(G)$ is a pgSPT invariant for a mirror reflection subgroup of $G$.

We now describe a procedure to obtain $\cC_0(G)$ based on the block-equivalence operations of Sec.~\ref{sec:2dcSPT}.  Suppose we have a state $\Psi \in {\cal B}_0$.  We can use the block-equivalence operations to deform this state to a canonical state, where $\Lambda$ has exactly one symmetry-related set of points for each Wyckoff class.  All points in the same Wyckoff class have symmetry-related charges.  We arbitrarily pick out a representative point in each class $w$, and specify its charge $q_w \in H^1(G_w, {\rm U}(1))$, which determines the charges of all points in the class.  The charges $q_w$ can be assigned independently for the different classes.  Therefore, canonical states are labeled by a charge $Q$ taking values in the direct product of $H^1$ factors for the different Wyckoff classes; that is, 
\begin{equation}
Q \in \cQ_c \equiv H^1(G_a) \times H^1(G_b) \times \cdots \text{,}
\end{equation}
where in the interest of compact notation we have defined
\begin{equation}
H^1(G_w) \equiv H^1(G_w, {\rm U}(1)) \text{.}
\end{equation}
Group addition in $\cQ_c$ corresponds physically to the operation of stacking two SPT states, \emph{i.e.} making a decoupled ``bilayer'' of the two states.

States with different values of $Q$ can be in the same phase.  We define a subgroup $\cQ_t \subset \cQ_c$ containing all $Q$'s such that the corresponding state is in the trivial phase.  The block-equivalence classification is then given by the quotient
\begin{equation}
\cC_0(G) =  \cQ_c / \cQ_t  \text{.}
\end{equation}

To proceed, the main task is to obtain $\cQ_t$.  In principle, $\cQ_t$ is the set of all canonical states that can be obtained from the trivial canonical state, using the block-equivalence operations. We conjecture that $\cQ_t$ is generated by \emph{splitting} and \emph{twisting} operations, as described below. This conjecture is clearly reasonable -- it may even appear obvious -- but we have not proved it rigorously. If this conjecture were incorrect, it would result in a classification that is too fine, if some generators of $\cQ_t$ were missed.  Therefore, the conjecture is verified \emph{a posteriori} by matching with the Thorngren-Else classification, and by decomposition of $\cC_0(G)$ into pgSPT and weak pgSPT invariants (see Appendix~\ref{app:decomposition}).

We now describe the operations generating $\cQ_t$:

\begin{itemize}

\item \emph{Splitting operations.} Starting from the trivial canonical state, we can split the points in a given Wyckoff class into collections of nearby points. One point in each collection is a point in the original Wyckoff class, and the other points are lower-symmetry points that can be brought arbitrarily close to the original point.   An example of splitting in two dimensions is shown in the center column of Fig.~\ref{fig:p3m1}.  The role of splitting operations in obtaining $\cC_0(G)$ is illustrated below for space group \#200.  More information about splitting operations is given in Appendix~\ref{app:splitting}.

\item \emph{Twisting operations.} These operations arise for certain one-dimensional Wyckoff classes in non-symmorphic space groups.  Twisting is a sequence of splitting, sliding and grouping operations that  involves points only in a single one-dimensional Wyckoff class.  This has a non-trivial effect on $\cQ_t$, and hence on $\cC_0(G)$, only when: (1) $G_w = C_3, C_4, C_6$, and the axis swept out by a Wyckoff point is contained in a glide plane with glide direction along the axis.  (2)  $G_w = C_{2v}$, and the Wyckoff axis coincides with a four-fold screw axis.  An example of twisting is discussed below for space group \#101.  Twisting operations and their effect on $\cC_0(G)$ are described more generally in Appendix~\ref{app:twisting}.

\end{itemize}

Computations using the splitting and twisting operations are simplified by a graph-theoretic representation that saves us from complicated geometrical visualization for many different space groups. First, represent the Wyckoff classes as vertices in a directed graph that we call the $W$-graph.  If $w$ has non-trivial twisting operations, then we denote its vertex with an open circle.  Otherwise, we used filled circles when twisting operations are trivial. We will add directed edges to the graph to represent splitting operations.  Algebraically, each vertex corresponds to a $H^1$ factor in $\cQ_c$.  Edges (and open vertices) are associated with sets of generators of $\cQ_t$.

If the points in class $w$ can be slid arbitrarily close to the higher-symmetry point $w_h$, so that they can be grouped together with $w_h$, draw a directed edge $w \to w_h$.  In the corresponding splitting operation, we split each point in $w_h$ to a collection that includes the original $w_h$ point, and nearby $w$ points.

We call the splitting operation trivial if it can generate all possible values of $q_w$, while always leaving $q_{w_h}$ unchanged.  Equivalently, the corresponding grouping operation always leaves $q_{w_h}$ unchanged.  When the splitting operation is trivial,  $H^1(G_w)$  is contained in $\cQ_t$.  We draw the directed edge as a dashed arrow for trivial splitting operations, and use solid arrows when the splitting operation is non-trivial.

The charge configurations generated by the splitting operation are a property only of the $G_{w_h}$ site symmetry. This is convenient, because this means it is enough to study splitting for the crystallographic point groups, and we do not have to start from scratch for every space group.  Given a point group $G_{w_h}$, the International Tables for Crystallography enumerate  the distinct collections of nearby symmetry-equivalent points $w$.\cite{inttables}  We can then determine which charge configurations $(q_{w_h}, q_{w}) \in H^1(G_{w_h}) \times H^1(G_w)$ can be generated by the splitting operation starting from the trivial state with $(q_{w_h}, q_w) = (0,0)$.  This is done for all the crystallographic point groups in Appendix~\ref{app:splitting}.  In this notation, the splitting operation is trivial when the generated charge configurations are the set $\{ (0, q_w) \}$, for all values of $q_w$. 

To simplify the calculations further, starting from the $W$-graph, we implement a cleaning procedure to construct a $W$-quasigraph.  First, we erase each dashed arrow together with its tail vertex.  Then we continue erasing all solid arrows (together with their tail vertices) whose head vertices are already erased, until no headless solid arrows remain.   In general the $W$-quasigraph is not a true graph, because there can now be arrows lacking a tail vertex.  The erased vertices have corresponding  $H^1(G_w)$ factors lying entirely in $\cQ_t$, which disappear from $\cQ_c$ when we take the quotient.  We let $\tcQ_c$ be the product of $H^1$ factors for the vertices remaining in the $W$-quasigraph, and $\tcQ_t \subset \tcQ_c$ is generated by the splitting operations associated with the remaining solid arrows, and twisting operations associated with open vertices.  Note that arrows with a missing tail can still contribute to $\tcQ_t$.

We then proceed to compute the quotient $\tcQ_c / \tcQ_t$.  The $W$-quasigraph often breaks into disconnected components, and the quotient can be computed component-by-component, then taking the product over components.

\begin{figure}
\includegraphics[width=\columnwidth]{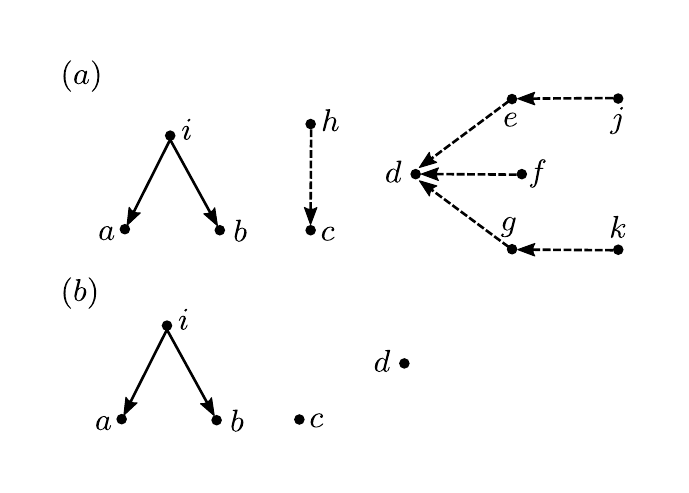}
\caption{$W$-graph (a) and $W$-quasigraph (b) for space group number 200. In the $W$-graph, some dashed arrows are omitted for clarity.}
\label{fig:sg200}
\end{figure}

We illustrate our general discussion with two example calculations of $\cC_0(G)$, beginning with the space group $P m \bar{3}$.  This is space group number 200 in the International Tables,\cite{inttables} and we refer to it as $G_{200}$. There are 12 Wyckoff classes, with letters $a, \dots, l$.  We ignore the $l$ points because their site symmetry is trivial.  Figure~\ref{fig:sg200} shows the $W$-graph and $W$-quasigraph for this space group.  To work out these graphs and understand the effect of the splitting operations, we used the entry for the space group in the International Tables and results obtained in Appendix~\ref{app:splitting}.  All twisting operations for this space group are trivial, but splitting plays a non-trivial role.

Examining the $W$-quasigraph, we see that $c$ and $d$ are isolated vertices, with $G_c = G_d = D_{2h}$.  Each vertex contributes a factor of $H^1(D_{2h}) = \zz^3$ to the classification. The non-trivial component of the $W$-quasigraph has three vertices $a,b,i$, with $G_a = G_b = T_h$ and $G_i = C_3$.  For this component, we write a general  element  $Q \in \tilde{\cal{Q}}_c$ as $Q = (q_a, q_b, q_i)$, where $q_a, q_b \in H^1(T_h) = \z_3 \times \zz$, and $q_i \in H^1(C_3) =  \z_3$.  We further write $q_a = (q^{C_3}_a, q^i_a)$, and similarly for $q_b$, where $q^{C_3}_a \in \z_3$ is the charge  associated with a $C_3$ subgroup of $T_h$, and $q^i_a \in \zz$ is the charge for the inversion subgroup of $T_h$.  We thus have the general form
\begin{equation}
Q = (q^{C_3}_a, q^i_a, q^{C_3}_b, q^i_b, q_i) \text{.}
\end{equation}

The splitting process for the $i \to a$ arrow generates $Q = (2,0,0,0,-1)=(2,0,0,0,2)$, while that for the $i \to b$ arrow generates $Q = (0,0,2,0,-1) = (0,0,2,0,2)$.  Therefore $\tilde{\cal{Q}}_t \simeq \z_3 \times \z_3$, and for this component we have the quotient $\tilde{\cal{Q}}_c / \tilde{\cal{Q}}_t = \z_3 \times \zz^2$.  Putting the results from the three components together, we have the classification
\begin{equation}
\cC_0(G_{200} ) = \z_3 \times \zz^8 \text{.}
\end{equation}

Now we describe the decomposition of $\cC_0(G_{200} )$ into pgSPT and weak pgSPT invariants. The Wyckoff classes $c$ and $d$ correspond to distinct centers of $D_{2h}$ symmetry. The pgSPT classification for $D_{2h}$ is $\zz^3$, and the $\zz^3$ factor associated with each of these vertices is a $D_{2h}$ pgSPT invariant.

Classes $a$ and $b$ have $T_h$ symmetry, where $H^1(T_h) = \z_3 \times \zz$, but where the $\z_3$ factor disappears from the pgSPT classification due to the adjoining operation.  The $T_h$ pgSPT classification is thus $\zz$, and two $\zz$ factors in $\cC_0(G_{200})$ are $T_h$ pgSPT invariants. So far, we have shown that the $\zz^8$ factor in $\cC_0(G_{200})$ is a product of pgSPT invariants.

The $\z_3$ factor in $\cC_0(G_{200})$ is a weak pgSPT invariant associated with stacking of $d=2$ pgSPT phases with $C_3$ symmetry.  To see this, we note that an element $q_3$ of the $\z_3$ factor of $\cC_0(G_{200})$ can be parametrized in terms of the canonical state charge configuration by
\begin{equation}
q_3 = q^{C_3}_a + q^{C_3}_b + 2 q_i \text{.}
\end{equation}
In fact $q_3$ measures the total $C_3$ charge in a primitive cell on the $[111]$ axis.  To see this, we have to examine the Wyckoff positions, and find all points on the $[111]$ axis within a single primitive cell.  There is a single $a$ point, a single $b$ point, and two $i$ points, corresponding to the factor of 2 in the last term.

Focusing on the $[111]$ $C_3$ rotation, and translation along the $[111]$ axis, we can view the block states we are describing as stacks of $d=2$ $C_3$ pgSPT states, with translation symmetry along the stacking direction.  $q_3$ measures the $\z_3$ pgSPT invariant per layer, which is a robust invariant in the presence of translation symmetry in the stacking direction.  This is the weak pgSPT invariant appearing as the  $\z_3$ factor in $\cC(G_{200})$.

Now we discuss an example that illustrates the role of non-trivial twisting operations.  We consider the space group $P 4_2 c m$, which we refer to as $G_{101}$ reflecting its numbering in the International Tables.  This is a non-symmorphic space group, and four-fold screw axes will play an important role.  There are four non-trivial Wyckoff classes.  One of them ($d$) has mirror site symmetry and can be trivially eliminated by grouping with $a$ points.  The remaining classes $a,b,c$ are zero-dimensional, so the $W$-quasigraph consists of three disconnected vertices.  Splitting operations are thus clearly trivial in this example.

First, considering class $c$, we have $G_c = C_2$.  Following the discussion of Appendix~\ref{app:twisting}, twisting operations are trivial for this class, because $H^1(C_2) \simeq \zz$ has no non-trivial automorphism.  Therefore class $c$ contributes a $H^1(C_2) = \zz$ factor to $\cC_0(G_{101})$.  This is a weak pgSPT invariant associated with stacking of $d=2$ $C_2$ pgSPT states.  The elementary ``translation'' symmetry along the stacking direction is actually a glide reflection.  Because this operation commutes with the $C_2$ rotation, the fact that it is a glide and not a pure translation plays no role.

We now turn to class $a$ (identical statements hold for class $b$).  Class $a$ has $G_a = C_{2v}$, and the one-dimensional axis swept out by a point in $a$ coincides with a four-fold screw axis.  As shown in Appendix~\ref{app:twisting}, twisting operations are non-trivial under these circumstances.

There are two $a$ points in a primitive cell with coordinates $(0,0,z)$ and $(0,0,z+1/2)$, which are related by the four-fold screw rotation.  This operation acts along the $z$-axis as a half translation, so we denote it by $t_h$.  Letting $\sigma_1$ and $\sigma_2$ be the two mirror reflections generating $G_a = C_{2v}$, we have
\begin{eqnarray}
t_h \sigma_1 t^{-1}_h &=& \sigma_2 \\
t_h \sigma_2 t^{-1}_h &=& \sigma_1 \text{.}
\end{eqnarray}
We denote by $q_z$ and $q_{z + 1/2}$ the $C_{2v}$ charges at the points $(0,0,z)$ and $(0,0,z+1/2)$, respectively.  Writing $q_z = (q^1_{z}, q^2_{z})$, where $q^1_{z}, q^2_{z} \in \zz$, the non-trivial action of $t_h$ on $C_{2v}$ implies
$q_{z+1/2} = (q^2_{z}, q^1_{z})$.

Charge configurations for the $a$ vertex are labeled by distinct elements $q_z \in \zz^2$, so the group of charge configurations is $\tcQ_c \simeq \zz^2$.  Because $a$ is an isolated vertex in the $W$-quasigraph, if we only considered splitting operations, we would incorrectly conclude that the $a$ class contributes a factor of $\zz^2$ to $\cC_0(G_{101})$.

\begin{figure}
\includegraphics[width=0.8\columnwidth]{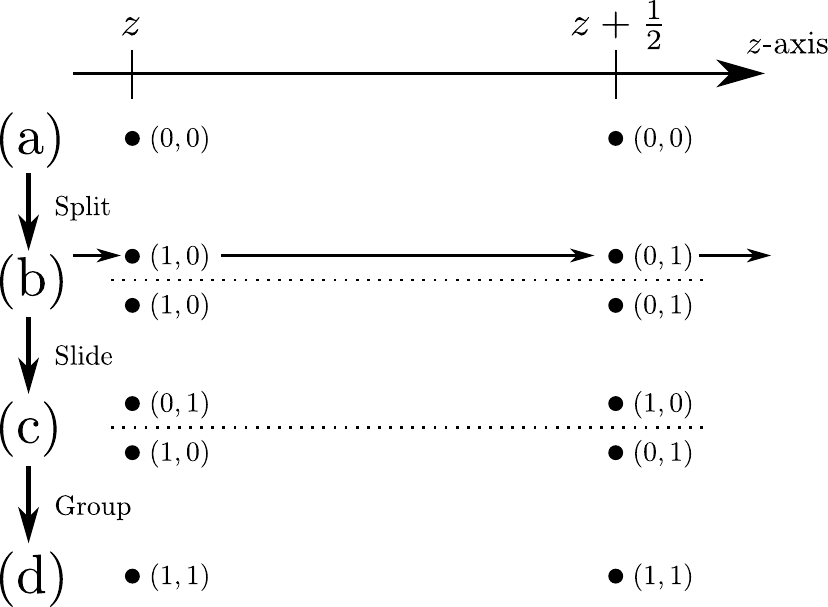}
\caption{Illustration of twisting operation for points in Wyckoff class $a$ of space group \#101.  Points $(0,0,z)$ and $(0,0,z+1/2)$ are shown as filled circles along the $z$-axis, with the $C_{2v}$ charge of each point indicated.  (a) Shows the initial state with trivial charges.  To obtain (b), each point is split into two with $C_{2v}$ charges as shown, resulting in two chains lying on the $z$-axis.  (c) is obtained by sliding the points in the top chain by $1/2$ as indicated in (b).  Finally, we obtain the final state (d) by grouping the points together and adding the $C_{2v}$ charges.}
\label{fig:twisting}
\end{figure}

We now start with the trivial charge configuration $q_z = q_{z+1/2} = (0,0) \in \zz^2$  (Fig.~\ref{fig:twisting}a), and apply block equivalence operations to obtain a non-zero element of $\tcQ_t$.  The sequence of block equivalence operations applied, taken together, is what we mean when referring to a twisting operation.  We write charge configurations as ordered pairs $[q_z, q_{z+1/2}]$, so the trivial configuration is denoted $[(0,0), (0,0)]$.  Strictly speaking, there is no need to specify $q_{z+1/2}$, as it is determined by $q_z$, but it is illustrative to keep track of both charges explicitly.

First, we split the block at $(0,0,z)$ into two blocks with charges $q_{z_1}$ and $q_{z_2}$, respectively, that we take to be $q_{z_1} = q_{z_2} =  (1,0)$.  To maintain symmetry, at the same time we must split the block at $(0,0,z+1/2)$ into two blocks with charges $q_{(z+1/2)_1} = q_{(z+1/2)_2} = (0,1)$.  This splitting operation, illustrated in Fig.~\ref{fig:twisting}b, takes a single chain of (trivial) charges on the $z$-axis to two chains of non-trivial charges.

Next, we slide the charges of the first chain along the $z$-axis until they fall into registry again with the second chain, to obtain the configuration shown in Fig.~\ref{fig:twisting}c.  This has the effect of transforming $[q_{z_1}, q_{(z+1/2)_1} ] \to [(0,1), (1,0)]$.  Finally, we group the two chains together, to again obtain a single chain of charges, which is now in the non-trivial configuration $[(1,1), (1,1)]$  (Fig.~\ref{fig:twisting}d).

Other similar operations do not produce configurations beyond $[(0,0), (0,0)]$ and $[(1,1), (1,1)]$, so we have shown these two configurations make up $\tcQ_t$, and thus $\tcQ_t \simeq \zz$.  Taking the quotient $\tcQ_c / \tcQ_t = \zz$, we see that class $a$ contributes a $\zz$ factor to the classification $\cC_0(C_{101})$.  Since the same is true for class $b$, we thus find
\begin{equation}
\cC_0(G_{101} ) = \zz^3 \text{.}
\end{equation}
In Appendix~\ref{app:twisting}, it is shown that the $\zz$ factors contributed by classes $a$ and $b$ are weak pgSPT invariants.

\section{Lieb-Schultz-Mattis constraint} 
\label{sec:LSM}

It has recently been understood there is an intimate connection between Lieb-Schultz-Mattis (LSM) constraints in $d$ dimensions, and SPT phases with crystalline symmetries in $d+1$ dimensions.\cite{cheng16translational}  Here we exploit this connection, which is a type of bulk-boundary correspondence, to obtain an LSM constraint for $d=2$ bosonic systems with wallpaper group symmetry.  This is related via the bulk-boundary correspondence to $d=3$ cSPT phases with block-dimension one.
Other LSM constraints involve a combination of internal and crystalline symmetries, and, to our knowledge, LSM constraints involving \emph{only} crystalline symmetry have not been obtained previously.  After using the bulk-boundary correspondence to obtain our LSM constraint, we give an independent argument for it based on dimensional reduction, working strictly in two dimensions.  We note that Qi, Fang and Fu have independently obtained the same LSM constraint.\cite{yangqiLSM}

By a LSM constraint, we mean a generalization of the celebrated Lieb-Schultz-Mattis theorem,\cite{lieb61} a version of which states that in a one-dimensional spin system with ${\rm SO}(3)$ spin and lattice translation symmetries, finite-range interactions, and half-odd-integer spin per primitive cell, the ground state becomes degenerate in the thermodynamic limit.  This implies that a symmetry-preserving short-range entangled ground state -- \emph{i.e.} a SPT state or other integer topological phase -- is impossible.  The LSM theorem and its generalizations are interesting in part because they show how a microscopic property -- the pattern of $S = 1/2$ projective representations in the unit cell -- constrains certain universal, infrared properties.

LSM constraints have been obtained in arbitrary spatial dimensions,\cite{oshikawa00commensurability, misguich02degeneracy, hastings04Lieb}  in systems with lattice translation combined with internal symmetry,\cite{chen11a,cheng16translational}  in systems with both space group and internal symmetry,\cite{parameswaran13topological, roy12space,watanabe15filling,po17lattice} and in systems with magnetic translation symmetry.\cite{lu17Lieb, yang17dyonic}  In all these cases, internal symmetry is involved.  It should be noted that, apart from the work of Hastings generalizing the LSM theorem to higher spatial dimensions,\cite{hastings04Lieb} these LSM constraints -- and the constraint we obtain here -- do not currently have the status of  rigorous mathematical theorems.

To illustrate the connection between LSM constraints and SPT phases, we follow the ideas of Ref.~\onlinecite{cheng16translational} and observe that a $S = 1/2$ chain can be viewed as the edge of a stack of $S = 1$ chains in the Haldane phase.  We assume translation symmetry along the stacking direction, and that the edge preserves both translation and ${\rm SO}(3)$ symmetries, so that the LSM theorem applies.  The $d=2$ bulk is a non-trivial SPT phase protected by the same symmetries, sometimes referred to as a ``weak'' SPT phase because translation symmetry is involved.  In the language of this paper, the bulk is a block-dimension one cSPT state. Then we see that the LSM constraint for the $S = 1/2$ chain is the same as the statement that a symmetric edge of this $d=2$ SPT phase cannot be gapped out trivially, \emph{i.e.} the edge cannot be in a symmetry-preserving, short-range entangled ground state.  This is what we mean in this section by bulk-boundary correspondence.

It is important to note that the statement that symmetric boundaries of SPT phases cannot be trivially gapped  is a conjecture.  Indeed, this statement is false for all block-dimension zero cSPT phases -- this is familiar from the study of reflection SPT phases in one dimension, which do not support gapless end states.  The conjecture is believed to hold for large classes of SPT phases, but in general the bulk-boundary correspondence should be viewed as a tool to obtain conjectured LSM constraints, and it is desirable to give independent supporting arguments.

To state our LSM constraint, we consider a $d=2$ spin system with wallpaper group symmetry.  Unlike in our discussions of SPT phases, we allow some spins to transform projectively under their site symmetry.  We find:
\begin{itemize}
\item[] \emph{If the system contains any spin transforming projectively under its site symmetry, a symmetry-preserving, gapped, short-range entangled ground state is impossible.}
\end{itemize}
We argue for this statement both using the bulk-boundary correspondence, viewing the $d=2$ system as the surface of a block-dimension one cSPT state, and also using an independent argument directly in two dimensions.  In addition, if \emph{only} wallpaper group symmetry is present, we use the bulk-boundary correspondence to argue the converse statement, namely that if no spins transform projectively, then a symmetry-preserving, gapped, short-range entangled ground state can occur for some choice of parameters.

We now obtain our LSM constraint from the bulk-boundary correspondence. 
 We let $G$ be a wallpaper group, and consider a $G$-symmetric surface of a $d=3$ bulk.  We take the surface normal to be along the $z$-axis.  The bulk space group is denoted $G_{3d}$ and determined by $G$ using a prescription we now describe.  Translations and rotations in $G$ correspond to translations and rotations in $G_{3d}$ in the obvious way.  Reflections and glides in $G$ correspond to vertical mirror or glide planes in $G_{3d}$.  Using this correspondence, $G_{3d}$ is generated by the operations in $G$ and by translations in the $z$-direction.  It follows that $G_{3d}$ is a product of $z$-axis translations and $G$, so that the surface termination only breaks translations along the surface normal.  In this sense, $G_{3d}$ can be viewed as a minimal ``extension'' of $G$ into three dimensions.   It should be emphasized that only block-dimension one bulk cSPT states and their classification by $\cC_1(G_{3d})$ are relevant for this discussion.

Centers of $C_n$ symmetry on the surface extend into the bulk as $C_n$ axes. Similarly, $D_n$ centers on the surface correspond to $C_{nv}$ axes in the bulk.  All these axes are parallel to the $z$-axis.  We consider block-dimension one bulk cSPT states, all of which can be obtained by placing $d=1$ SPT phases on $C_{nv}$ axes for $n = 2,4,6$.  These $d=1$ SPT phases have effective internal symmetry $\z_n \rtimes \zz \simeq C_{nv} \simeq D_n$, and obey a $\zz$ classification.    The corresponding classification of cSPT phases is given by a product of $\zz$ factors, one for each $C_{nv}$ ($n =2,4,6$) Wyckoff class in $G_{3d}$.  $C_{3v}$ and $C_n$ axes play no role.

Each symmetry-equivalent family of $C_{nv}$ axes corresponds on the surface to a Wyckoff class with $D_n$ site symmetry.  Placing non-trivial $d=1$ SPT states on the $C_{nv}$ axes corresponds to placing non-trivial $D_n$ projective representations at the points of the corresponding Wyckoff class.  We note that $D_n$ ($n = 2,4,6$) is the only two-dimensional crystallographic point group admitting non-trivial projective representations.  Moreover, $H^2(D_n, {\rm U}(1))  = \zz$ for even $n$, so there is only a single type of non-trivial $D_n$ projective representation, corresponding to the single non-trivial $d=1$ SPT phase on the $C_{nv}$ axis.

This discussion shows that a two-dimensional $G$ symmetric system can be viewed as the surface of a non-trivial $G_{3d}$ cSPT phase if and only if the two-dimensional system contains some spins transforming as non-trivial projective representations under site symmetry.  Assuming that symmetric surfaces of the relevant cSPT phases cannot be trivially gapped, our LSM constraint follows.  Moreover, a $d=2$ $G$-symmetric system in which no spins transform projectively under their site symmetry can be viewed as a surface of a trivial $d=3$ SPT phase.  We therefore expect there is no obstruction to entering a symmetry-preserving, gapped and short-range entangled phase.  This means that such a phase should occur for some choice of parameters in a Hamiltonian governing the $d=2$ system.

We now give an alternative argument for our LSM constraint, working in $d=2$ and using dimensional reduction. We note that, while Ref.~\onlinecite{song17topological} introduced dimensional reduction to classify pgSPT phases in systems with only integer spins, dimensional reduction can be carried out for any pgSPT state, whether or not some spins transform projectively.  It is enough to consider a system with $D_n$ point group symmetry ($n = 2,4,6$).  We suppose that there is a spin at the center of $D_n$ symmetry transforming as a non-trivial projective representation of $D_n$.  We will assume that a symmetry-preserving, gapped, short-range entangled state is possible, and obtain a contradiction.

The only known possibilities for the desired short-range entangled state are (1) an $E_8$ state or (2) a $D_n$ SPT state.  We are not aware of rigorous arguments showing these are indeed the only possible states, but we will assume this to be the case.  The $E_8$ state is easily excluded:  it has chiral edge modes and is thus incompatible with $D_n$ symmetry.

Therefore, we consider a $D_n$ SPT state.  We can apply dimensional reduction as in Appendix~\ref{app:pgreview} to reduce the ground state to a $D_n$-symmetric zero-dimensional region containing the center of $D_n$ symmetry.   Because spins away from the origin come in pairs, this entire zero-dimensional region must transform as a non-trivial projective representation of $D_n$.  Therefore if its ground state is symmetric, it is degenerate, which contradicts our assumption of an SPT state, and we conclude a $D_n$ SPT state is impossible under these circumstances.

More carefully, we can apply the same argument in a finite but large system with periodic boundary conditions, and then take the thermodynamic limit.   In this situation, there will generally be a finite number of centers of $D_n$ symmetry, separated from one another by lengths on the order of the system size.  At least some of these centers have spins transforming projectively under $D_n$.  If it happens that the total many-body wave function transforms projectively under $D_n$, then there is a degenerate ground state even for finite size.  We assume instead that the many-body wave function transforms linearly, so that the finite-size ground state can be unique.  Assuming a $D_n$ SPT state and applying dimensional reduction, the system reduces to a few well-separated projective spins lying at the symmetry centers, which are embedded within a trivial gapped medium.  This medium mediates exponentially decaying interactions among the projective spins.  This splits their degeneracy, but the splitting is exponentially small in the system size and vanishes in the thermodynamic limit, where the ground state becomes degenerate. This establishes our LSM constraint.

Finally, we remark that our LSM constraint winds up only involving point group symmetry in an essential way; the full wallpaper group symmetry does not play an important role.  This is the case even though the the bulk-boundary correspondence arguments leading to the constraint do include wallpaper group symmetry.  We can explain this by noting that the bulk cSPT phases involved in obtaining the LSM constraint can be understood as $C_{nv}$ pgSPT phases.

\section{Can all crystalline SPT phases be built from lower-dimensional states?}
\label{sec:general}

In this section, we argue that if a certain reasonable but unproven assumption holds, then all cSPT phases can be built from lower-dimensional invertible topological states.  We would like to be able to apply the dimensional reduction procedure of Ref.~\onlinecite{song17topological}, reviewed in Appendix~\ref{app:pgreview}, in the presence of space group symmetry.  We will see that a na\"{\i}ve application of this procedure fails, but it can be fixed if we add an extra step, which requires making a certain assumption.

We begin with a cSPT ground state $|\psi \rangle$ protected by space group symmetry. To keep the discussion simple, we assume that only space group symmetry is present.  The system may be either bosonic or fermionic.  By definition, there is a finite-depth quantum circuit $U^{loc}$ such that $\tilde{U}^{loc} | \psi \rangle = | T \rangle$, where $| T \rangle$ is a trivial product state (or atomic insulator, in a fermionic system).  In general, $\tilde{U}^{loc}$ does not respect symmetry.  

\begin{figure}
\includegraphics[width=0.7\columnwidth]{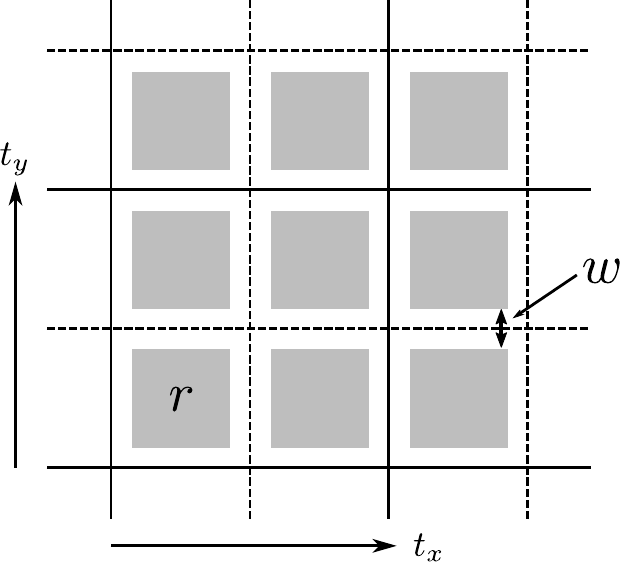}
\caption{Illustration of dimensional reduction for the wallpaper group $p2mm$.  Solid and dashed lines are reflection axes, and the elementary translations $t_x$ and $t_y$ are shown.  Region $r$ is copied using symmetry to obtain region $R$ as the union of gray-shaded squares.  The distance between neighboring squares is $w$.  Upon trivializing $R$, the system is reduced to a network of intersecting one-dimensional regions.}
\label{fig:sgtriv}
\end{figure}

To proceed, we find the largest possible spatial region so that no two points in the region are related by symmetry ($r$ in Fig.~\ref{fig:sgtriv}), and then copy this region throughout space using the symmetry, to obtain a region $R$.  An example is shown in Fig.~\ref{fig:sgtriv} for the wallpaper group $p2mm$.  We denote by $w$ the characteristic distance between connected components of $R$, as illustrated in Fig.~\ref{fig:sgtriv}.  

Next, we follow Ref.~\onlinecite{song17topological} to find a new finite-depth circuit $U^{loc}$ that locally trivializes the system in region $R$ and respects symmetry (see Appendix~\ref{app:pgreview} and Ref.~\onlinecite{song17topological} for more details).  $U^{loc}$ is constructed by first cutting the circuit $\tilde{U}^{loc}$ to obtain a new circuit supported on a region containing one of the components of $R$, and then copying the resulting circuit throughout space using the symmetry.  As discussed in Ref.~\onlinecite{song17topological} for the simple example of mirror reflection symmetry, this procedure requires that $w \gg \xi$, where $\xi$ is some characteristic correlation length of the state $| \psi \rangle$.  For point group symmetry, the region $R$ can be chosen so that $w$ is as large as desired.  However, in the present case, we have $w < a$, where $a$ is the lattice constant, and typically $a < \xi$.  Therefore we cannot follow Ref.~\onlinecite{song17topological} to construct a quantum circuit with the desired properties.

To circumvent this problem, we modify the original state $|\psi \rangle$.  First, we add a fine mesh of trivial degrees of freedom.  The mesh can be as fine as desired, and we need the mesh spacing to be much smaller than the lattice constant.  For the purposes of classifying phases, this step is certainly legitimate.

Second, we change parameters of the Hamiltonian, preserving symmetry, to entangle the new degrees of freedom with the original state, obtaining a state $| \psi' \rangle$.  Crucially, we assume that this can be done so that, by choosing a fine enough mesh, we can make the correlation length of $| \psi' \rangle$ as small as desired, and in particular $\xi \ll a$.  We believe this assumption is physically reasonable and we expect it to hold, but we do not have an argument that it is true, so it should be viewed as an unproven assumption.  We note that if this assumption is not true, it would mean there is some cSPT state with entanglement on the scale of the lattice spacing that cannot be removed, which seems unnatural.

With the correlation length of $| \psi' \rangle$ as small as desired, there is no longer an obstruction to constructing the finite-depth circuit $U^{loc}$.  We have
\begin{equation}
U^{loc} | \psi' \rangle = |T \rangle_R \otimes | \psi'' \rangle_{\bar{R}} \text{,}
\end{equation}
where $|T \rangle_R$ is a trivial product state on region $R$, and $| \psi'' \rangle_{\bar{R}}$ is some state on the complement $\bar{R}$.  This latter region can be viewed as a network of lower-dimensional systems with effective internal symmetry, and we expect that cSPT phases reduced to $\bar{R}$ as above can be constructed and classified by putting down (and perhaps gluing together) lower-dimensional invertible topological phases on various subregions of $\bar{R}$.

\section{Discussion}
\label{sec:discussion}

In this paper, we considered bosonic  crystalline SPT (cSPT) phases protected by space group or point group symmetry, and classified a subset of such phases built from lower-dimensional SPT blocks.  Our classification matches that of Thorngren and Else, obtained by very different methods, for  wallpaper groups in $d=2$ and space groups in $d=3$.  This allows us to clarify the physical properties of the states classified by Thorngren and Else, and, combined with a general argument based on a reasonable but unproven assumption, is evidence that all SPT phases protected by crystalline symmetry can be built from lower-dimensional blocks of invertible topological states.  Moreover, for the states we classified, there are no new SPT invariants beyond point group SPT (pgSPT) invariants, in the sense that the classifications  can be decomposed into point group SPT (pgSPT) and weak pgSPT invariants.  Finally, we obtained a Lieb-Schultz-Mattis (LSM) type constraint for $d=2$ spin systems that only involves crystalline symmetry, as opposed to the interplay between internal and crystal symmetries.

We conclude with a discussion of some possible extensions of the results presented here, and remarks on the connection between our results and the approach to LSM constraints in Ref.~\onlinecite{po17lattice}.  For simplicity, we focused in this paper on phases where the building blocks are lower-dimensional SPT states.  This ignores $d=3$ bosonic cSPT phases that can be built from $E_8$ states.\cite{song17topological, lu17classification}  We announce some preliminary results on the classification of these states that will be presented in a separate paper.  Let $G$ be a $d=3$ point group or space group.  If $G$ has only orientation-preserving symmetries, there are no $E_8$ based states.  This is consistent with the conjecture of Thorngren and Else that their classification is complete for such $G$.\cite{thorngren16gauging}  If $G$ has any orientation-reversing symmetries, then there are non-trivial $E_8$ based states, which add a single $\zz$ factor to the classification of cSPT phases.  We conjecture that complete classifications for $d=3$ bosonic cSPT phases are obtained by combining these results with the classifications obtained in Ref.~\onlinecite{thorngren16gauging} and here.

The approach developed here can be extended to treat SPT phases in $d$ dimensions with both internal and crystalline symmetries.  The key modification is that now blocks of dimension $d$ are needed, which have no effective internal symmetry coming from the crystalline symmetry, but do have true internal symmetry and can thus host $d$-dimensional internal-symmetry SPT states.  These blocks then need to be glued together consistent with the crystal symmetry.  Another modification is that the structure of adjoining and grouping/splitting/sliding operations should become richer, because internal-symmetry SPT states of various dimensionalities can exist away from high symmetry subspaces.  We conjecture that a generalization of our approach along these lines can produce complete classifications of SPT phases with both internal and crystal symmetries.  It will also be interesting to extend our approach to fermionic SPT phases in future work, both with and without internal symmetry.

As noted above, our block equivalence operations are closely related to the lattice homotopy operations introduced in Ref.~\onlinecite{po17lattice} in connection with LSM constraints.  We now describe the precise relationship and comment on some possible implications.  Ref.~\onlinecite{po17lattice} considered $d$-dimensional bosonic systems on a lattice $\Lambda$ with symmetry $G = G_s \times G_i$, with $G_s$ is a space group and $G_i$ an internal symmetry.  To each lattice site in $\Lambda$ is associated an element of $H^2(G_i, {\rm U}(1))$, which characterizes the $G_i$ representation of degrees of freedom at that site.  Spins are assumed to transform linearly under $G_s$, and, moreover, if $g_i \in G_i$ and $g_s \in G_s$, the action of $g_i$ and $g_s$ commutes on spins.  Lattice homotopy operations were introduced, where lattice sites can be slid, grouped and split, and where grouping and splitting respects the $H^2(G_i, {\rm U}(1))$ group operation.  These operations define equivalence classes of lattices $[ \Lambda ]$.  Ref.~\onlinecite{po17lattice} conjectured that a LSM type constraint holds whenever $[ \Lambda ]$ is non-trivial, \emph{i.e.} whenever the lattice cannot be deformed to the trivial lattice.  They established this conjecture in a wide range of cases using arguments based on flux insertion.

The lattice homotopy operations of Ref.~\onlinecite{po17lattice} are a special case of block-equivalence operations.  We consider $d+1$-dimensional SPT states also with symmetry $G = G_s \times G_i$, where now $G_s$ is a $d+1$-dimensional space group that is preserved at a $d$-dimensional surface.\footnote{Strictly speaking, we should include in $G_s$ translations along the surface normal, which are  broken by the surface, but these operations play no role in our discussion.}  We restrict to SPT states built by placing one-dimensional $G_i$-symmetric SPT phases on axes normal to the surface.  These SPT states are labeled by elements of $H^2(G_i, {\rm U}(1))$, and their block equivalence operations are precisely the lattice homotopy operations of Ref.~\onlinecite{po17lattice}.  Indeed, the surface terminations of these one-dimensional SPT states are precisely projective representations labeled by the same element of $H^2(G_i, {\rm U}(1))$, so these operations are really physically identical.

These observations allow us to rephrase the conjecture of Ref.~\onlinecite{po17lattice} in terms of a bulk-boundary correspondence, in the spirit of Ref.~\onlinecite{cheng16translational} and our results of Sec.~\ref{sec:LSM}.  We see that $[ \Lambda ]$ is non-trivial precisely when the corresponding  block-equivalence class of $d+1$-dimensional SPT block states is non-trivial.  Then the conjecture of Ref.~\onlinecite{po17lattice} becomes the statement that a non-trivial block-equivalence class implies the corresponding SPT phase is non-trivial, and that symmetry-preserving surfaces of this SPT phase are not trivially gappable.

It should be emphasized that this statement is also a conjecture that needs to be shown.  The first part of the statement -- non-trivial block-equivalence class implies non-trivial SPT phase -- can likely be shown in particular cases, and perhaps in general, by decomposing the block-equivalence classification into invariants associated with point groups, along the lines of Appendix~\ref{app:decomposition}.  Such invariants, including internal symmetry, can be obtained via the dimensional reduction approach of Ref.~\onlinecite{song17topological}.  It may be possible to establish the second part of the statement -- symmetry-preserving surfaces are non-trivial -- by generalizing and perhaps combining the flux-insertion arguments of Ref.~\onlinecite{po17lattice} and the dimensional reduction argument of Sec.~\ref{sec:LSM}.

The above discussion leads immediately to a host of new conjectured LSM constraints.  An axis $a$ penetrating into the SPT bulk has effective internal symmetry $G_a \subset G_s$, and, taking advantage of this, we can place one-dimensional SPT phases classified by $H^2(G_a \times G_i, {\rm U}(1))$ on the axis.  This allows for corresponding spin systems where lattice symmetries act projectively, and/or where some internal symmetry operations do not commute with site symmetries.  We are naturally led to the conjecture that a LSM constraint holds for this spin system if the corresponding block equivalence class is non-trivial.  Section~\ref{sec:LSM} establishes this conjecture in the special case of two-dimensional spin systems with no internal symmetry.

\acknowledgments{M.H. would like to thank Sid Parameswaran and Michael Zaletel for useful discussions.  We are grateful to Dominic Else for useful correspondence, and especially grateful to Liang Fu for collaboration on related prior work.  S.-J.H., Y.-P.H. and M.H. were supported by the U.S. Department of Energy, Office of Science, Basic Energy Sciences (BES) under Award number DE-SC0014415.   H.S. acknowledges financial support from the Spanish MINECO grants FIS2012-33152, FIS2015-67411, and the CAM research consortium QUITEMAD+  Grant No. S2013/ICE-2801.}

\appendix

\section{Dimensional reduction approach to point group SPT classification}
\label{app:pgreview}

Here, we review and illustrate the dimensional reduction approach to pgSPT classification given in Ref.~\onlinecite{song17topological}.  We focus on the illustrative examples of $C_2$ symmetry in $d=2$, and $C_i$ (inversion) symmetry in $d=3$, which we treat simultaneously.  These examples allow us to highlight the key points and illuminate a more general statement about dimensional reduction.

\begin{figure}
\includegraphics[width=\columnwidth]{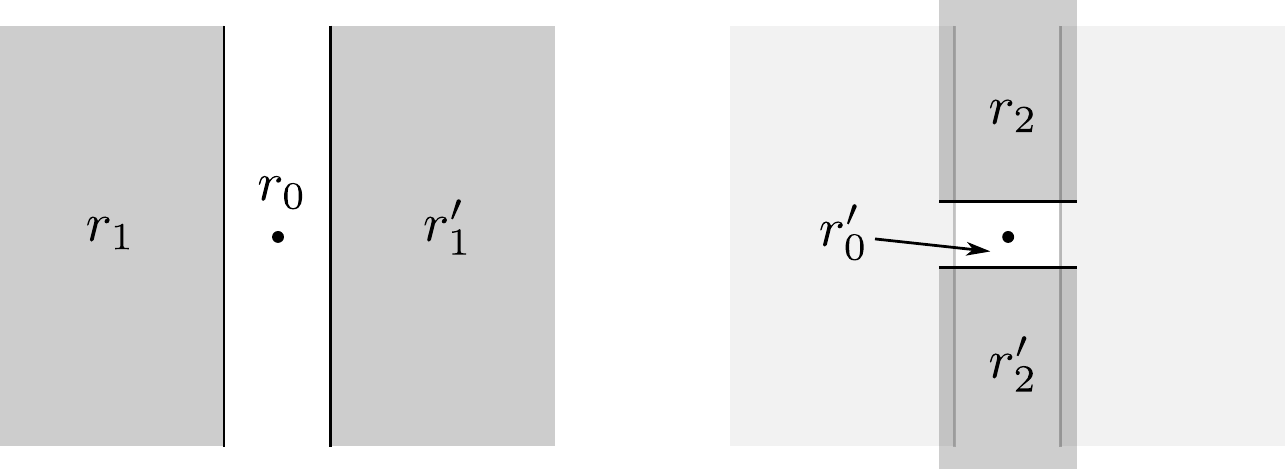}
\caption{Regions for dimensional reduction of a $d=2$ pgSPT state protected by two-fold rotation ($C_2$), or a $d=3$ pgSPT state protected by inversion ($C_i$).  In the $d=3$ case, only the left panel is relevant, and should be interpreted as a cross section through the origin.  The black dot is the symmetry center. The left panel illustrates the first dimensional reduction step, and the right panel illustrates the second step for the $d=2$ case.}
\label{fig:dimr1}
\end{figure}

Figure~\ref{fig:dimr1} shows two-dimensional space (for the $C_2$ example), or a cross section through the origin in three-dimensional space (for $C_i$).  In the left panel of the figure, space is divided into three regions $r_0$, $r_1$ and $r'_1$.  The latter two regions are semi-infinite and are images of one another under the $C_2$ or $C_i$ symmetry.  The region $r_0$ is a strip (in two dimensions) or a slab (in three dimensions), that is invariant under the symmetry and contains the origin.  The thickness $w$ of $r_0$ should be taken much larger than any correlation length $\xi$, but still finite when taking the thermodynamic limit.

If $| \psi \rangle$ is a pgSPT ground state under the appropriate symmetry, the arguments of Ref.~\onlinecite{song17topological} show that the ground state is adiabatically connected (preserving symmetry) to a state of the form $| T \rangle_{r_1} \otimes | \psi \rangle_{r_0} \otimes | T \rangle_{r'_1}$, where $| T \rangle_{r_1}$ and $| T \rangle_{r'_1}$ are trivial product states related to one another by symmetry, and $| \psi \rangle_{r_0}$ is a possibly non-trivial state defined in $r_0$ that is invariant under the symmetry.

Ref.~\onlinecite{song17topological} describes how to construct a finite-depth, symmetry-preserving quantum circuit achieving this dimensional reduction, that is
\begin{equation}
U^{loc} | \psi \rangle = | T \rangle_{r_1} \otimes | \psi \rangle_{r_0} \otimes | T \rangle_{r'_1} \text{.} \label{eqn:dimr}
\end{equation}
The finite-depth circuit $U^{loc}$ is constructed starting from the non-symmetry preserving circuit $\tilde{U}^{loc}$ that trivializes the state $|\psi\rangle$, and which must exist by the assumption that we have a SPT phase.  That is,
\begin{equation}
\tilde{U}^{loc} | \psi \rangle = | T \rangle \text{,}
\end{equation}
where $T$ is a trivial product state.  To construct $U^{loc}$ from $\tilde{U}^{loc}$, we first cut $\tilde{U}^{loc}$ to obtain a new circuit $U^{loc}_{r_1}$ with support in a region containing $r_1$, and extending slightly into $r_0$.  Then we conjugate $U^{loc}_{r_1}$ by the symmetry operation to obtain a similar circuit in region $r'_1$, $U^{loc}_{r'_1}$.  We have
\begin{equation}
U^{loc} = U^{loc}_{r_1} U^{loc}_{r'_1} \text{,}
\end{equation}
and the action of $U^{loc}$ on $|\psi \rangle$ is as given in Eq.~(\ref{eqn:dimr}).  For a more detailed discussion, the reader should consult Ref.~\onlinecite{song17topological}.

So far, we have reduced a $d$-dimensional pgSPT state to some state in $d-1$ dimensions.  However, we have not yet finished with dimensional reduction; we would like to reduce the state to a space where the symmetry acts only as an internal symmetry.  In both our examples, this means reducing down to a zero-dimensional region centered at the origin.  

Let us first consider $C_2$ symmetry in two dimensions.  Proceeding as before, we divide the strip $r_0$ into three regions -- two semi-infinite strips $r_2$ and $r'_2$ that are related by $C_2$ rotation, and a region $r'_0$ centered on the origin, as shown in the right panel of Fig.~\ref{fig:dimr1}.  Focusing on $r_2$, we have an effectively one-dimensional topological state with no symmetry.
In a bosonic system, such a state is trivial, so we can trivialize $r_2$ away from the origin by acting with a one-dimensional quantum circuit, and we can trivialize $r'_2$ at the same time by copying this circuit using the $C_2$ symmetry.

To describe the resulting state in language that generalizes to arbitrary point groups, we recall that the subset $S \subset {\mathbb R}^d$ was defined to be the union of all points in space fixed by at least one non-trivial point group operation $g \in G$.  In the present case, $S$ is just a single point at the origin.  Then we define $S_t$ to be a thickened version of $S$ that remains invariant under symmetry.  In the present example, we can take $S_t = r'_0$.  Finally, let $\bar{S_t}$ be the complement of $S_t$ in ${\mathbb R}^d$.  The second step of the dimensional reduction procedure then shows 
\begin{equation}
|\psi \rangle \to | T \rangle_{\bar{S_t}} \otimes | \psi \rangle_{S_t} \text{,} \label{eqn:S-state}
\end{equation}
where the arrow denotes adiabatic continuity,  where $ | T \rangle_{\bar{S_t}} $ is a trivial product state on $\bar{S_t}$, and $| \psi \rangle_{S_t}$ is a state on $S_t$ that may be non-trivial.

In our two-dimensional example, arriving at Eq.~(\ref{eqn:S-state}) did not require any assumptions beyond $| \psi \rangle$ being a pgSPT state.  The situation is different in $d=3$, where we do have to make an additional assumption, which amounts to excluding certain pgSPT phases from consideration.  In three dimensions, the slab $r_0$ is an effectively two-dimensional system with $C_2$ rotation symmetry.  If we zoom in and look at a piece of $r_0$ away from the origin, we have a two-dimensional system with no symmetry at all.  Unlike in the previous example, such a system can be in an $E_8$ state, which is robust in the absence of symmetry.  Indeed, the $E_8$ state is compatible with $C_2$ rotation symmetry, and the whole slab $r_0$ can be in an $E_8$ state.  If this happens, the second dimensional reduction step, where we attempt to reduce $r_0$ down to a lower-dimensional region, fails, because a two-dimensional quantum circuit cannot trivialize the $E_8$ state.

In this paper, we are primarily interested in crystalline SPT phases built from lower-dimensional SPT building blocks. We encountered an obstruction to continuing the dimensional reduction in a pgSPT state built from an $E_8$ state, which is not an SPT state, so we should exclude it from consideration in keeping with our focus.  Therefore we assume that $E_8$ states do not appear at any stage of the dimensional reduction procedure.  Because $E_8$ states (and multiple copies thereof) are believed to be the only bosonic invertible topological phases that are not SPT phases, this amounts to considering only those pgSPT phases built from lower-dimensional SPT blocks, as desired.  It is straightforward to extend our analysis to include pgSPT phases built from $E_8$ states, and indeed this was done for mirror reflection and $C_{2v}$ in Ref.~\onlinecite{song17topological}, but for other point groups we leave consideration of such states for future work.

Once we assume that an $E_8$ state does not appear, we can continue the dimensional reduction in our $d=3$ example to obtain a state of the form Eq.~(\ref{eqn:S-state}).  (We actually need two more steps, first to reduce $r_0$ to a quasi-one-dimensional strip, then to a zero-dimensional region centered on the origin.)  In general, with the present assumptions, any pgSPT state can be reduced to a state of the form Eq.~(\ref{eqn:S-state}).

A state of the form Eq.~(\ref{eqn:S-state}) can be understood in terms of SPT blocks with effective internal symmetry.  To see this, we work in $d=3$, and assume for concreteness that $S$ has some points whose neighborhood in $S$ (intersection of a ball containing the point with $S$) is two-dimensional.  Such a two-dimensional portion of $S$ is a mirror plane, and zooming in on some two-dimensional portion of $S$, we have an effectively $d=2$ system with $\zz$ effective internal symmetry.  This system can either be in a non-trivial Ising SPT phase, or it can be trivial.  (It cannot be an $E_8$ state by the assumption we made above.)

If some of the planes in $S$ host non-trivial states, we can construct a reference state with $G$ symmetry and the same pattern of Ising SPT states on mirror planes, and then make a bilayer of this state with the original ground state.  This makes all the planes in $S$ trivial.  Next, we can find one-dimensional portions of $S$, consisting of points whose neighborhood is one-dimensional, or that lie at the intersection of two or more planes.  These one-dimensional portions of $S$ can be in one-dimensional SPT states.  Proceeding along these lines, we see that states of the form  Eq.~(\ref{eqn:S-state}) can be understood in terms of lower-dimensional SPT blocks.

\section{The first cohomology group $H^1(G, {\rm U}(1))$}
\label{app:coho}

Here, we define the first cohomology group $H^1(G, {\rm U}(1))$, which is used throughout the paper in the description of block-dimension zero states.  This is standard material; we provide it here in the interest of making our paper more accessible and self-contained.

Let $G$ be a group, and let $\omega : G \to {\rm U}(1)$ be a one-dimensional representation of $G$.  This means that $\omega(g_1) \omega(g_2) = \omega(g_1 g_2)$.  As a set, $H^1(G, {\rm U}(1))$ is the set of one-dimensional representations of $G$.  We give this set an Abelian group structure via the tensor product operation; that is, if $\omega_1$ and $\omega_2$ are one-dimensional representations, their product $\omega_1 \omega_2$ is defined by:
\begin{equation}
(\omega_1 \omega_2)(g) = \omega_1(g) \omega_2(g) \text{.}
\end{equation}

We make two notational comments.  First, because we only use the first cohomology group with ${\rm U}(1)$ coefficients in this paper, we sometimes omit the coefficient group and write $H^1(G) \equiv H^1(G, {\rm U}(1))$.  Second, in this appendix, we use multiplicative notation to define $H^1(G, {\rm U}(1))$, but we use additive notation for cohomology groups in the rest of the paper.

\section{Details of block dimension zero states}
\label{app:zero}

Block dimension zero states are introduced in Sec.~\ref{sec:2dcSPT}.  Here, we consider some technical details of such states.  First, we show a statement made in Sec.~\ref{sec:2dcSPT}, that knowing the charges $q_p$ completely determines $\lambda(g,p)$ up to some gauge-like freedom.   This is why it is enough to specify $q_p$ in the data characterizing a state.  Second, if $|\Psi\rangle$ is a block dimension zero state invariant under a point group $G$, we describe how to compute $U_g | \Psi \rangle$, for $g \in G$.  The latter result is used to work out the splitting operations described in Appendix~\ref{app:splitting}.

\begin{figure}
\includegraphics[width=0.9\columnwidth]{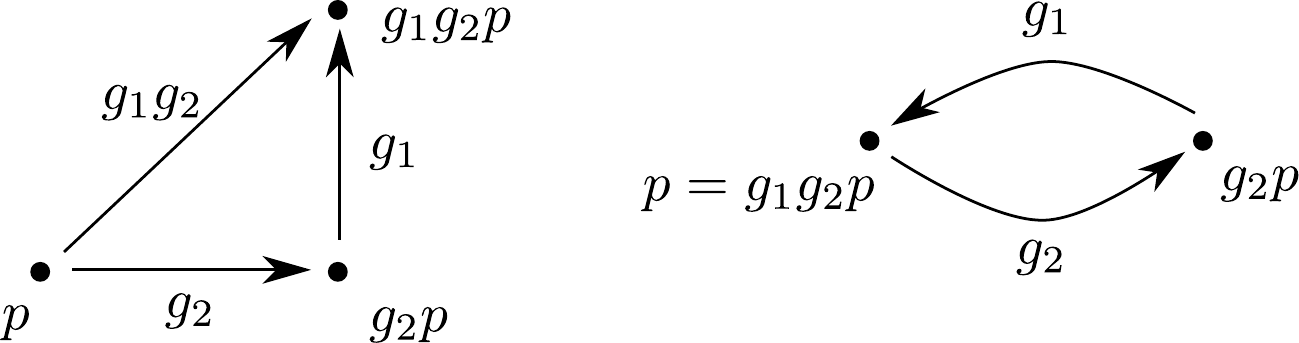}
\caption{Graphical interpretation of Eq.~\ref{eqn:lambda-appendix}.  Vertices are  points $p \in \Lambda$ and directed edges are associated with group elements $g$ joining $p$ to $g p$.  On the left, we show the case where $p$, $g_2 p$ and $g_1 g_2 p$ are all different.  Knowing $\lambda$ on any two of the edges determines it on the third, via Eq.~\ref{eqn:lambda-appendix}.  On the right, we have the case  $g_1 g_2 p = p$, \emph{i.e.} $g_1 g_2 \in G_p$.  In this case, $\lambda$ on one edge is determined by its value on the other edge, and by $\lambda(g_1 g_2, p) = D_{q_p}(g_1 g_2)$.}
\label{fig:lambda}
\end{figure}

To show $\lambda(g,p)$ is determined by the charges $q_p$, we introduce  in Fig.~\ref{fig:lambda} a graphical representation of the relation
\begin{equation}
\lambda(g_1 g_2, p) = \lambda(g_1, g_2 p) \lambda(g_2, p) \text{.} \label{eqn:lambda-appendix}
\end{equation}
This graphical representation allows us to think about $\Lambda$ as the vertices of a directed multi-graph, where each directed edge joining $p$ to $g p$ is labeled by the group element $g$.

First, suppose we know $\lambda(g,p)$ for all $p \in \Lambda$ but only for $g \in G_p$; that is, we specify $q_p$.  Aided by the graphical representation, we can build up the rest of $\lambda(g,p)$.  The graph associated with $\Lambda$ 
will in general have some number of disconnected components, because not all points are related by symmetry.  For each component, we choose a connected subgraph that is a tree, and for each edge in the tree we set the corresponding $\lambda(g,p) = 1$.  We can then use Eq.~(\ref{eqn:lambda-appendix}) to uniquely determine all the other $\lambda(g,p)$'s, corresponding to the edges we left out.
 
Next, suppose we have a function $\lambda(g,p)$ satisfying Eq.~(\ref{eqn:lambda-appendix}).  Again we choose the same tree structure, and we observe that making the change of basis
\begin{equation}
| \psi_p \rangle \to \alpha(p) | \psi_p \rangle 
\end{equation}
induces the transformation
\begin{equation}
\lambda(g,p) \to \alpha(gp) \lambda(g,p) \alpha^{-1}(p) \text{.}
\end{equation}
It is clear that we can make such a transformation to set $\lambda(g,p) = 1$ on the edges of the tree.  Once in this ``gauge,'' the other values of $\lambda(g,p)$ with $gp \neq p$ are then determined by the $D_{q_p}$'s using Eq.~(\ref{eqn:lambda-appendix}).  We have thus shown that $\lambda(g,p)$ is complete determined by the charges $q_p$, up to gauge-like freedom that physically corresponds simply to a site-dependent change of basis.

Now we consider a different question.  Suppose that $\Psi \in {\cal B}_0$ is invariant under the point group $G$.  We would like to compute $U_g | \Psi \rangle$ for some $g \in G$.  Clearly $U_g | \Psi \rangle = \lambda_{\Psi} | \Psi \rangle$, and our task is to determine the phase factor $\lambda_{\Psi}$.  To do this, we divide $\Lambda$ into its orbits $O_1, \dots, O_k$ under the action of $g$.  For each orbit we define 
$| \Psi_{O_i} \rangle \equiv \bigotimes_{p \in O_i} | \psi_{p} \rangle$
so that $| \Psi \rangle = \bigotimes_{i = 1}^k | \Psi_{O_i} \rangle$.  Clearly $U_g | \psi_{O_i} \rangle = \lambda_{O_i} | \psi_{O_i} \rangle$.  Therefore,
\begin{equation}
\lambda_{\Psi} = \prod_{i = 1}^k \lambda_{O_i} \text{,}
\end{equation}
and we need to determine the $\lambda_{O_i}$ phase factors.

If $O_i$ consists of a single point $p$, then $\lambda_{O_i} = \lambda(g,p)$.  Now suppose $O_i$ contains $n > 1$ points.  Using the definition $U_g | \psi_p \rangle = \lambda(g,p) | \psi_{g p} \rangle$, and using Eq.~(\ref{eqn:lambda-appendix}) repeatedly, we obtain
\begin{equation}
U_g | \psi_{O_i} \rangle = \lambda(g^n, p_1) | \psi_{O_i} \rangle \text{,}
\end{equation}
so $\lambda_{O_i} = \lambda(g^n, p_1)$.  If $g^n = 1$, then $\lambda_{O_i} = 1$.  This is always the case if $g$ is a rotation, mirror reflection, or inversion operation, so that for these symmetries only points fixed by $g$ contribute to the total $g$ charge of $| \Psi \rangle$.  For rotation-reflections $g = S_3, S_4$ or $S_6$, points on the axis form orbits of size two.  In these three cases, since $g^2$ is $C_3^2, C_2$ and $C_3$, respectively, pairs of points on the axis give a contribution determined by the rotation charge of one point in the pair.  For points off the axis and away from the origin, orbits of rotation-reflections still satisfy $g^n = 1$.

\section{Block dimension factorization}
\label{app:block-dim-factor}

Here, we describe the general structure of how $\cC(G)$ decomposes into states of fixed block dimension, and give arguments that
\begin{equation}
\cC(G) = \cC_0(G) \times \cC_1(G) \times \cC_2(G) \label{eqn:app-dim-factor}
\end{equation}
 for $d=3$ bosonic cSPT phases built from lower-dimensional SPT blocks.  We also discuss an example where the factorization does not hold, which illustrates the general structure.

The general structure is as follows.  We let ${\cal D}_{d_b}(G)$ be the classification of cSPT phases with block dimension \emph{less than or equal to} $d_b$.  These phases clearly form a group under the usual stacking operation, because adding two states in ${\cal D}_{d_b}(G)$ cannot produce a state with higher block dimension.  Moreover, we have a sequence of subgroups ${\cal D}_{d_b - 1}(G) \subset {\cal D}_{d_b}(G)$.  We also have ${\cal D}_0(G) = \cC_0(G)$, and ${\cal D}_2(G) = \cC(G)$.

States with fixed block dimension $d_b > 0$ need not form a group, but they do form a group up to stacking with lower-dimensional block states.  That is, we can define
\begin{equation}
\cC_{d_b}(G) = \frac{{\cal D}_{d_b}(G)}{{\cal D}_{d_b - 1}(G)} \text{.}
\end{equation}
We would like to show that ${\cal D}_{d_b}(G) \simeq \cC_{d_b}(G) \times {\cal D}_{d_b - 1}(G)$, which is the desired factorization.

We will consider stacking of $d_b = 1$ and $d_b = 2$ blocks, and show these states form a group under the stacking operation. It is enough to consider a single $d_b = 1$ or $d_b = 2$ state, and show that a trivial state results when it is stacked with itself.

We start with $d_b = 1$.  It is sufficient to focus on a single block $b$, which is a $C_{nv}$ axis with $n = 2,4,6$.  $b$ is invariant under a symmetry group $G_{1d}$ containing the effective internal symmetry $G_b \simeq C_{nv}$ as a subgroup.  We consider a specific model of the non-trivial $d=1$ SPT state:  we label lattice sites along the $d=1$ axis by $i$, and at each site we place a tensor product of two $S = 1/2$ spins, with spin operators $\vec{S}_{L i}$ and $\vec{S}_{R i}$.  $G_{1d}$ may contain $d=1$ inversion symmetry; in that case, we choose all sites $i$ to lie away from inversion centers.  The Hamiltonian is
\begin{equation}
H_{1d} = \sum_i \vec{S}_{R i} \cdot \vec{S}_{L, i+1} \text{.}
\end{equation}
If we project onto the $S = 1$ subspace at each site, the ground state becomes the Affleck-Kennedy-Lieb-Tasaki (AKLT) state.\cite{affleck87rigorous}  Even before projection the ground state is in the Haldane phase, \emph{i.e.} if we consider  full ${\rm SO}(3)$ spin symmetry, the ground state is in the non-trivial SPT phase.  Recalling that $C_{nv} \simeq \z_n \rtimes \zz$, we associate the $\z_n$ factor with $2 \pi / n$ rotations about some axis in spin space, and the $\zz$ factor with $\pi$ rotations about a perpendicular axis.  The ground state is also in the single non-trivial SPT phase under this lower symmetry.

\begin{figure}
\includegraphics[width=\columnwidth]{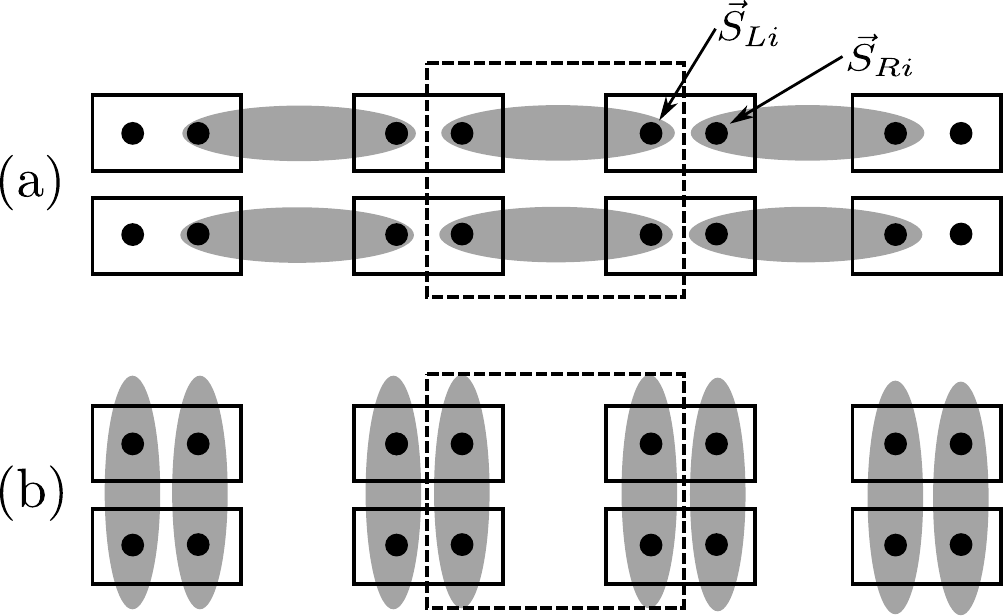}
\caption{(a) Stack of two AKLT-type spin chains on a $d_b = 1$ block $b$ with $G_b \simeq C_{nv}$, as described in the text.  Squares represent lattice sites comprised of a tensor product of $S = 1/2$ spins (black dots), with spin operators $\vec{S}_{R i}$ and $\vec{S}_{L i}$.  Gray ovals represent singlet pairs.  The dashed-line box shows where unitaries act to transform the state to the manifestly trivial state shown in (b).}
\label{fig:1dstack}
\end{figure}

Now we stack two such spin chains on $b$; the resulting state $| \psi_{{\rm stack}} \rangle$ is represented in Fig.~\ref{fig:1dstack}a.  We act on $| \psi_{{\rm stack}} \rangle$ with a product of unitaries, where each unitary acts on the tensor product space of the four $S = 1/2$ spins participating in the bond joining $i$ to $i+1$ (\emph{i.e.}, the $\vec{S}_{R i}$ and $\vec{S}_{L, i+1}$ spins in each chain).  It is clear that these four spins can be transformed into the singlet state shown in Fig.~\ref{fig:1dstack}b by a symmetry-preserving unitary, taking into account that the bond can be a center of inversion in $G_{1d}$.  The resulting state is a trivial block-dimension zero state; each site is fixed only by the $C_{nv}$ subgroup of $G_{1d}$, and carries trivial $C_{nv}$ charge.

To conclude this discussion, we consider stacking two identical $d_b = 2$ blocks on a mirror plane, each hosting an Ising SPT state.  We do not specify the Hamiltonian for these blocks, but focus on the ground state wave functions.  For layer $i$ ($i = 1,2$), we consider the wave function\cite{levin12braiding}
\begin{equation}
| \psi_i \rangle = C \sum_{D_i} (-1)^{N(D_i)} | D_i \rangle \text{,}
\end{equation}
where the sum is over all Ising domain wall configurations, $N(D_i)$ is the number of closed domain wall loops in $D_i$, and $C$ is a normalization constant.  Such a wave function can be implemented at the lattice scale consistent with any spatial symmetries of the $d_b = 2$ plane.

Stacking the two blocks together results in the wave function
\begin{equation}
| \psi_{{\rm stack}} \rangle = C^2 \sum_{D_1, D_2} (-1)^{N(D_1) + N(D_2) } | D_1 \rangle \otimes | D_2 \rangle \text{.}
\end{equation}
We now add a ferromagnetic Ising exchange coupling the two layers.  This interaction has the effect of ``lining up'' the domain walls, and as the strength of the interaction is increased, configurations with $D_1 = D_2$ will dominate the wave function.  We expect that the coupling can be made strong without passing through a phase transition, and in the limit of strong coupling the wave function becomes
\begin{equation}
| \psi_{{\rm stack}} \rangle = C' \sum_{D} | D \rangle \otimes |D \rangle \text{,}
\end{equation}
where $C'$ is a normalization constant.  This wave function is a trivial product state, with sites carrying trivial site symmetry charge.  

While Eq.~(\ref{eqn:app-dim-factor}) holds for the bosonic cSPT phases studied in this paper, it does not hold in general.  To illustrate this, we briefly discuss an example\cite{yoshida15correlation} of fermionic SPT phases where the factorization does not hold.  We consider electron systems in $d=3$ with $[{\rm U}(1) \rtimes \zz^T ] \times \zz^P$ symmetry, where $\zz^P$ is mirror reflection, and $\zz^T$ is time reversal, which squares to fermion parity.  We consider a SPT state whose symmetry-preserving surface has a single massless Dirac fermion.  This state can of course be viewed as the familiar topological band insulator if we ignore the $\zz^P$ symmetry.  Similarly, if we ignore  $\zz^T$, it is a non-trivial topological crystalline insulator.

Because this state is non-trivial even ignoring the spatial symmetry, it should be viewed as a $d_b = 3$ state.  Now, stacking two of these states together produces a state whose surface is two massless Dirac fermions.  This state is trivial if we ignore the $\zz^P$, but it is a non-trivial topological crystalline insulator that can be dimensionally reduced to the mirror plane.\cite{isobe15, song17topological}  Therefore, in this example, we stacked two $d_b = 3$ states to obtain a non-trivial $d_b = 2$ state.  This implies the classification does not factorize over block dimensions.

\section{Splitting operations and point group SPT classification for block dimension zero states}
\label{app:splitting}

This appendix pertains to the classification of block-dimension zero cSPT phases in two and three dimensions, both for point group symmetry and space group symmetry.  In particular, we consider splitting operations for point groups as discussed in Sec.~\ref{sec:3dcSPT}.  We develop a formalism to describe splitting operations, and use this to explain how splitting operations are related to the adjoining operation in the classification of pgSPT phases. We show that $\cC_0(G) = H^1(G) / {\rm Adj}(G)$, where $G$ is a point group with zero-dimensional fixed space, and ${\rm Adj}(G)$ is a subgroup of $H^1(G)$ that we define.  Then, we enumerate those crystallographic point groups with non-trivial splitting operations, give the charge configurations generated by splitting, and determine ${\rm Adj}(G)$.

Let $G$ be a crystallographic point group, and let $w_0$ be a Wyckoff class containing a single center $p_0$ of $G$ symmetry, so that $G_{w_0} = G$.  Moreover, let $w$ be a Wyckoff class containing a collection of symmetry-equivalent points that can be slid arbitrarily close to $p_0$.  Each point in $w$ has site symmetry $G_w$, and taken together, the points in $w$ form a pattern with $G$ symmetry.  For each $G$, the distinct possible classes $w$ can be found by consulting the International Tables for Crystallography.

We place zero-dimensional blocks at $p_0$ and at the points of $w$, so that a block state is specified by the charge configuration $(q_{w_0}, q_w) \in H^1(G_{w_0}) \times H^1(G_w)$.  Here, $q_w$ is the charge of some arbitrarily chosen representative point in $w$.

The block state with charge configuration $(0, q_w)$ transforms as a one-dimensional representation of $G$, with $G$-charge given by $g_w(q_w)$.  That is, applying the grouping block-equivalence operation to this state, we get a new state labeled by $(g_w(q_w), 0)$.  Formally, there is a group homomorphism
\begin{equation}
g_w : H^1(G_w) \to H^1(G_{w_0}) \text{.}
\end{equation}
More generally, the total $G$ charge of a block state labeled by $(q_{w_0}, q_w)$ is given by
$q_{w_0} + g_w(q_w)$.  The homomorphism $g_w$ can be computed by following the discussion in the latter part of Appendix~\ref{app:zero}.

We are interested in knowing which block states can be obtained from the trivial state labeled by $(0,0)$ via splitting operations.  The formalism developed above gives a simple answer to this question:  the most general charge configuration that can be obtained via splitting from the trivial state is $(-g_w(q_w), q_w)$, where the negative sign in the first entry denotes the inverse operation, and $q_w$ runs over all possible values in $H^1(G_w)$.  Recall that in Sec.~\ref{sec:3dcSPT}, the splitting operation was defined to be trivial if it can generate all possible values of $q_w$, while always leaving $q_{w_0}$ unchanged.  We see that this is the same as the statement that the homomorphism $g_w = 0$, \emph{i.e.} it is the trivial homomorphism.

Charge configurations of the form $(-g_w(q_w), q_w)$ are referred to as splitting configurations.  The splitting configurations form a group isomorphic to $H^1(G_w)$, and can be conveniently specified in terms of generators.  This information is presented below for $d=2$ and $d=3$ point groups.

The adjoining operation that appears in pgSPT classification can be described simply in this formalism, and we use this to obtain a simple result for the classification of block-dimension zero pgSPT phases.  If we start with the state labeled by $(q_{w_0}, 0)$ as a pgSPT state, we can adjoin zero-dimensional blocks at the points of $w$.  That is, adjoining transforms the state by $(q_{w_0}, 0) \to (q_{w_0}, q_w)$,
for any $q_w$.  We can then group the $w$ points together with the center of symmetry at $p_0$.  The net result is that we transform the original state by
\begin{equation}
(q_{w_0}, 0) \to (q_{w_0} + g_w(q_w) , 0 ) \text{.}
\end{equation}

More generally, we need to consider adjoining zero-dimensional blocks in more than one Wyckoff class.  Let $w_1, \dots, w_k$ be Wyckoff classes labeling the distinct possibilities for symmetry-equivalent points near the center of symmetry at $p_0$.  As usual, we ignore the Wyckoff class containing general points with trivial site symmetry.  Adopting the short-hand notation $H^1_i \equiv H^1(G_{w_i})$, a general block state is labeled by a charge configuration
\begin{equation}
(q_0, q_1, \dots,q_k) \in H^1_0 \times H^1_1 \times \cdots \times H^1_k \text{.}
\end{equation}

\begin{table*}
\begin{tabular}{c|c|c|c|c|c}
$G$ / $H^1$ & $q \in H^1(G)$ & $G_w$ / $H^1$ & $q \in H^1(G_w)$ & Coordinates of $w$ points & $g_w$ on generators of $H^1(G_w)$  \\
\hline
$S_4$ / $\z_4$ & $q^{S_4}$ & $C_2$ & $q^{C_2}$ & $(0,0,\pm z)$ & $g_w(1) = 2$   \\
\hline
$C_{4h}$ / $\z_4 \times \zz$ & $(q^{C_4}, q^m)$ & $C_4$ / $\z_4$ & $q^{C_4}$ & $(0,0,\pm z)$ & $g_w(1) = (2,0)$  \\
\hline
$D_{2d}$ / $\zz^2$ & $(q^{S_4}, q^{C_2})$ & $C_{2v}$ / $\zz^2$ & $(q^{m_1}, q^{m_2})$ & $(0,0,\pm z)$ & 
$g_w(1,0) = g_w(0,1) = (1,0)$ \\
\hline
$C_{3i}$ / $\z_3 \times \zz$ & $(q^{C_3}, q^i)$ & $C_3$ / $\z_3$ & $q^{C_3}$ & $(0,0,\pm z)$ & $g_w(1) = (2,0)$  \\
\hline
$D_3$ / $\zz$ & $q^{C_2}$ & $C_2$ / $\zz$ & $q^{C_2}$ & $(x,0,0);(0,x,0);(\bar{x},\bar{x},0)$ & $g_w(1) = 1$ \\
\hline
$C_{3v}$ / $\zz$ & $q^m$ & $C_s$ / $\zz$ & $q^m$ & $(x,\bar{x},z);(x,2x,z);(2\bar{x},\bar{x},z)$ & $g_w(1) = 1$ \\
\hline
$C_{3h}$ / $\z_3 \times \zz$ & $(q^{C_3}, q^m)$ & $C_3$ / $\z_3$ & $q^{C_3}$ & $(0,0,\pm z)$ & $g_w(1) = (2,0)$ \\
\hline
$C_{3h}$ / $\z_3 \times \zz$ & $(q^{C_3}, q^m)$  & $C_s$ / $\zz$ & $q^{m}$ & \parbox[t]{3.3cm}{ $(x,y,0);(\bar{y},x-y,0)$ \\
$(\bar{x}+y,\bar{x},0)$ } & $g_w(1) = (0,1)$ \\
\hline
$C_{6h}$ / $\z_6 \times \zz$ & $(q^{C_6}, q^i)$ & $C_6$ / $\z_6$ & $q^{C_6}$ & $(0,0,\pm z)$ & $g_w(1) = (2,0)$ \\
\hline
$D_{3h}$ / $\zz^2$ & $(q^{m_v}, q^{m_h})$ & $C_{2v}$ / $\zz^2$ & $(q^{m_v}, q^{m_h})$ &
 \parbox[t]{3.3cm}{ $(x,\bar{x},0);(x,2x,0)$ \\ $(2\bar{x},\bar{x},0)$ } &  \parbox[t]{3.3cm}{ $g_w(1,0) = (1,0)$ \\
 $g_w(0,1) = (0,1)$ } \\
\hline
$T$ / $\z_3$ & $q^{C_3}$ & $C_3$ / $\z_3$ & $q^{C_3}$ &   \parbox[t]{3.3cm}{ $(x,x,x); (\bar{x},\bar{x},x)$ \\  $(\bar{x},x,\bar{x}); (x, \bar{x}, \bar{x})$ }  & $g_w(1) = 1$  \\
\hline 
$T_h$ / $\z_3 \times \zz$ & $(q^{C_3}, q^i)$  & $C_3$ / $\z_3$ & $q^{C_3}$ & $(\pm x, \pm x, \pm x)$ & $g_w(1) = (2,0)$ \\
\hline
$T_d$ / $\zz$ & $q^m$ & $C_{2v}$ / $\zz^2$ & $(q^{m_1}, q^{m_2})$ & \parbox[t]{3.3cm}{  $(\pm x,0,0); (0,\pm x, 0)$ \\
 $(0,0, \pm x)$ } & $g_w(1,0) = g_w(0,1) = 1$
\end{tabular}
\caption{Splitting operations for three-dimensional crystallographic point groups.  Only non-trivial splitting operations are shown.  The first column gives the point group $G$ and $H^1(G)$.  The second column gives the form of an element of $H^1(G)$.  $q^{C_n}$ is a rotation charge, $q^m$ a mirror reflection charge, $q^i$ an inversion charge, and $q^{S_4}$ a four-fold roto-reflection charge.  See the text for further explanation. The third column gives the site symmetry $G_w$ (and $H^1(G_w)$) of a Wyckoff class $w$ whose points can be brought arbitrarily close to a center of $G$ symmetry, and the fourth column gives the form of an element of $H^1(G_w)$.  The coordinates of the center of $G$ symmetry are taken to be $(0,0,0)$, and the fifth column gives the coordinates of the points in $w$.  For trigonal and hexagonal point groups, coordinates are given in a hexagonal system of primitive vectors.  A bar over a coordinate denotes a minus sign.  The last column specifies the homomorphism $g_w$ by its action on the generators of $H^1(G_w)$.  \label{tab:splitting}}
\end{table*}

\begin{table*}
\begin{tabular}{c|c|c|c}
$G$ & $H^1(G)$ & ${\rm Adj}(G)$ & $H^1(G) / {\rm Adj}(G)$ \\
\hline
$S_4$ & $\z_4$ & $\zz$ & $\zz$ \\
\hline
$C_{4h}$ & $\z_4 \times \zz$  & $\zz \subset \z_4$ & $\zz^2$ \\
\hline
$D_{2d}$ & $\zz^a \times \zz^b$ & $\zz^a$ & $\zz$ \\
\hline
$C_{3i}$ & $\z_3 \times \zz$ & $\z_3$ & $\zz$ \\
\hline
$D_3$ & $\zz$  & $\zz$ & Trivial \\
\hline
$C_{3v}$ & $\zz$ & $\zz$ & Trivial \\
\hline
$C_{3h}$ & $\z_3 \times \zz$ & $\z_3 \times \zz$ & Trivial \\
\hline
$C_{6h}$ & $\z_6 \times \zz$ & $\z_3 \subset \z_6$ & $\zz^2$ \\
\hline
$D_{3h}$ & $\zz^2$ & $\zz^2$ & Trivial \\
\hline
$T$ & $\z_3$ & $\z_3$ & Trivial \\
\hline
$T_h$ & $\z_3 \times \zz$ & $\z_3$ & $\zz$ \\
\hline
$T_d$ & $\zz$ & $\zz$ & Trivial
\end{tabular}
\caption{${\rm Adj}(G)$ for three-dimensional crystallographic point groups.  Only those groups for which ${\rm Adj}(G)$ is non-trivial are shown.  ${\rm Adj}(G)$ is computed using the information in Table~\ref{tab:splitting}.  The last column gives the quotient $H^1(G) / {\rm Adj} (G)$, which is equal to $\cC_0(G)$ for those point groups with zero-dimensional fixed space (all those in the table except $G = C_{3v}$).  \label{tab:adjG}}
\end{table*}

Starting with the state $(q_0, 0, \dots, 0)$, we adjoin arbitrary charges in the nearby blocks to obtain the state $(q_0, q_1, \dots, q_k)$.  Then we group the blocks together at the center of symmetry, resulting in the transformation
\begin{equation}
(q_0, 0, \dots, 0) \to (q_0 + g_{w_1}(q_1) + \cdots + g_{w_k}(q_k), 0,\dots,0) \text{.}
\end{equation}
Therefore we have a map
\begin{equation}
{\cal A} : H^1_1 \times \cdots \times H^1_k \to H^1_0 \text{,}
\end{equation}
given by
\begin{equation}
{\cal A}(q_1, \dots, q_k) = g_{w_1}(q_1) + \cdots + g_{w_k}(q_k) \text{.}
\end{equation}
The image of this map is precisely the set of all one-dimensional representations that can be obtained by the adjoining operation, which was the definition of ${\rm Adj}(G) \subset H^1_0$ given in Sec.~\ref{subsec:pgspt3d}.  Therefore we define
\begin{equation}
{\rm Adj}(G) = {\rm Im} {\cal A} \text{.}
\end{equation}
Taking the quotient of $H^1(G)$ by ${\rm Adj}(G)$ gives precisely the information about a $G$ charge that is stable under adjoining.  Therefore, when the fixed space of $G$ is a single point, the classification of block-dimension zero pgSPT phases is
\begin{equation}
\cC_0(G) = \frac{ H^1(G) }{{\rm Adj}(G)} \text{.}
\end{equation}

Now we proceed to describe splitting operations and ${\rm Adj}(G)$ for all $d=2$ and $d=3$ crystallographic point groups.  We have considered all possible splitting operations, but only describe those that are non-trivial.

We begin in $d=2$, where $D_3$ is the only point group with a non-trivial splitting operation.  We recall that $D_3$ is algebraically isomorphic to $\z_3 \rtimes \zz$, and we have $H^1(D_3) = \zz$. $D_3$ is generated by three mirror reflections as shown in Fig.~\ref{fig:D3}.  There is a nontrivial splitting operation where the Wyckoff class $w$ contains three points on reflection axes related by three-fold rotation symmetry.  These points have $D_1$ site symmetry, and $H^1(D_1) = \zz$.  The splitting operation can be described by giving $g_w : H^1(D_1) \to H^1(D_3)$ on the single generator of its domain, and we find $g_w(1) = 1$.  This implies that ${\rm Adj}(D_3) = \zz$, and $\cC_0 (D_3)$ is thus trivial.

In three dimensions, the following 19 point groups have only trivial splitting operations: $C_i$, $C_s$, $C_{2h}$, $C_n$ ($n=2,3,4,6$), $C_{nv}$ ($n=2,4,6$), $D_n$ ($n=2,4,6$), $D_{nh}$ ($n=2,4,6$), $D_{3d}$, $O$, $O_h$.  For these groups, ${\rm Adj}(G)$ is trivial, and for those groups with fixed-space dimension zero, $\cC_0(G) = H^1(G, {\rm U}(1))$.

This leaves 12 point groups with non-trivial splitting operations, which we give in Table~\ref{tab:splitting}.  All these point groups have only one non-trivial splitting operation, except $C_{3h}$, which has two.  We present the splitting operations by specifying the homomorphism $g_w : H^1(G_w) \to H^1(G)$ via its action on the generators $G_w$.  This information is then used to compute ${\rm Adj}(G)$; the results are given in Table~\ref{tab:adjG}.

In order to specify $g_w$, Table~\ref{tab:splitting} also fixes conventions for writing the elements of $H^1(G)$ and $H^1(G_w)$, which we now explain.  In general, it is possible and convenient to specify elements $H^1(G)$ in terms of symmetry charges of certain subgroups of $G$.  For example, the group $C_{3h}$ has both a $C_3$ subgroup (three-fold rotations about the $z$-axis) and a $C_s$ subgroup (mirror reflection in the $xy$ plane).  It can be shown that $q \in H^1(C_{3h})$ can be written $q = (q^{C_3}, q^m)$, where $q^{C_3} \in H^1(C_3) = \z_3$ is the $C_3$ rotation charge, and $q^m \in H^1(C_s) = \zz$ is the $C_s$ mirror reflection charge.  For most point groups in Table~\ref{tab:splitting}, it is clear which subgroup is being referred to. In some cases there are multiple isomorphic subgroups that are conjugate to one another, and in such cases one of these subgroups can be chosen arbitrarily; for example, the group $T$ has four $C_3$ subgroups.

For a few point groups, more explanation is needed to clarify the forms of symmetry charges given in Table~\ref{tab:splitting}.  In the case of $D_{2d}$, $q^{S_4}$ is the charge of $z$-axis roto-reflections, which is constrained to take values of $\zz$ due to the properties of $D_{2d}$.  There, $q^{C_2}$ is the charge of a $C_2$ rotation perpendicular to the $z$-axis.

There are a few splitting operations where $G_w = C_{2v}$, which is generated by two perpendicular mirror planes.  A $C_{2v}$ charge can be specified by giving the mirror reflection charges for these two planes separately, and we label them $q^{m_1}$ and $q^{m_2}$.

The group $D_{3h}$ is generated by the three vertical mirror planes of $C_{3v}$, and a horizontal mirror plane (the $xy$ plane).  There, $q^{m_v}$ refers to the mirror charge of a vertical mirror operation, while $q^{m_h}$ is the charge of the horizontal mirror reflection.  In the non-trivial splitting operation where $G_w = C_{2v}$, the $C_{2v}$ site symmetry of each point in $w$ is generated by the horizontal mirror reflection, and one of the horizontal mirror reflections, so it is natural to use the same notation to specify the $C_{2v}$ symmetry charge.

Finally, the group $T_d$ contains six mirror operations, where the normals to the mirror planes lie in the $\langle 110 \rangle$ directions.  The $T_d$ charge $q \in H^1(T_d) = \zz$ can be specified by giving the mirror charge $q^m \in \zz$ for any of these mirror planes.

\section{Twisting operations for block dimension zero crystalline SPT states in three dimensions}
\label{app:twisting}

Here we give a detailed discussion of twisting operations for block dimension zero cSPT states with space group symmetry in $d=3$.  We give a general discussion and enumerate those cases with non-trivial twisting operations.

For certain one-dimensional Wyckoff classes in non-symmorphic space groups, the axis swept out by a Wyckoff point can coincide  with a screw axis, or be contained in a glide plane, where the glide direction is along the axis.  
The screw  or glide operation becomes a half translation on the Wyckoff axis, and can act non-trivially on the site-symmetry $G_{w}$, if $G_{w}$ has at least one non-trivial automorphism.  If, in addition, $H^1(G_w) \equiv H^1(G_w, {\rm U}(1))$ has a non-trivial automorphism, the half translations can have non-trivial action on the $G_w$ charge, which results in non-trivial twisting operations.  

We find that non-trivial twisting operations arise in two types of situations. (1) A Wyckoff  axis with $G_{w}=C_{n}$ for $n=3,4,6$ is contained in a glide plane, with glide direction along the axis. (2) A Wyckoff axis with $G_{w}=C_{2v}$ coincides with a four-fold screw axis.

We now describe the action of translations on $G_w$ in these cases, denoting by $t_h$ the half translation arising from the glide or screw operation.  For type (1), the half translation acts on the $C_{n}$ rotation by
\begin{equation}
t_{h} C_n t_{h}^{-1} = C_n^{-1} \text{.}
\label{eqn:cn_action}
\end{equation}
For type (2), we have
\begin{eqnarray}
t_{h} \sigma_1 t_{h}^{-1} &=& \sigma_2
\label{eqn:r_action1} \\
t_{h} \sigma_2 t_{h}^{-1} &=& \sigma_{1} \text{,}
\label{eqn:r_action2}
\end{eqnarray}
where $\sigma_1$ and $\sigma_2$ are the two mirror reflections generating $C_{2v}$.

These non-trivial group actions restrict the allowed $G_w$ charges within a unit cell.  We let $q_z$ be the $G_w$ charge of a point $p_z$ on the Wyckoff axis, and $q_{z + 1/2}$ the charge at the point $t_h p_z$, \emph{i.e.} by acting on the first point with a half translation.  We have $q_z, q_{z+1/2} \in H^1(G_w)$.  Applying Eq.~(\ref{eqn:related-charges}), these charges are related by
\begin{equation}
D_{q_{z + 1/2}}(g) = D_{q_z}( t_h g t^{-1}_h ) \text{,}  \label{eqn:twisting-relation}
\end{equation}
for all $g \in G_w$, and where $D_q (g)$ is the one-dimensional representation of $G_w$ labeled by $q \in H^1(G_w)$.

The relation Eq.~(\ref{eqn:twisting-relation}) induces an automorphism 
\begin{equation}
{\mathfrak t}_h : H^1(G_w) \to H^1(G_w) \text{,}
\end{equation}
where ${\mathfrak t}_h (q_z) = q_{z + 1/2}$.
In case (1), ${\mathfrak t}_h (q) = - q$, \emph{i.e.} ${\mathfrak t}_h$ is the inversion automorphism.  This implies that $q_{z + 1/2} = - q_z$.  In case (2), an element $q \in H^1(C_{2v}) = \zz^2$ can be written $q = (q_1, q_2)$, for $q_1, q_2 \in \zz$.  The automorphism acts by ${\mathfrak t}_h [ (q_1, q_2) ]=  (q_2, q_1)$.  This implies that if $q_z = (q_1^z, q_2^z)$, then $q_{z + 1/2} =  (q_2^z, q_1^z )$.

We let $\cQ_c$ be the group of $G_w$ charge configurations. We have $\cQ_c \simeq H^1(G_w)$, because elements of $\cQ_c$ are of the form $(q_z, q_{z+1/2}) = (q_z, {\mathfrak t}_h(q_z) )$.  We then define $\cQ_t \subset \cQ_c$ to be the set of charge configurations that can be obtained from the trivial configuration $(0,0) \in \cQ_c$ by applying the block equivalence operations.  More specifically, we apply twisting operations.     We then obtain $H^{1}_{\circ}(G) = \cQ_c / \cQ_t$, which is the contribution of the Wyckoff class to the cSPT classification $\cC_0(G)$ for the space group $G$.  Moreover, in each case we show $H^{1}_{\circ}(G)$ is a weak pgSPT invariant.

Below, we obtain $\cQ_c$, $\cQ_t$, and $H^1_\circ(G_w)$ for each case where non-trivial twisting operations arise.  These results are summarized in Table~\ref{tab:twisting}.

\begin{table}
\begin{tabular}{c|c|c}
$G_w$ & $\cQ_c \simeq H^1(G_w)$ & $H^1_{\circ}(G_w)$ \\
\hline
$C_{3}$ & $\z_3$ & -- \\
\hline
$C_{4}$ & $\z_4$ & $\z_2$ \\
\hline
$C_{6}$ & $\z_6$ & $\z_2$ \\
\hline
$C_{2v}$ & $\z_{2}^{2}$ & $\z_{2}$ \\
\hline
\end{tabular}
\caption{Effect of non-trivial twisting operations on the classification block dimension zero cSPT phases.  The first column is the site symmetry of a one-dimensional Wyckoff class.  The results of this table apply when the corresponding one-dimensional Wyckoff axis is either contained in a glide plane ($G_w = C_3, C_4, C_6$) or coincides with a four-fold screw axis ($G_w = C_{2v}$), as described in the text.  $\cQ_c$ is the group labeling $G_w$ charge configurations on the Wyckoff axis, and $H^1_{\circ}(G_w)$ is the contribution of the Wyckoff class to $\cC_0(G)$ for the space group $G$, taking twisting operations into account. \label{tab:twisting}}
\end{table}

\subsection{$G_{w}=C_{3}$}

Here, $q_z, q_{z+1/2} \in H^1(C_3) = \z_3$.  $\cQ_c$ contains three charge configurations, which are $(0,0)$, $(1,2)$, and $(2,1)$.  To obtain $\cQ_t$, we describe the effect of twisting operations on the $(0,0)$ configuration. We note that the $(0,0)$ configuration describes a chain of zero-dimensional blocks lying on the Wyckoff axis.  First, we split $(0,0)$ into the product of charge configurations $(1,2)_1 \times (2,1)_2$; this can be thought of as splitting the original chain into two new chains.  Next, we slide the charges on the second chain along the Wyckoff axis by half a lattice constant, which transforms the state by
\begin{equation}
(1,2)_1 \times (2,1)_2 \to (1,2)_1 \times (1,2)_2 \text{.}
\end{equation}
Finally, grouping the chains back together, we obtain the configuration $(1,2) \in \cQ_c$.  If instead we slide the charges of the first chain by half a lattice constant before grouping the chains back together, we obtain $(2,1) \in \cQ_c$.  Therefore, we have shown $\cQ_t = \cQ_c$, and $H^1_{\circ}(C_3)$ is trivial.

\subsection{$G_{w}=C_{4}$}

Now $q_z, q_{z+1/2} \in H^1(C_4) = \z_4$, and the charge configurations in $\cQ_c$ are $(0,0)$, $(1,3)$, $(2,2)$, and $(3,1)$.  We split $(0,0)$ to $(1,3)_1 \times (3,1)_2$, and then slide the charges of the first chain by half a lattice constant to obtain $(3,1)_1 \times (3,1)_2 \simeq (2,2)$.  Other twisting operations either also produce $(2,2)$, or leave the $(0,0)$ state invariant.  Therefore $\cQ_t = \{ (0,0), (2,2) \} \simeq \zz$, and $H^1_{\circ}(C_4) = \z_4 / \zz = \zz$.

We would like to show that $H^1_{\circ}(C_4) = \zz$ is a weak pgSPT invariant.  We do this by focusing on the symmetry generated by the $C_2$ subgroup of $C_4$, and by $t_h$.  We note that $C_2$ rotations commute with $t_h$.  Considering the non-trivial state with $(q_z, q_{z+1/2}) = (1,3)$, the $C_2$ charge configuration is $(1,1)$.  On the Wyckoff axis we therefore have a chain of non-trivial $C_2$ charges, and $t_h$ plays the role of a translation symmetry along the stacking direction.  Therefore we can think of this state as a stack of $d=2$ $C_2$ pgSPT layers, with non-trivial $\zz$ invariant per layer.

\subsection{$G_{w}=C_{6}$}

Here, $q_z, q_{z+1/2} \in H^1(C_6) = \z_6$, and the charge configurations in $\cQ_c$ are $(0,0)$, $(1,5)$, $(2,4)$, $(3,3)$, $(4,2)$ and $(5,1)$.  We split $(0,0)$ to $(1,5)_1 \times (5,1)_2$, and slide the charges of the first chain by half a lattice constant to obtain $(5,1)_1 \times (5,1)_2 \simeq (4,2)$.  If instead we slide the charges of the second chain by half a lattice constant, we get $(1,5)_1 \times (1,5)_2 \simeq (2,4)$.  Considering other twisting operations does not lead to more states, and we find $\cQ_t = \{ (0,0), (2,4), (4,2) \} \simeq \z_3$.  Taking the quotient $\cQ_c / \cQ_t$, we find $H^1_{\circ}(C_6) = \zz$.

It can be shown that this is a weak pgSPT invariant by focusing on the symmetry generated by the $C_2$ subgroup of $C_6$ and $t_h$, and following the analysis given above for $G_w = C_4$.

\subsection{$G_{w}=C_{2v}$}

Here, $q_z, q_{z+1/2} \in H^1(C_{2v}) = \zz^2$, and the charge configurations in $\cQ_c$ are $[ (0,0), (0,0) ]$,
$[ (1,0), (0,1) ]$, $[ (0,1), (1,0) ]$ and $[ (1,1), (1,1) ]$.  We split  $[ (0,0), (0,0) ]$ to obtain
\begin{equation}
\left[ (1,0), (0,1) \right]_1 \times \left[ (1,0), (0,1) \right]_2 \text{,}
\end{equation}
and slide the charges of the first chain by half a lattice constant to obtain
\begin{equation}
\left[ (0,1), (1,0) \right]_1 \times \left[ (1,0), (0,1) \right]_2 \simeq \left[ (1,1), (1,1) \right] \text{.}
\end{equation}
Other twisting operations do not lead to additional states, so we find $\cQ_t \simeq \zz$, with
\begin{equation}
\cQ_t = \{ [ (0,0), (0,0) ], [ (1,1), (1,1)] \} \text{.}
\end{equation}
Taking the quotient, we have $H^1_{\circ} (C_{2v} ) = \zz$.

To show that $H^1_{\circ} (C_{2v} ) = \zz$ is a weak pgSPT invariant, we focus on the symmetry generated by the $C_2$ rotation subgroup of $C_{2v}$, and by $t_h$.  Considering the non-trivial state with
\begin{equation}
(q_z, q_{z+1/2}) = [ (1,0), (0,1) ] \text{,}
\end{equation}
the $C_2$ charge configuration is $(1,1)$.  As above in the discussion of the case $G_w = C_4$, this state can be viewed as a non-trivial stack of $d=2$ $C_2$ pgSPT states, with $t_h$ playing the role of translation symmetry in the stacking direction.

\section{Completeness of pgSPT and weak pgSPT invariants}
\label{app:decomposition}

Here, we consider three-dimensional cSPT phases protected by space group symmetry.  The block-equivalence classification for space group $G$ is $\cC(G) = \cC_0(G) \times \cC_1(G) \times \cC_2(G)$.  We show that any two distinct elements of $\cC(G)$ can be distinguished by pgSPT and weak pgSPT invariants.  This implies that these two elements are different phases, so the block-equivalence classification indeed gives a classification of phases.  It also follows that the cSPT phases we classify (and, equivalently, those classified in Ref.~\onlinecite{thorngren16gauging}) can be fully characterized by pgSPT and weak pgSPT invariants.

We recall the definitions of pgSPT and weak pgSPT invariants.  A pgSPT invariant is a SPT invariant associated with some site symmetry subgroup of $G$.  Given a $G$-symmetric cSPT state, we can focus on a site symmetry subgroup, view the state as a pgSPT phase protected by the site symmetry, and compute the resulting invariant.  A weak pgSPT invariant is obtained by compactifying one or more dimensions of space, viewing the resulting system as a lower-dimensional point group SPT phase, and characterizing the dependence of the lower-dimensional pgSPT invariant on the length in the finite dimensions.

As explained in Sec.~\ref{sec:3dcSPT}, $\cC_1(G) \times \cC_2(G)$ can be factored into pgSPT invariants associated with $C_{nv}$ axes and mirror planes.  Therefore, here, it is enough to concentrate on $\cC_0(G)$.  We will establish the following claim:
\begin{claim}
Consider $\Psi \in \cC_0(G)$.  $\Psi \neq 0$ implies that $\Psi$ has a non-trivial pgSPT or weak pgSPT invariant.
\end{claim}
It follows from this claim that non-zero elements of $\cC_0(G)$ are non-trivial cSPT phases.

It also follows from Claim 1 that two distinct elements of $\cC_0(G)$ have different pgSPT or weak pgSPT invariants.  This is easily established by contradiction:  We consider non-zero elements $\Psi, \Psi' \in \cC_0(G)$, with $\Psi \neq \Psi'$.  We suppose $\Psi$ and $\Psi'$ have the same pgSPT and weak pgSPT invariants.  Then the difference $\Psi - \Psi'$ is non-zero but has trivial pgSPT and weak pgSPT invariants, a contradiction.

To establish Claim 1, we expose some structure of $\cC_0(G)$ that will be useful.  Given an element $\Psi \in \cC_0(G)$, we define an integer $D(\Psi) \in \{0, 1,2,3 \}$ as follows.  We consider a state representing $\Psi$, and apply block equivalence operations to remove points with the lowest Wyckoff dimension until this can no longer be done.  $D(\Psi)$ is defined to be the lowest Wyckoff dimension of a point in the resulting state.  For $\Psi = 0$, we define $D(0) \equiv 3$.

For example, $D(\Psi) = 0$ means that there is some point $p$ with Wyckoff dimension zero (\emph{i.e.} the position of $p$ is fixed), such that $p$ carries non-trivial $G_p$ charge that cannot be removed by applying block equivalence operations.  $D(\Psi) = 1$ means that block equivalence operations can be applied to remove all points with Wyckoff dimension zero, but it is not possible to remove all points with Wyckoff dimension less than two.

We also define subgroups of $\cC_0(G)$ by
\begin{equation}
W_n = \{ \Psi \in \cC_0(G) |  D(\Psi) \geq n \} \text{.}
\end{equation}
We have the sequence of subgroups
\begin{equation}
0 = W_3 \subset W_2 \subset W_1 \subset W_0 = \cC_0(G) \text{.}
\end{equation}
We can also define quotients
\begin{equation}
V_n = \frac{W_n}{W_{n+1}} \text{.}
\end{equation}
It will follow from the discussion below that $V_0$ corresponds to pgSPT invariants, while $V_1$ and $V_2$ correspond to weak pgSPT invariants. Given $G$, it is possible to decompose $\cC_0(G)$ into $V_0$, $V_1$ and $V_2$, which is a decomposition of the cSPT classification into pgSPT and weak pgSPT invariants.  In general, this decomposition is not simply a product; that is,
\begin{equation}
\cC_0(G) \neq V_0 \times V_1 \times V_2 \text{,}
\end{equation}
although such a factorization does hold in many cases.  For instance, for space group \#200 (see Sec.~\ref{sec:3dcSPT}), $\cC_0(G) = \z_3 \times \zz^8$, $V_0 = \zz^8$, and $V_1 = \z_3$  ($V_2$ is trivial).  We have $\cC_0(G) = V_0 \times V_1$.   We give an example below (space group \#82) where the decomposition into $V_0$, $V_1$ and $V_2$ is not simply a product.

Now we turn to establishing Claim 1.  First, we consider $\Psi \in \cC_0(G)$ with $D(\Psi) = 0$.  Then in any state representing $\Psi$, there is some point $p$ with Wyckoff dimension zero whose $G_p$ charge cannot be removed by applying block equivalence operations.  It follows that the $G_p$ charge cannot be removed by the adjoining operation used to classify pgSPT phases, and $\Psi$ is a non-trivial  $G_p$ pgSPT phase.

Next, we consider $\Psi$ with $D(\Psi) = 1$.  We fix a state representing $\Psi$ where all points have Wyckoff dimension one and higher.  Let $p$ be a point of Wyckoff dimension one that cannot be split to points of Wyckoff dimension two or eliminated completely.  At least one such point exists because $D(\Psi) = 1$.  The site symmetry $G_p$ can be $C_n$ ($n = 2,3,4,6$) or $C_{nv}$ ($n = 2,4,6$).  Let $A$ be the axis swept out by $p$ as it is slid along its symmetry axis, and let $G_A$ be the subgroup of $G$ taking $A$ into itself.  The line $A$ can be viewed as a one-dimensional system with symmetry group $G_A$ and on-site symmetry $G_p$, and it is thus clear that $G_p$ is a normal subgroup of $G_A$.  The quotient $\tilde{G}_A = G_A / G_p$ can be viewed as a one-dimensional space group of $A$.  There are only two one-dimensional space groups, which means there are two cases to consider:  Case 1: $\tilde{G}_A$ acts on $A$ only by translation.  Case 2: The action of  $\tilde{G}_A$ on $A$ is generated by translation and inversion.

\emph{Case 1}.  We choose $t \in G_A$ so that the corresponding element $[t] \in \tilde{G}_A$ is the elementary one-dimensional translation.  As a three-dimensional operation, $t$ can be chosen to be a pure translation, a glide reflection, or a screw rotation.  We can apply block equivalence operations to group points along $A$, so that $A$ contains only a lattice points separated by elementary translations, each carrying non-trivial $G_p$ charge.  We make the system finite along the axis $A$ with length $L$ and periodic boundary conditions so that $t^L = 1$.

First, we suppose that $t$ commutes with $G_p$.  Taking periodic boundary conditions is compatible with $G_p$ symmetry, and we can view the finite system as a two-dimensional pgSPT state, with point group corresponding to $G_p$.  Because the $G_p$ charge per unit cell along $A$ is non-zero, the $d=2$ pgSPT index has non-trivial dependence on $L$, and the state has a non-trivial weak pgSPT invariant.

Second, we suppose that $t$ does not commute with $G_p$.  This is precisely the situation studied in Appendix~\ref{app:twisting}; we need only recapitulate the results obtained there in the context of the present discussion.  There are only three non-trivial possibilities, and in each case the block-equivalence classes of charge configurations along $A$ are labeled by a $\zz$ invariant.  Two of the cases are $G_p = C_4$ or $G_p = C_6$ with $t$ a glide reflection.  The other case is $G_p = C_{2v}$, with $t$ a four-fold screw rotation.  In general, taking periodic boundary conditions here is not compatible with $G_p$ symmetry, because $t^L$ need not commute with $G_p$.  However, in all these cases, $G_p$ has a $C_2$ subgroup that commutes with $t$, so we can view the compactified system as a $d=2$ pgSPT state with $C_2$ symmetry.  In Appendix~\ref{app:twisting} it is shown that the corresponding weak pgSPT invariant resolves those charge configurations on $A$ that are non-trivial under block equivalence.

\emph{Case 2}.  Choose $a, b \in G_A$ so that $[a], [b] \in \tilde{G}_A$ are one-dimensional inversion operations at neighboring inversion centers.  Then, letting $t = ba$, $[t] \in \tilde{G}_A$ is the elementary one-dimensional translation that defines a primitive cell along the axis $A$.  As in Case 1, we compactify the system along $A$, taking periodic boundary conditions with $t^L = 1$.

We observe that $a^2$ acts on $A$ as an on-site operation, so $a^2 \in G_p$.  Moreover, $a^2$ is orientation preserving, so the only possibilities are $a^2 = 1,2,3$, where $2$ and $3$ denote two-fold and three-fold rotations.  If $a^2 = 3$, we can redefine $a$ so that $a^2 = 1$.  It follows that we can always choose $a$ to be one of four operations, $a = \bar{1}, m, 2', \bar{4}$.  Here, $\bar{1}$ is inversion, $m$ is mirror reflection with the normal of the mirror plane along $A$, $2'$ is two-fold rotation about an axis perpendicular to $A$, and $\bar{4}$ is a four-fold roto-reflection along $A$.  Similarly, we can take $b = \bar{1}, m, 2', \bar{4}$.

We let $G_a$ and $G_b$ be the site symmetry groups at the two one-dimensional inversion centers.  Given a fixed $G_p$, only certain choices for $G_a$ are consistent.  Clearly, $G_a$ must contain $G_p$ as a subgroup, and must contain $a = \bar{1}, m, 2', \bar{4}$.  In addition, we can always move $p$ and its inversion image $a p$ to the $a$ inversion center and group these two points together.  Doing so must not result in a trivial $G_a$ charge, which would contradict $D(\Psi) = 1$.  Therefore, the corresponding splitting/grouping operation must be non-trivial, \emph{i.e.} the homomorphism $g_w$ defined in Appendix~\ref{app:splitting} must be non-trivial.  Corresponding statements hold for $G_b$.  Using these restrictions on $G_a$ and $G_b$, we now proceed case-by-case through the different possibilities for $G_p$.

$G_p = C_2$.  Here, $G_a = G_b = S_4$.  We can take $a = \bar{4}$ and $b = \bar{4}^{-1}$, so that $t$ is a pure translation.  The compactified system has two-dimensional $C_4$ point group symmetry, because the $S_4$ rotation-reflection acts on the two-dimensional system as a four-fold rotation.  The corresponding weak pgSPT invariant is non-trivial.

$G_p = C_3$.  Here, $G_a = C_{3i}, T_h, C_{3h}$, and similarly for $G_b$.  In the first two cases, $a = \bar{1}$, and in the third case $a = m$.  Depending on the choices of $G_a$ and $G_b$, $t$ is either a pure translation or a two-fold screw rotation, both of which commute with $G_p = C_3$.  Therefore compactifying with length $L$ always preserves $C_3$ symmetry.  There is a non-trivial weak pgSPT invariant associated with two-dimensional $C_3$ symmetry.

$G_p = C_{4}$.  Here, $G_a = G_b = C_{4h}$, with $a = b = m$, so that $t$ is a pure translation.  The compactified system has two-dimensional $C_4$ symmetry, and the corresponding weak pgSPT invariant is non-trivial.

$G_p = C_{6}$.  Here, $G_a = G_b = C_{6h}$, with $a = b = m$, so that $t$ is a pure translation.  The compactified system has two-dimensional $C_6$ symmetry, and the corresponding weak pgSPT invariant is non-trivial.

$G_p = C_{2v}$.  Here, $G_a, G_b = D_{2d}, T_d$.  For both these point groups we can take $a = 2'$ or $a = \bar{4}$.  Choosing $a = b = 2'$, $t$ is a pure translation.  There is a non-trivial $C_{2v}$ charge per unit cell.  Upon compactifying to two dimensions, $C_{2v}$ becomes the $d=2$ point group $D_2$, and the corresponding weak pgSPT invariant is non-trivial.

$G_p = C_{4v}, C_{6v}$.  There are no possible $G_a, G_b$ satisfying the restrictions.  Therefore, Case 2 does not arise for these choices of $G_p$.  This completes the discussion of Case 2.

Finally, we consider $\Psi$ with $D(\Psi) = 2$.  We fix a state representing $\Psi$ where all points have Wyckoff dimension two.  Let $p$ be such a point that cannot be eliminated completely by applying block equivalence operations.  The site symmetry of $p$ is mirror reflection, \emph{i.e.} $G_p = C_s$.  We let $P$ be the mirror plane swept out by $p$, and $G_P$ the subgroup of $G$ taking $P$ into itself.  The quotient $\tilde{G}_P = G_P / G_p$ is a wallpaper group of the plane $P$.  We define $H_P \subset G_P$ to be the subgroup of orientation-preserving operations (as three-dimensional rigid motions).  It is straightforward to show that $G_P \simeq H_P \times G_p$.  Therefore, we can view the mirror plane as a two-dimensional system with $G_p \simeq \zz$ on-site symmetry that commutes with the wallpaper group $\tilde{G}_P \simeq H_P$.

The wallpaper group $\tilde{G}_P$ cannot contain any two-fold rotation centers or reflection axes, because the point $p$ and its images under symmetry can be slid to these centers/axes, grouped together there, and eliminated.  The only possibilities are therefore $\tilde{G}_P = p1, pg, p3$.

Treating $\tilde{G}_P = p1, p3$ together, we let $t_1$ and $t_2$ be two elementary translations in $\tilde{G}_P$.  $D(\Psi) = 2$ implies each primitive cell carries a non-trivial $G_p$ charge.  Compactifying the system in both directions so that $t_1^{L_1} = t_2^{L_2} = 1$, we have a one-dimensional pgSPT state where the one-dimensional inversion corresponds to the $G_p$ mirror reflection.  The one-dimensional pgSPT invariant is non-trivial when $L_1 L_2$ is odd, and trivial when $L_1 L_2$ is even, and we have a non-trivial weak pgSPT invariant.

Next, taking $\tilde{G}_P = pg$, we let $t_1$ be the glide reflection, and $t_2$ be an elementary translation normal to the glide axis.  Using these operations to define an effective unit cell, the $G_p$ charge per unit cell is non-trivial.  We compactify by first setting $t_2^{L_2} = 1$.  Next, we compactify in the $t_1$ direction by setting $t_1^{L_1} = 1$.  If $L_1$ is odd, this is a twisted boundary condition.  Odd $L_1$ breaks $t_2$ symmetry, but there is no need to preserve $t_2$ symmetry after first using it to define the periodic boundary condition in the $t_2$ direction. Crucially, the choice of boundary conditions is compatible with $G_p$ mirror symmetry, so that the mirror plane on the compactified system carries trivial (non-trivial) $G_p$ charge when $L_1 L_2$ is even (odd).  Therefore, there is a non-trivial weak pgSPT invariant.  This completes the proof of Claim 1. $\qed$

\begin{figure}
\includegraphics[width=0.6\columnwidth]{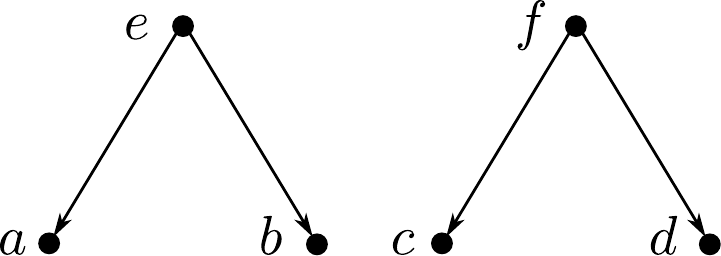}
\caption{$W$-quasigraph for space group $I \bar{4}$ (number \#82).  Vertices $a,b,c,d$ have site symmetry $S_4$, and vertices $e,f$ have site symmetry $C_2$.}
\label{fig:82}
\end{figure}

We close this Appendix by discussing an example where the decomposition of $\cC_0(G)$ into pgSPT and weak pgSPT invariants is not simply a product.  We consider the space group $I\bar{4}$, which is \#82 in the International Tables,\cite{inttables} and we refer to it as $G_{82}$.  The $W$-quasigraph is shown in Fig.~\ref{fig:82}.  We focus on the component with vertices $a,b$ and $e$, which have site symmetry $S_4$, $S_4$ and $C_2$, respectively.  Applying block equivalence operations produces a classification $\z_4 \times \zz$ for this component of the graph.  (The other component behaves identically, and the full classification is $\cC_0(G_{82}) = \z_4^2 \times \zz^2$.)

  The point group $S_4$ has a $\zz$ pgSPT classification, and six of the seven non-zero elements of $\z_4 \times \zz$ have non-trivial $S_4$ pgSPT invariants.  The non-zero element with trivial pgSPT invariants generates the subgroup $\zz \subset \z_4$.  We identify this $\zz$ subgroup with $V_1$, as its non-trivial element is characterized by a weak pgSPT invariant (see below).  Then we have $V_0 = (\z_4 \times \zz) / V_1 \simeq \zz \times \zz$, which is the group of $S_4$ pgSPT invariants for the two $S_4$ centers.

To see that $V_1$ is associated with a weak pgSPT invariant, we note that a representative state for the non-trivial element has non-trivial $S_4$ charge of 2 at $a$, where we use additive notation and write $H^1(S_4, {\rm U}(1)) = \z_4 = \{ 0,1,2,3 \}$.  $b$ and $e$ points carry trivial charge.  The unit cell coordinates of $a$ are $(0,0,0)$, with the roto-reflection axis in the $z$-direction.  We compactify the system in the $z$-direction with length $L$, then the $S_4$ symmetry (at $a$ points) becomes a two-dimensional $C_4$ point group symmetry.  The $C_4$ charge at the origin (\emph{i.e.} the projection of the $z$-axis to a point) is trivial when $L$ is even, and is 2 when $L$ is odd, so this state has a non-trivial weak pgSPT invariant.

\section{Classifications of crystalline SPT phases for space group symmetry in three dimensions}
\label{app:3dclassification}

Here, in Table~\ref{table:B-cSPT}, we give $\cC_0(G)$, $\cC_1(G)$ and $\cC_2(G)$ for all 230 space groups in three dimensions.  The classification $\cC(G) = \cC_0(G) \times \cC_1(G) \times \cC_2(G)$ can be obtained by taking the product of the given factors, and agrees with the Thorngren-Else classification, which was obtained in Ref.~\onlinecite{thorngren16gauging} for all space groups except numbers  227, 228 and 230.

\renewcommand*\arraystretch{1.45}

\begin{table*}[h]
\begin{tabular}{|c|cccc|}
\hline 
No. & symbol & $\cC_0(G)$ & $\cC_1(G)$ & $\cC_2(G)$\tabularnewline
\hline 
1 & $P1$ &  &  & \tabularnewline
\hline
2 & $P\overline{1}$ & $\mathbb{Z}_{2}^{8}$ &  & \tabularnewline \hline
3 & $P2$ & $\mathbb{Z}_{2}^{4}$ &  & \tabularnewline \hline
4 & $P2_{1}$ &  &  & \tabularnewline \hline
5 & $C2$ & $\mathbb{Z}_{2}^{2}$ &  & \tabularnewline \hline
6 & $Pm$ & $\mathbb{Z}_{2}^{2}$ &  & $\mathbb{Z}_{2}^{2}$\tabularnewline \hline
7 & $Pc$ &  &  & \tabularnewline \hline
8 & $Cm$ & $\mathbb{Z}_{2}$ &  & $\mathbb{Z}_{2}$\tabularnewline \hline
9 & $Cc$ &  &  & \tabularnewline \hline
10 & $P2/m$ & $\mathbb{Z}_{2}^{16}$ &  & $\mathbb{Z}_{2}^{2}$\tabularnewline \hline
11 & $P2_{1}/m$ & $\mathbb{Z}_{2}^{5}$ &  & $\mathbb{Z}_{2}$\tabularnewline \hline
12 & $C2/m$ & $\mathbb{Z}_{2}^{10}$ &  & $\mathbb{Z}_{2}$\tabularnewline \hline
13 & $P2/c$ & $\mathbb{Z}_{2}^{6}$ &  & \tabularnewline \hline
14 & $P2_{1}/c$ & $\mathbb{Z}_{2}^{4}$ &  & \tabularnewline \hline
15 & $C2/c$ & $\mathbb{Z}_{2}^{5}$ &  & \tabularnewline \hline
16 & $P222$ & $\mathbb{Z}_{2}^{16}$ &  & \tabularnewline \hline
17 & $P222_{1}$ & $\mathbb{Z}_{2}^{4}$ &  & \tabularnewline \hline
18 & $P2_{1}2_{1}2$ & $\mathbb{Z}_{2}^{2}$ &  & \tabularnewline \hline
19 & $P2_{1}2_{1}2_{1}$ &  &  & \tabularnewline \hline
20 & $C222_{1}$ & $\mathbb{Z}_{2}^{2}$ &  & \tabularnewline \hline
21 & $C222$ & $\mathbb{Z}_{2}^{9}$ &  & \tabularnewline \hline
22 & $F222$ & $\mathbb{Z}_{2}^{8}$ &  & \tabularnewline \hline
23 & $I222$ & $\mathbb{Z}_{2}^{8}$ &  & \tabularnewline \hline
24 & $I2_{1}2_{1}2_{1}$ & $\mathbb{Z}_{2}^{3}$ &  & \tabularnewline \hline
25 & $Pmm2$ & $\mathbb{Z}_{2}^{8}$ & $\mathbb{Z}_{2}^{4}$ & $\mathbb{Z}_{2}^{4}$\tabularnewline \hline
26 & $Pmc2_{1}$ & $\mathbb{Z}_{2}^{2}$ &  & $\mathbb{Z}_{2}^{2}$\tabularnewline \hline
27 & $Pcc2$ & $\mathbb{Z}_{2}^{4}$ &  & \tabularnewline \hline
28 & $Pma2$ & $\mathbb{Z}_{2}^{3}$ &  & $\mathbb{Z}_{2}$\tabularnewline \hline
29 & $Pca2_{1}$ &  &  & \tabularnewline \hline
30 & $Pnc2$ & $\mathbb{Z}_{2}^{2}$ &  & \tabularnewline \hline
31 & $Pmn2_{1}$ & $\mathbb{Z}_{2}$ &  & $\mathbb{Z}_{2}$\tabularnewline \hline
32 & $Pba2$ & $\mathbb{Z}_{2}^{2}$ &  & \tabularnewline \hline
33 & $Pna2_{1}$ &  &  & \tabularnewline \hline
34 & $Pnn2$ & $\mathbb{Z}_{2}^{2}$ &  & \tabularnewline \hline
35 & $Cmm2$ & $\mathbb{Z}_{2}^{5}$ & $\mathbb{Z}_{2}^{2}$ & $\mathbb{Z}_{2}^{2}$\tabularnewline \hline
36 & $Cmc2_{1}$ & $\mathbb{Z}_{2}$ &  & $\mathbb{Z}_{2}$\tabularnewline \hline
37 & $Ccc2$ & $\mathbb{Z}_{2}^{3}$ &  & \tabularnewline \hline
38 & $Amm2$ & $\mathbb{Z}_{2}^{4}$ & $\mathbb{Z}_{2}^{2}$ & $\mathbb{Z}_{2}^{3}$\tabularnewline
\hline 
\end{tabular}\quad{}%
\begin{tabular}{|c|cccc|}
\hline 
No. & symbol & $\cC_0(G)$ & $\cC_1(G)$ & $\cC_2(G)$\tabularnewline
\hline 
39 & $Aem2$ & $\mathbb{Z}_{2}^{3}$ &  & $\mathbb{Z}_{2}$\tabularnewline \hline
40 & $Ama2$ & $\mathbb{Z}_{2}^{2}$ &  & $\mathbb{Z}_{2}$\tabularnewline \hline
41 & $Aea2$ & $\mathbb{Z}_{2}$ &  & \tabularnewline \hline
42 & $Fmm2$ & $\mathbb{Z}_{2}^{3}$ & $\mathbb{Z}_{2}$ & $\mathbb{Z}_{2}^{2}$\tabularnewline \hline
43 & $Fdd2$ & $\mathbb{Z}_{2}$ &  & \tabularnewline \hline
44 & $Imm2$ & $\mathbb{Z}_{2}^{4}$ & $\mathbb{Z}_{2}^{2}$ & $\mathbb{Z}_{2}^{2}$\tabularnewline \hline
45 & $Iba2$ & $\mathbb{Z}_{2}^{2}$ &  & \tabularnewline \hline
46 & $Ima2$ & $\mathbb{Z}_{2}^{2}$ &  & $\mathbb{Z}_{2}$\tabularnewline \hline
47 & $Pmmm$ & $\mathbb{Z}_{2}^{24}$ & $\mathbb{Z}_{2}^{12}$ & $\mathbb{Z}_{2}^{6}$\tabularnewline \hline
48 & $Pnnn$ & $\mathbb{Z}_{2}^{10}$ &  & \tabularnewline \hline
49 & $Pccm$ & $\mathbb{Z}_{2}^{16}$ &  & $\mathbb{Z}_{2}$\tabularnewline \hline
50 & $Pban$ & $\mathbb{Z}_{2}^{10}$ &  & \tabularnewline \hline
51 & $Pmma$ & $\mathbb{Z}_{2}^{12}$ & $\mathbb{Z}_{2}^{2}$ & $\mathbb{Z}_{2}^{3}$\tabularnewline \hline
52 & $Pnna$ & $\mathbb{Z}_{2}^{4}$ &  & \tabularnewline \hline
53 & $Pmna$ & $\mathbb{Z}_{2}^{9}$ &  & $\mathbb{Z}_{2}$\tabularnewline \hline
54 & $Pcca$ & $\mathbb{Z}_{2}^{5}$ &  & \tabularnewline \hline
55 & $Pbam$ & $\mathbb{Z}_{2}^{8}$ &  & $\mathbb{Z}_{2}^{2}$\tabularnewline \hline
56 & $Pccn$ & $\mathbb{Z}_{2}^{4}$ &  & \tabularnewline \hline
57 & $Pbcm$ & $\mathbb{Z}_{2}^{4}$ &  & $\mathbb{Z}_{2}$\tabularnewline \hline
58 & $Pnnm$ & $\mathbb{Z}_{2}^{8}$ &  & $\mathbb{Z}_{2}$\tabularnewline \hline
59 & $Pmmn$ & $\mathbb{Z}_{2}^{6}$ & $\mathbb{Z}_{2}^{2}$ & $\mathbb{Z}_{2}^{2}$\tabularnewline \hline
60 & $Pbcn$ & $\mathbb{Z}_{2}^{3}$ &  & \tabularnewline \hline
61 & $Pbca$ & $\mathbb{Z}_{2}^{2}$ &  & \tabularnewline \hline
62 & $Pnma$ & $\mathbb{Z}_{2}^{3}$ &  & $\mathbb{Z}_{2}$\tabularnewline \hline
63 & $Cmcm$ & $\mathbb{Z}_{2}^{7}$ & $\mathbb{Z}_{2}$ & $\mathbb{Z}_{2}^{2}$\tabularnewline \hline
64 & $Cmce$ & $\mathbb{Z}_{2}^{6}$ &  & $\mathbb{Z}_{2}$\tabularnewline \hline
65 & $Cmmm$ & $\mathbb{Z}_{2}^{16}$ & $\mathbb{Z}_{2}^{6}$ & $\mathbb{Z}_{2}^{4}$\tabularnewline \hline
66 & $Cccm$ & $\mathbb{Z}_{2}^{12}$ &  & $\mathbb{Z}_{2}$\tabularnewline \hline
67 & $Cmme$ & $\mathbb{Z}_{2}^{14}$ & $\mathbb{Z}_{2}$ & $\mathbb{Z}_{2}^{2}$\tabularnewline \hline
68 & $Ccce$ & $\mathbb{Z}_{2}^{7}$ &  & \tabularnewline \hline
69 & $Fmmm$ & $\mathbb{Z}_{2}^{14}$ & $\mathbb{Z}_{2}^{3}$ & $\mathbb{Z}_{2}^{3}$\tabularnewline \hline
70 & $Fddd$ & $\mathbb{Z}_{2}^{6}$ &  & \tabularnewline \hline
71 & $Immm$ & $\mathbb{Z}_{2}^{13}$ & $\mathbb{Z}_{2}^{6}$ & $\mathbb{Z}_{2}^{3}$\tabularnewline \hline
72 & $Ibam$ & $\mathbb{Z}_{2}^{9}$ &  & $\mathbb{Z}_{2}$\tabularnewline \hline
73 & $Ibca$ & $\mathbb{Z}_{2}^{5}$ &  & \tabularnewline \hline
74 & $Imma$ & $\mathbb{Z}_{2}^{10}$ & $\mathbb{Z}_{2}$ & $\mathbb{Z}_{2}^{2}$\tabularnewline \hline
75 & $P4$ & $\mathbb{Z}_{4}^{2}\times\mathbb{Z}_{2}$ &  & \tabularnewline \hline
76 & $P4_{1}$ &  &  & \tabularnewline
\hline 
\end{tabular}\quad{}%
\begin{tabular}{|c|cccc|}
\hline 
No. & symbol & $\cC_0(G)$ & $\cC_1(G)$ & $\cC_2(G)$\tabularnewline
\hline 
77 & $P4_{2}$ & $\mathbb{Z}_{2}^{3}$ &  & \tabularnewline \hline
78 & $P4_{3}$ &  &  & \tabularnewline \hline
79 & $I4$ & $\mathbb{Z}_{4}\times\mathbb{Z}_{2}$ &  & \tabularnewline \hline
80 & $I4_{1}$ & $\mathbb{Z}_{2}$ &  & \tabularnewline \hline
81 & $P\overline{4}$ & $\mathbb{Z}_{4}^{2}\times\mathbb{Z}_{2}^{3}$ &  & \tabularnewline \hline
82 & $I\overline{4}$ & $\mathbb{Z}_{4}^{2}\times\mathbb{Z}_{2}^{2}$ &  & \tabularnewline \hline
83 & $P4/m$ & $\mathbb{\mathbb{Z}}_{4}^{2}\times\mathbb{Z}_{2}^{10}$ &  & $\mathbb{Z}_{2}^{2}$\tabularnewline \hline
84 & $P4_{2}/m$ & $\mathbb{Z}_{2}^{10}$ &  & $\mathbb{Z}_{2}$\tabularnewline \hline
85 & $P4/n$ & $\mathbb{Z}_{4}^{2}\times\mathbb{Z}_{2}^{3}$ &  & \tabularnewline \hline
86 & $P4_{2}/n$ & $\mathbb{Z}_{4}\times\mathbb{Z}_{2}^{4}$ &  & \tabularnewline \hline
87 & $I4/m$ & $\mathbb{Z}_{4}\times\mathbb{Z}_{2}^{7}$ &  & $\mathbb{Z}_{2}$\tabularnewline \hline
88 & $I4_{1}/a$ & $\mathbb{Z}_{4}\times\mathbb{Z}_{2}^{3}$ &  & \tabularnewline \hline
89 & $P422$ & $\mathbb{Z}_{2}^{12}$ &  & \tabularnewline \hline
90 & $P42_{1}2$ & $\mathbb{Z}_{4}\times\mathbb{Z}_{2}^{4}$ &  & \tabularnewline \hline
91 & $P4_{1}22$ & $\mathbb{Z}_{2}^{3}$ &  & \tabularnewline \hline
92 & $P4_{1}2_{1}2$ & $\mathbb{Z}_{2}$ &  & \tabularnewline \hline
93 & $P4_{2}22$ & $\mathbb{Z}_{2}^{12}$ &  & \tabularnewline \hline
94 & $P4_{2}2_{1}2$ & $\mathbb{Z}_{2}^{5}$ &  & \tabularnewline \hline
95 & $P4_{3}22$ & $\mathbb{Z}_{2}^{3}$ &  & \tabularnewline \hline
96 & $P4_{3}2_{1}2$ & $\mathbb{Z}_{2}$ &  & \tabularnewline \hline
97 & $I422$ & $\mathbb{Z}_{2}^{8}$ &  & \tabularnewline \hline
98 & $I4_{1}22$ & $\mathbb{Z}_{2}^{5}$ &  & \tabularnewline \hline
99 & $P4mm$ & $\mathbb{Z}_{2}^{6}$ & $\mathbb{Z}_{2}^{3}$ & $\mathbb{Z}_{2}^{3}$\tabularnewline \hline
100 & $P4bm$ & $\mathbb{Z}_{4}\times\mathbb{Z}_{2}^{2}$ & $\mathbb{Z}_{2}$ & $\mathbb{Z}_{2}$\tabularnewline \hline
101 & $P4_{2}cm$ & $\mathbb{Z}_{2}^{3}$ & $\mathbb{Z}_{2}^{2}$ & $\mathbb{Z}_{2}$\tabularnewline \hline
102 & $P4_{2}nm$ & $\mathbb{Z}_{2}^{3}$ & $\mathbb{Z}_{2}$ & $\mathbb{Z}_{2}$\tabularnewline \hline
103 & $P4cc$ & $\mathbb{Z}_{2}^{3}$ &  & \tabularnewline \hline
104 & $P4nc$ & $\mathbb{Z}_{4}\times\mathbb{Z}_{2}$ &  & \tabularnewline \hline
105 & $P4_{2}mc$ & $\mathbb{Z}_{2}^{4}$ & $\mathbb{Z}_{2}^{3}$ & $\mathbb{Z}_{2}^{2}$\tabularnewline \hline
106 & $P4_{2}bc$ & $\mathbb{Z}_{2}^{2}$ &  & \tabularnewline \hline
107 & $I4mm$ & $\mathbb{Z}_{2}^{3}$ & $\mathbb{Z}_{2}^{2}$ & $\mathbb{Z}_{2}^{2}$\tabularnewline \hline
108 & $I4cm$ & $\mathbb{Z}_{2}^{2}$ & $\mathbb{Z}_{2}$ & $\mathbb{Z}_{2}$\tabularnewline \hline
109 & $I4_{1}md$ & $\mathbb{Z}_{2}^{2}$ & $\mathbb{Z}_{2}$ & $\mathbb{Z}_{2}$\tabularnewline \hline
110 & $I4_{1}cd$ & $\mathbb{Z}_{2}$ &  & \tabularnewline \hline
111 & $P\overline{4}2m$ & $\mathbb{Z}_{2}^{10}$ & $\mathbb{Z}_{2}^{2}$ & $\mathbb{Z}_{2}$\tabularnewline \hline
112 & $P\overline{4}2c$ & $\mathbb{Z}_{2}^{10}$ &  & \tabularnewline \hline
113 & $P\overline{4}2_{1}m$ & $\mathbb{Z}_{4}\times\mathbb{Z}_{2}^{3}$ & $\mathbb{Z}_{2}$ & $\mathbb{Z}_{2}$\tabularnewline \hline
114 & $P\overline{4}2_{1}c$ & $\mathbb{Z}_{4}\times\mathbb{Z}_{2}^{2}$ &  & \tabularnewline
\hline 
\end{tabular}

\caption{The classification of those $d=3$ bosonic cSPT phases built from lower-dimensional SPT states, for all 230 space groups. The
first and second columns list the number and short international symbol
of the space groups, followed by the classification of  phases built from zero-, one- and two-dimensional blocks, 
labeled respectively by $\cC_0(G)$, $\cC_1(G)$ and $\cC_2(G)$.  Trivial classifications are denoted by blank space.
(Continued on next page.)}

\label{table:B-cSPT}
\end{table*}

\addtocounter{table}{-1}

\begin{table*}[h]
\begin{tabular}{|c|cccc|}
\hline 
No. & symbol & $\cC_0(G)$ & $\cC_1(G)$ & $\cC_2(G)$\tabularnewline
\hline 
115 & $P\overline{4}m2$ & $\mathbb{Z}_{2}^{8}$ & $\mathbb{Z}_{2}^{3}$ & $\mathbb{Z}_{2}^{2}$\tabularnewline \hline
116 & $P\overline{4}c2$ & $\mathbb{Z}_{2}^{7}$ &  & \tabularnewline \hline
117 & $P\overline{4}b2$ & $\mathbb{Z}_{4}\times\mathbb{Z}_{2}^{5}$ &  & \tabularnewline \hline
118 & $P\overline{4}n2$ & $\mathbb{Z}_{4}\times\mathbb{Z}_{2}^{5}$ &  & \tabularnewline \hline
119 & $I\overline{4}m2$ & $\mathbb{Z}_{2}^{6}$ & $\mathbb{Z}_{2}^{2}$ & $\mathbb{Z}_{2}$\tabularnewline \hline
120 & $I\overline{4}c2$ & $\mathbb{Z}_{2}^{6}$ &  & \tabularnewline \hline
121 & $I\overline{4}2m$ & $\mathbb{Z}_{2}^{6}$ & $\mathbb{Z}_{2}$ & $\mathbb{Z}_{2}$\tabularnewline \hline
122 & $I\overline{4}2d$ & $\mathbb{Z}_{4}\times\mathbb{Z}_{2}^{2}$ &  & \tabularnewline \hline
123 & $P4/mmm$ & $\mathbb{Z}_{2}^{18}$ & $\mathbb{Z}_{2}^{9}$ & $\mathbb{Z}_{2}^{5}$\tabularnewline \hline
124 & $P4/mcc$ & $\mathbb{Z}_{2}^{12}$ &  & $\mathbb{Z}_{2}$\tabularnewline \hline
125 & $P4/nbm$ & $\mathbb{Z}_{2}^{11}$ & $\mathbb{Z}_{2}$ & $\mathbb{Z}_{2}$\tabularnewline \hline
126 & $P4/nnc$ & $\mathbb{Z}_{2}^{8}$ &  & \tabularnewline \hline
127 & $P4/mbm$ & $\mathbb{Z}_{4}\times\mathbb{Z}_{2}^{9}$ & $\mathbb{Z}_{2}^{3}$ & $\mathbb{Z}_{2}^{3}$\tabularnewline \hline
128 & $P4/mnc$ & $\mathbb{Z}_{4}\times\mathbb{Z}_{2}^{7}$ &  & $\mathbb{Z}_{2}$\tabularnewline \hline
129 & $P4/nmm$ & $\mathbb{Z}_{2}^{9}$ & $\mathbb{Z}_{2}^{2}$ & $\mathbb{Z}_{2}^{2}$\tabularnewline \hline
130 & $P4/ncc$ & $\mathbb{Z}_{2}^{5}$ &  & \tabularnewline \hline
131 & $P4_{2}/mmc$ & $\mathbb{Z}_{2}^{14}$ & $\mathbb{Z}_{2}^{7}$ & $\mathbb{Z}_{2}^{3}$\tabularnewline \hline
132 & $P4_{2}/mcm$ & $\mathbb{Z}_{2}^{12}$ & $\mathbb{Z}_{2}^{4}$ & $\mathbb{Z}_{2}^{2}$\tabularnewline \hline
133 & $P4_{2}/nbc$ & $\mathbb{Z}_{2}^{8}$ &  & \tabularnewline \hline
134 & $P4_{2}/nnm$ & $\mathbb{Z}_{2}^{11}$ & $\mathbb{Z}_{2}$ & $\mathbb{Z}_{2}$\tabularnewline \hline
135 & $P4_{2}/mbc$ & $\mathbb{Z}_{2}^{7}$ &  & $\mathbb{Z}_{2}$\tabularnewline \hline
136 & $P4_{2}/mnm$ & $\mathbb{Z}_{2}^{9}$ & $\mathbb{Z}_{2}^{3}$ & $\mathbb{Z}_{2}^{2}$\tabularnewline \hline
137 & $P4_{2}/nmc$ & $\mathbb{Z}_{2}^{5}$ & $\mathbb{Z}_{2}^{2}$ & $\mathbb{Z}_{2}$\tabularnewline \hline
138 & $P4_{2}/ncm$ & $\mathbb{Z}_{2}^{8}$ & $\mathbb{Z}_{2}$ & $\mathbb{Z}_{2}$\tabularnewline \hline
139 & $I4/mmm$ & $\mathbb{Z}_{2}^{12}$ & $\mathbb{Z}_{2}^{5}$ & $\mathbb{Z}_{2}^{3}$\tabularnewline \hline
140 & $I4/mcm$ & $\mathbb{Z}_{2}^{10}$ & $\mathbb{Z}_{2}^{2}$ & $\mathbb{Z}_{2}^{2}$\tabularnewline \hline
141 & $I4_{1}/amd$ & $\mathbb{Z}_{2}^{7}$ & $\mathbb{Z}_{2}$ & $\mathbb{Z}_{2}$\tabularnewline \hline
142 & $I4_{1}/acd$ & $\mathbb{Z}_{2}^{5}$ &  & \tabularnewline \hline
143 & $P3$ & $\mathbb{Z}_{3}^{3}$ &  & \tabularnewline \hline
144 & $P3_{1}$ &  &  & \tabularnewline \hline
145 & $P3_{2}$ &  &  & \tabularnewline \hline
146 & $R3$ & $\mathbb{Z}_{3}$ &  & \tabularnewline \hline
147 & $P\overline{3}$ & $\mathbb{Z}_{3}^{2}\times\mathbb{Z}_{2}^{4}$ &  & \tabularnewline \hline
148 & $R\overline{3}$ & $\mathbb{Z}_{3}\times\mathbb{Z}_{2}^{4}$ &  & \tabularnewline \hline
149 & $P312$ & $\mathbb{Z}_{2}^{2}$ &  & \tabularnewline \hline
150 & $P321$ & $\mathbb{Z}_{3}\times\mathbb{Z}_{2}^{2}$ &  & \tabularnewline \hline
151 & $P3_{1}12$ & $\mathbb{Z}_{2}^{2}$ &  & \tabularnewline \hline
152 & $P3_{1}21$ & $\mathbb{Z}_{2}^{2}$ &  & \tabularnewline \hline
153 & $P3_{2}12$ & $\mathbb{Z}_{2}^{2}$ &  & \tabularnewline
\hline 
\end{tabular}\quad{}%
\begin{tabular}{|c|cccc|}
\hline 
No. & symbol & $\cC_0(G)$ & $\cC_1(G)$ & $\cC_2(G)$\tabularnewline
\hline 
154 & $P3_{2}21$ & $\mathbb{Z}_{2}^{2}$ &  & \tabularnewline \hline
155 & $R32$ & $\mathbb{Z}_{2}^{2}$ &  & \tabularnewline \hline
156 & $P3m1$ & $\mathbb{Z}_{2}$ &  & $\mathbb{Z}_{2}$\tabularnewline \hline
157 & $P31m$ & $\mathbb{Z}_{3}\times\mathbb{Z}_{2}$ &  & $\mathbb{Z}_{2}$\tabularnewline \hline
158 & $P3c1$ &  &  & \tabularnewline \hline
159 & $P31c$ & $\mathbb{Z}_{3}$ &  & \tabularnewline \hline
160 & $R3m$ & $\mathbb{Z}_{2}$ &  & $\mathbb{Z}_{2}$\tabularnewline \hline
161 & $R3c$ &  &  & \tabularnewline \hline
162 & $P\overline{3}1m$ & $\mathbb{Z}_{2}^{8}$ &  & $\mathbb{Z}_{2}$\tabularnewline \hline
163 & $P\overline{3}1c$ & $\mathbb{Z}_{2}^{3}$ &  & \tabularnewline \hline
164 & $P\overline{3}m1$ & $\mathbb{Z}_{2}^{8}$ &  & $\mathbb{Z}_{2}$\tabularnewline \hline
165 & $P\overline{3}c1$ & $\mathbb{Z}_{2}^{3}$ &  & \tabularnewline \hline
166 & $R\overline{3}m$ & $\mathbb{Z}_{2}^{8}$ &  & $\mathbb{Z}_{2}$\tabularnewline \hline
167 & $R\overline{3}c$ & $\mathbb{Z}_{2}^{3}$ &  & \tabularnewline \hline
168 & $P6$ & $\mathbb{Z}_{3}^{2}\times\mathbb{Z}_{2}^{2}$ &  & \tabularnewline \hline
169 & $P6_{1}$ &  &  & \tabularnewline \hline
170 & $P6_{5}$ &  &  & \tabularnewline \hline
171 & $P6_{2}$ & $\mathbb{Z}_{2}^{2}$ &  & \tabularnewline \hline
172 & $P6_{4}$ & $\mathbb{Z}_{2}^{2}$ &  & \tabularnewline \hline
173 & $P6_{3}$ & $\mathbb{Z}_{3}^{2}$ &  & \tabularnewline \hline
174 & $P\overline{6}$ & $\mathbb{Z}_{3}^{3}\times\mathbb{Z}_{2}^{2}$ &  & $\mathbb{Z}_{2}^{2}$\tabularnewline \hline
175 & $P6/m$ & $\mathbb{Z}_{3}^{2}\times\mathbb{Z}_{2}^{8}$ &  & $\mathbb{Z}_{2}^{2}$\tabularnewline \hline
176 & $P6_{3}/m$ & $\mathbb{Z}_{3}^{2}\times\mathbb{Z}_{2}^{3}$ &  & $\mathbb{Z}_{2}$\tabularnewline \hline
177 & $P622$ & $\mathbb{Z}_{2}^{8}$ &  & \tabularnewline \hline
178 & $P6_{1}22$ & $\mathbb{Z}_{2}^{2}$ &  & \tabularnewline \hline
179 & $P6_{5}22$ & $\mathbb{Z}_{2}^{2}$ &  & \tabularnewline \hline
180 & $P6_{2}22$ & $\mathbb{Z}_{2}^{8}$ &  & \tabularnewline \hline
181 & $P6_{4}22$ & $\mathbb{Z}_{2}^{8}$ &  & \tabularnewline \hline
182 & $P6_{3}22$ & $\mathbb{Z}_{2}^{2}$ &  & \tabularnewline \hline
183 & $P6mm$ & $\mathbb{Z}_{2}^{4}$ & $\mathbb{Z}_{2}^{2}$ & $\mathbb{Z}_{2}^{2}$\tabularnewline \hline
184 & $P6cc$ & $\mathbb{Z}_{2}^{2}$ &  & \tabularnewline \hline
185 & $P6_{3}cm$ & $\mathbb{Z}_{2}$ &  & $\mathbb{Z}_{2}$\tabularnewline \hline
186 & $P6_{3}mc$ & $\mathbb{Z}_{2}$ &  & $\mathbb{Z}_{2}$\tabularnewline \hline
187 & $P\overline{6}m2$ & $\mathbb{Z}_{2}^{4}$ & $\mathbb{Z}_{2}^{2}$ & $\mathbb{Z}_{2}^{3}$\tabularnewline \hline
188 & $P\overline{6}c2$ & $\mathbb{Z}_{2}^{2}$ &  & $\mathbb{Z}_{2}$\tabularnewline \hline
189 & $P\overline{6}2m$ & $\mathbb{Z}_{3}\times\mathbb{Z}_{2}^{4}$ & $\mathbb{Z}_{2}^{2}$ & $\mathbb{Z}_{2}^{3}$\tabularnewline \hline
190 & $P\overline{6}2c$ & $\mathbb{Z}_{3}\times\mathbb{Z}_{2}^{2}$ &  & $\mathbb{Z}_{2}$\tabularnewline \hline
191 & $P6/mmm$ & $\mathbb{Z}_{2}^{12}$ & $\mathbb{Z}_{2}^{6}$ & $\mathbb{Z}_{2}^{4}$\tabularnewline \hline
192 & $P6/mcc$ & $\mathbb{Z}_{2}^{8}$ &  & $\mathbb{Z}_{2}$\tabularnewline
\hline 
\end{tabular}\quad{}%
\begin{tabular}{|c|cccc|}
\hline 
No. & symbol & $\cC_0(G)$ & $\cC_1(G)$ & $\cC_2(G)$\tabularnewline
\hline 
193 & $P6_{3}/mcm$ & $\mathbb{Z}_{2}^{6}$ & $\mathbb{Z}_{2}$ & $\mathbb{Z}_{2}^{2}$\tabularnewline \hline
194 & $P6_{3}/mmc$ & $\mathbb{Z}_{2}^{6}$ & $\mathbb{Z}_{2}$ & $\mathbb{Z}_{2}^{2}$\tabularnewline \hline
195 & $P23$ & $\mathbb{Z}_{3}\times\mathbb{Z}_{2}^{4}$ &  & \tabularnewline \hline
196 & $F23$ & $\mathbb{Z}_{3}$ &  & \tabularnewline \hline
197 & $I23$ & $\mathbb{Z}_{3}\times\mathbb{Z}_{2}^{2}$ &  & \tabularnewline \hline
198 & $P2_{1}3$ & $\mathbb{Z}_{3}$ &  & \tabularnewline \hline
199 & $I2_{1}3$ & $\mathbb{Z}_{3}\times\mathbb{Z}_{2}$ &  & \tabularnewline \hline
200 & $Pm\overline{3}$ & $\mathbb{Z}_{3}\times\mathbb{Z}_{2}^{8}$ & $\mathbb{Z}_{2}^{4}$ & $\mathbb{Z}_{2}^{2}$\tabularnewline \hline
201 & $Pn\overline{3}$ & $\mathbb{Z}_{3}\times\mathbb{Z}_{2}^{4}$ &  & \tabularnewline \hline
202 & $Fm\overline{3}$ & $\mathbb{Z}_{3}\times\mathbb{Z}_{2}^{4}$ & $\mathbb{Z}_{2}$ & $\mathbb{Z}_{2}$\tabularnewline \hline
203 & $Fd\overline{3}$ & $\mathbb{Z}_{3}\times\mathbb{Z}_{2}^{2}$ &  & \tabularnewline \hline
204 & $Im\overline{3}$ & $\mathbb{Z}_{3}\times\mathbb{Z}_{2}^{5}$ & $\mathbb{Z}_{2}^{2}$ & $\mathbb{Z}_{2}$\tabularnewline \hline
205 & $Pa\overline{3}$ & $\mathbb{Z}_{3}\times\mathbb{Z}_{2}^{2}$ &  & \tabularnewline \hline
206 & $Ia\overline{3}$ & $\mathbb{Z}_{3}\times\mathbb{Z}_{2}^{3}$ &  & \tabularnewline \hline
207 & $P432$ & $\mathbb{Z}_{2}^{6}$ &  & \tabularnewline \hline
208 & $P4_{2}32$ & $\mathbb{Z}_{2}^{6}$ &  & \tabularnewline \hline
209 & $F432$ & $\mathbb{Z}_{2}^{4}$ &  & \tabularnewline \hline
210 & $F4_{1}32$ & $\mathbb{Z}_{2}$ &  & \tabularnewline \hline
211 & $I432$ & $\mathbb{Z}_{2}^{5}$ &  & \tabularnewline \hline
212 & $P4_{3}32$ & $\mathbb{Z}_{2}$ &  & \tabularnewline \hline
213 & $P4_{1}32$ & $\mathbb{Z}_{2}$ &  & \tabularnewline \hline
214 & $I4_{1}32$ & $\mathbb{Z}_{2}^{4}$ &  & \tabularnewline \hline
215 & $P\overline{4}3m$ & $\mathbb{Z}_{2}^{4}$ & $\mathbb{Z}_{2}^{2}$ & $\mathbb{Z}_{2}$\tabularnewline \hline
216 & $F\overline{4}3m$ & $\mathbb{Z}_{2}^{2}$ & $\mathbb{Z}_{2}^{2}$ & $\mathbb{Z}_{2}$\tabularnewline \hline
217 & $I\overline{4}3m$ & $\mathbb{Z}_{2}^{3}$ & $\mathbb{Z}_{2}$ & $\mathbb{Z}_{2}$\tabularnewline \hline
218 & $P\overline{4}3n$ & $\mathbb{Z}_{2}^{4}$ &  & \tabularnewline \hline
219 & $F\overline{4}3c$ & $\mathbb{Z}_{2}^{2}$ &  & \tabularnewline \hline
220 & $I\overline{4}3d$ & $\mathbb{Z}_{4}\times\mathbb{Z}_{2}$ &  & \tabularnewline \hline
221 & $Pm\overline{3}m$ & $\mathbb{Z}_{2}^{10}$ & $\mathbb{Z}_{2}^{5}$ & $\mathbb{Z}_{2}^{3}$\tabularnewline \hline
222 & $Pn\overline{3}n$ & $\mathbb{Z}_{2}^{5}$ &  & \tabularnewline \hline
223 & $Pm\overline{3}n$ & $\mathbb{Z}_{2}^{6}$ & $\mathbb{Z}_{2}^{3}$ & $\mathbb{Z}_{2}$\tabularnewline \hline
224 & $Pn\overline{3}m$ & $\mathbb{Z}_{2}^{8}$ & $\mathbb{Z}_{2}$ & $\mathbb{Z}_{2}$\tabularnewline \hline
225 & $Fm\overline{3}m$ & $\mathbb{Z}_{2}^{7}$ & $\mathbb{Z}_{2}^{4}$ & $\mathbb{Z}_{2}^{2}$\tabularnewline \hline
226 & $Fm\overline{3}c$ & $\mathbb{Z}_{2}^{5}$ & $\mathbb{Z}_{2}$ & $\mathbb{Z}_{2}$\tabularnewline \hline
227 & $Fd\overline{3}m$ & $\mathbb{Z}_{2}^{5}$ & $\mathbb{Z}_{2}$ & $\mathbb{Z}_{2}$\tabularnewline \hline
228 & $Fd\overline{3}c$ & $\mathbb{Z}_{2}^{3}$ &  & \tabularnewline \hline
229 & $Im\overline{3}m$ & $\mathbb{Z}_{2}^{8}$ & $\mathbb{Z}_{2}^{3}$ & $\mathbb{Z}_{2}^{2}$\tabularnewline \hline
230 & $Ia\overline{3}d$ & $\mathbb{Z}_{2}^{4}$ &  & \tabularnewline \hline
 &  &  &  & \tabularnewline
\hline 
\end{tabular}

\caption{(Continued from previous page.) The classification of those $d=3$ bosonic cSPT phases built from lower-dimensional SPT states, for all 230 space groups. The
first and second columns list the number and short international symbol
of the space groups, followed by the classification of  phases built from zero-, one- and two-dimensional blocks, 
labeled respectively by $\cC_0(G)$, $\cC_1(G)$ and $\cC_2(G)$.  Trivial classifications are denoted by blank space.}

\end{table*}

\bibliography{cSPT_packing}

\end{document}